\documentclass[aps,pre,twocolumn,superscriptaddress,floatfix,amsmath,amssymb,]{revtex4} 
\usepackage{placeins}
\usepackage{graphicx}
\usepackage{xcolor}
\usepackage{dcolumn}
\usepackage{bm}

\usepackage[utf8]{inputenc}
\usepackage[T1]{fontenc}
\usepackage{mathptmx}
\usepackage{etoolbox}
\usepackage{soul}
\usepackage{caption}
\usepackage{subcaption}
\usepackage[normalem]{ulem}

\begin{document}

\title{Exploring  Self-Organization of Charged Dust Dimers in Plasma}

\author{Aman Singh Katariya}
\email {amansk362@gmail.com} 
\affiliation{Department of Physics, Indian Institute of Technology Delhi, Hauz Khas, New Delhi 110016, India}

\author{Amita Das}
\email {amita@iitd.ac.in}
\affiliation{Department of Physics, Indian Institute of Technology Delhi, Hauz Khas, New Delhi 110016, India}

\author{Mamta Yadav}
\affiliation{Department of Physics, Indian Institute of Technology Delhi, Hauz Khas, New Delhi 110016, India}

\author{Bibhuti Bhusan Sahu}
\affiliation{Department of Energy Sciences and Engineering, Indian Institute of Technology Delhi, Hauz Khas, New Delhi 110016, India}

\begin{abstract} 

We investigate the self-organization of charged dust dimers in plasma using Molecular Dynamics (MD) simulations, with emphasis on both positional and orientational ordering. For a finite number of dimers confined by a radial electric field, the system evolves from simple arrangements to ring-like structures as the particle number increases. These rings exhibit diverse orientational states, including radial, transverse, and mixed alignments of the dimer axis, reflecting a strong coupling between spatial confinement and orientational degrees of freedom. For larger systems studied under periodic boundary conditions, bulk-like behavior emerges with coupled positional and orientational correlations. The results highlight the significance of anisotropy in determining equilibrium structures and demonstrate that orientational order plays a crucial role alongside positional ordering in complex plasmas with shaped particles. This work provides motivation for experimental studies involving shaped dust particles to explore orientational ordering phenomena beyond conventional spherical dust particle systems.
\end{abstract}
\maketitle

\section{ Introduction}
The dusty plasma system is an ideal and simple test bed for studying the emergence of macroscopic order from intrinsic microscopic interactions. The phenomena of crystallization and melting can be visually seen to be occurring in such a setup \cite{thomas1994plasma, thomas1996melting, ichiki2004}. In systems far from thermal equilibrium, a self-organization process occurs as the system tends to move toward the equilibrium \cite{nicolis1977self}. As self-organization happens, structural changes occur, leading to many fascinating phenomena, such as pattern formation and phase transitions \cite{HAKEN198426}, defects in liquid crystals \cite{de1993physics}, and collective behavior \cite{VICSEK201271}, to name a few. Its impact spans a variety of fields, including natural sciences as well as social sciences. Complex systems showing chaotic or turbulent behavior can undergo self-organization to reach an equilibrium in their deterministic space \cite{DeterministicNonperiodicFlow}. In biological ecosystems, many such examples can be seen, such as pattern formation in ant colonies \cite{antcolonies}, swarm behavior in insects \cite{Beekman2008, szopekhoneybees}, bird flocking \cite{birdflock_reynolds, bajec2009organizedbirds}, etc. Examples can be found in even cellular structures in  protein systems \cite{Whitesides2002SelfAssemblyAA}, bacterial colonies \cite{cizrok1996,sokolov2007,cizrok2001}, and more. In chemical systems, self-organization can lead to the formation of turing patterns \cite{turing, Tompkins2014}, oscillation reactions such as the Belousov-Zhabotinsky Reaction \cite{BZR, BZRinLPD}, pH oscillators \cite{Epstein_Pojman_1998}, autocatalytic reactions \cite{prigogine&lefever}, self-organized criticality \cite{PBak1987}, liquid crystals \cite{de1993physics}, etc. More often, self-organization in chemical systems appears in the form of self-assemblies, such as in molecular assembly \cite{Whitesides2002SelfAssemblyAA}, layer assemblies\cite{decher1997fuzzy}, colloid crystals \cite{pieranski1983colloidal}, grid complexes \cite{ruben2004grid}, etc. It can also be seen in animal social behavior through large organized networks in insects foraging, protecting colonies, etc. \cite{Seeley1995TheWO, szopekhoneybees}, and even in human societies in trade routes \cite{krugman1996self}, traffic networks \cite{Helbing2001traffic}, etc.

Self-organization happens when the internal order of the systems increases with time, and as a result, different properties start to emerge \cite{nicolis1977self}. Since the order begins to increase, the entropy will likely go down, violating the second law of thermodynamics \cite{glansdorff1971thermodynamic}. Instead, the system draws energy from the surroundings. This results in a decrease of free energy in the system, allowing for self-organization to happen in the system \cite{Whitesides2002SelfAssemblyAA}. In physical systems, such changes are often accompanied by phase transitions or changes in the symmetry of the system \cite{HAKEN198426}. A wonderful example of this is the formation of snowflake crystals with very symmetrical patterns, which can change depending on different conditions \cite{Libbrecht2005snow}. More examples can be found in magnetization \cite{stanley1971introduction}, heat convection cells (B\'enard cells) \cite{koschmieder1993benard}, geophysical disturbances \cite{Hallet1990-qa}, formation of plasma structures \cite{ivlev2000anisotropic,moefillivlev, deshwal2022chaotic, maity2020dynamical}, etc. Self-organization, although ubiquitous, yet remains fully understood. Various models have been developed to explain this phenomenon on different scales \cite{cross&hohenberg, degond2014collective}.

A complex system, such as the dusty plasma, can provide suitable conditions to study self-organization and its after-effects. The formation of coulomb clusters has been observed in atomic systems by trapping ions in a Paul rf-trap \cite{wineland1987atomic, diedrich1987observation, mortensen2006observation}. In 1994, experimental observations of crystal formation in dusty plasmas were made by Chu and Lin in silane plasma \cite{chu1994direct}, by Thomas, Morfill, and Demmel in argon plasma using melamine dust particles \cite{thomas1994plasma}, and by Hayashi and Tachibana in methane plasma, where carbon particles grow to form coulombic crystals \cite{hayashi1994observation}. Later, many studies have also been done on dusty plasma \cite{Feng2016, feng2010viscoelasticity, melzer2021physics, sato2001dynamics, PhysRevB.51.7700, KATARIYA2025134692, melzer2003mode, lisina2019dynamic, melzer2019cluster}. Studies have been done to study the formation of strongly coupled dust crystals in a dusty plasma medium \cite{lai-lin, yadav2023structure, yadav2025bulk, YADAV2024134326, maity2020dynamical, maity2019molecular, deshwal2022chaotic}. In \cite{lai-lin}, topological defects that arise due to the tight packaging of the coulomb crystals are studied. A study of ground state configurations and their eigenmodes with respect to the screening parameter has been made in reference \cite{astrakharchik1999properties}. Spectral properties have been studied in two-dimensional Coulomb clusters in references\cite{PhysRevB.51.7700, melzer2003mode}. Dynamical properties and phase transition studies for two-dimensional clusters have been presented in references \cite{deshwal2022chaotic, astrakharchik1999two, lisina2019dynamic}. The above studies have considered point-like dust particles resembling spherical dust particles. However, there are observations of differently shaped dust particles in experiments \cite{lowen1994charged, molotkov2000liquid, annaratone2001levitate}. Theoretical models and numerical models have been discussed for cluster formation in rod-like/cylindrical dust particles in plasma \cite{ivlev2003rodtheory, lisina2016spatial, vaulina2016formation}. 

In this work, we focus on the orientational order of dust particles with an anisotropic body symmetry using molecular dynamics simulations. We consider dumbbell-shaped dimers in a strongly coupled regime and confine them using a radial electric field profile. The occurrence of dimer-shaped particles has been observed in several experimental studies ranging from normal plasma setups to thin film deposition devices. Vekeslman et. al. reported the formation of C2 dimers in a DC arc discharge plasma between graphite electrodes \cite{vekselman2018quantitative}. The C2 dimer is formed near the anode surface and acts as a strong precursor for the formation of nanotubes. The formation of C2 dimers using a pyrolytic graphite rod when laser-ablated using a KrF excimer laser in an argon environment has also been reported \cite{krajnovich1995laserc2, yamagata1999optical, nica2020formation}. Oohara and Hatakeyama reported the formation of fullerene dimers inside a fullerene injection plasma \cite{oohara2003pair}, and Wang et. al.\cite{wang1997synthesis} and Shvartsburg et. al. \cite{shvartsburg1999ball} show that they form ball and chain-like structures that resemble dumbbell shapes. Dimers were also observed forming in sputtering devices \cite{hippler2017pressure, hippler2018formation, curda2023role, bogaerts1999role}. 

We present a detailed study of the self-organization of charged dust dimers immersed in a plasma, using Molecular Dynamics (MD) simulations. The simulation details are presented in Section II. The work focuses on understanding both positional (translational) and orientational (rotational) ordering in such systems, extending beyond the conventional emphasis on positional correlations in complex plasmas.

Two distinct configurations are investigated. In the first, a finite number of dust dimers is confined by an external radial electric field. For small dimer populations, the system exhibits well-defined structural arrangements, which evolve into ring-like configurations as the number of dimers increases. Within these ring structures, the dimers display a wide variety of orientational behavior, including predominantly radial or transverse alignment of the dimer axis, as well as mixed orientational states that emerge with increasing system size and the formation of multiple rings. This demonstrates a strong interplay between confinement, inter-particle interactions, and orientational degrees of freedom. These studies have been presented in Section III.

In the second configuration, presented in Section IV, a larger ensemble of dimers is studied under periodic boundary conditions, allowing exploration of bulk behavior. The results reveal the emergence of collective ordering characterized by coupled positional and orientational correlations, highlighting the role of anisotropic interactions in determining the equilibrium structure. These findings underscore the importance of incorporating orientational degrees of freedom in the study of complex plasmas and provide a theoretical basis for future experimental investigations involving shaped or anisotropic dust particles. The work thus opens new directions for exploring rich phase behavior and self-organization in strongly coupled plasma systems and has been summarized in Section V.

\section{ MD Simulation Details}
\label{mdsim}
\begin{figure}[h!]
    \centering
    \includegraphics[width=0.75\linewidth]{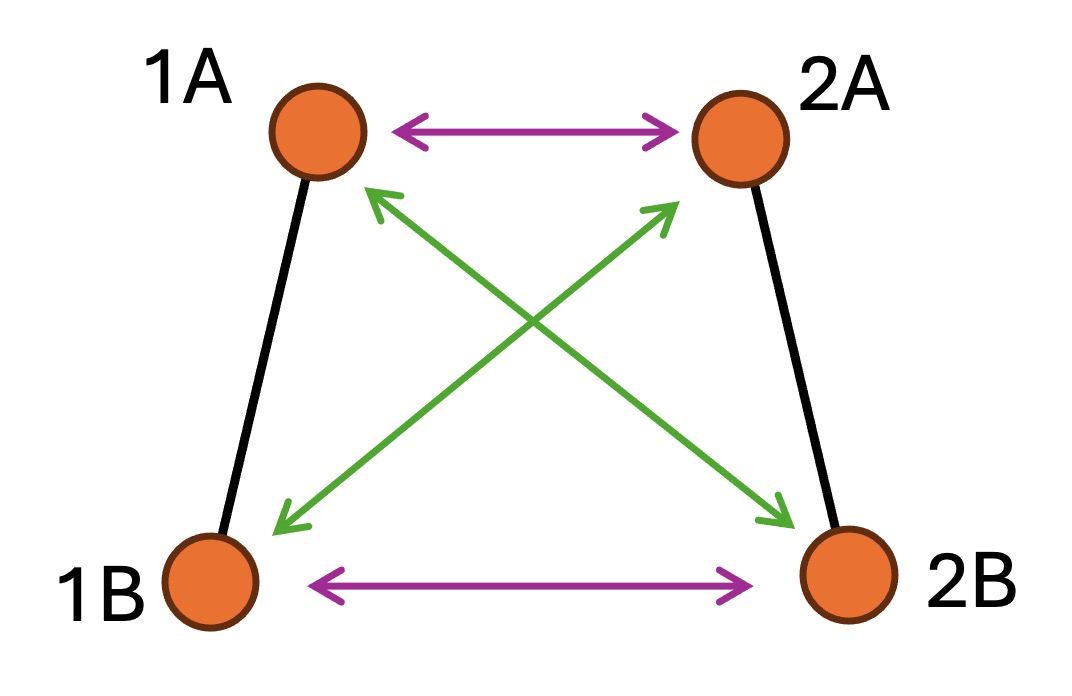}
    \caption{A schematic representation of interactions amongst two dimers. Figure shows two dimers $1$ and $2$. The two ends are represented by $A$ and $B$. The distance between the ends of each dimers are fixed and is depicted by the thick black lines. The double headed green and magenta arrows shows the paired Yukawa interaction operative between the ends of two distinct dimers. }
    \label{fig:dimer_int}
\end{figure}
In this study, we carried out a two-dimensional (2D) Molecular Dynamics (MD) simulation using the open-source code LAMMPS \cite{plimpton1995fast} to simulate the dynamical evolution of shaped dust particles. For simplicity, we consider here charged dust dimers, which break the symmetry of the spherical dust particles that have so far been conventionally studied. The dimers are like  dumbbell having charge accumulation at the ends of the dumbbell. This, in a sense, emulates a cylindrical dust particle with a sharp edge at the two ends where charge can preferentially accumulate.  The charge ($Q_d$) of each bell is chosen to be $11940e$ (where $e$ is the charge of an electron), and the mass ($M_d$) of the dimer is set as $6.99\times10^{-13}$ kg \cite{nosenko2004shear}. The separation between the bells is chosen as $0.3mm$, which is comparable to the debye length of a single bell of the dimer. To avoid any pre-correlations among the dimers, they are placed randomly in the simulation box. The negatively charged dimers interact with each other through a Yukawa potential or the screened Coulomb potential, incorporating the response of the background lighter electron and ion species of the plasma medium \cite{Konopka2000Measurement, ShuklaMamun2002Intro}. The Yukawa potential is of the form $U(r)= (Q_d/4\pi\epsilon_or)exp(-\kappa r)$. Where $Q_d$ and $\kappa$ are the charges on the dust particles and the inverse of screening length for the dust charge, respectively. A typical screening length for Yukawa pair interaction between dust particles is of the order of plasma Debye length, $\lambda_D$. All the length scales are normalized by the inter-dust grain distance ($a= 2.285\times10^{-3}$m). The normalized value for our screening parameter thus becomes $\kappa = a/\lambda_D$. The simulation box expands for $L_x = L_y= 12.7943a$ in the x and y direction. A confining potential is applied radially to the simulation box to confine the repelling dimers.

The total force acting on any $l^{th}$ particle at any time is the sum of the Yukawa interactions by all other particles and the external force due to boundary potential  ($F_E$) 
 \begin{equation}
     \textbf{F}_l = -Q_d \sum_{\substack{i,j=1 \\ i\neq j}}^{N} \nabla U(r_i,r_j) + F_E
     \label{equation of motion}
 \end{equation}
where $N$ is the total number of dust particles, $r_i$ and $r_j$ define the positions of the $i^{th}$ and $j^{th}$ particle. 'Particle' here corresponds to the individual charged bell of the dimer. The bells of the same dimer do not interact with each other through the Yukawa potential but are bound to each other at a rigid distance of $l_d = 0.3mm$. The schematic figure shows the possible interactions between two dimers $1$ and $2$, respectively. The two points of each dimer has been identified with symbols $A$ and $B$. The pair of particles on which the Yukawa potential acts has also been shown in Fig. \ref{fig:dimer_int}. 

As mentioned earlier, we study the organization pattern (both positional and orientational) first for a finite number of dust dimers, increasing them one by one. In this particular case,    
a force $F_E$  radially confines the dimers in the center of the simulation box. This force is represented by an electric field,
\begin{equation}
    E = K\left(x-\frac{L_x}{2}\right) \hat{x} + K\left(y-\frac{L_y}{2}\right) \hat{y}
\end{equation}
where $K$ is the strength of the field. To track the dynamics of the dimers, the simulation time step is set as $0.01\omega_{pd}^{-1}$.   For simulation and analysis purposes, all the time scales are normalized with a typical response time scale of  $\omega_{pd}^{-1}$.  Thermal equilibrium is achieved by using the Nose-Hoover thermostat (NVT) \cite{nose1984molecular, PhysRevA.31.1695}. The Coulomb coupling parameter is much higher than unity.  

 \section {Organization in  small clusters of dimers under a confining electric field}

Molecular Dynamics (MD) simulations were performed to investigate the structural organization of strongly coupled charged dimers confined to a two-dimensional plane. The dimers were initially distributed randomly within a background plasma environment, whose screening effects were modeled through a Yukawa (screened-Coulomb) interaction potential. The system was then allowed to relax under an NVT thermostat until equilibrium was attained. An externally applied radial electric field provided confinement, with its potential minimum located at the center of the simulation domain. The resulting equilibrium configuration emerged from the competition between the mutual Yukawa repulsion, which tends to disperse the dimers, and the confining electric field, which drives them toward the center. This interplay leads to the formation of well-ordered structural arrangements corresponding to states of minimum potential energy. In addition to characterizing these equilibrium structures, we also investigate whether the resulting configurations remain stationary or exhibit persistent dynamical behavior in the equilibrated state, analogous to the collective rotational and oscillatory dynamics reported previously in certain monomeric systems.

\begin{figure}[htbp!]
  \centering
  \begin{subfigure}[h!]{0.45\textwidth}
    \centering
    \includegraphics[width=\textwidth]{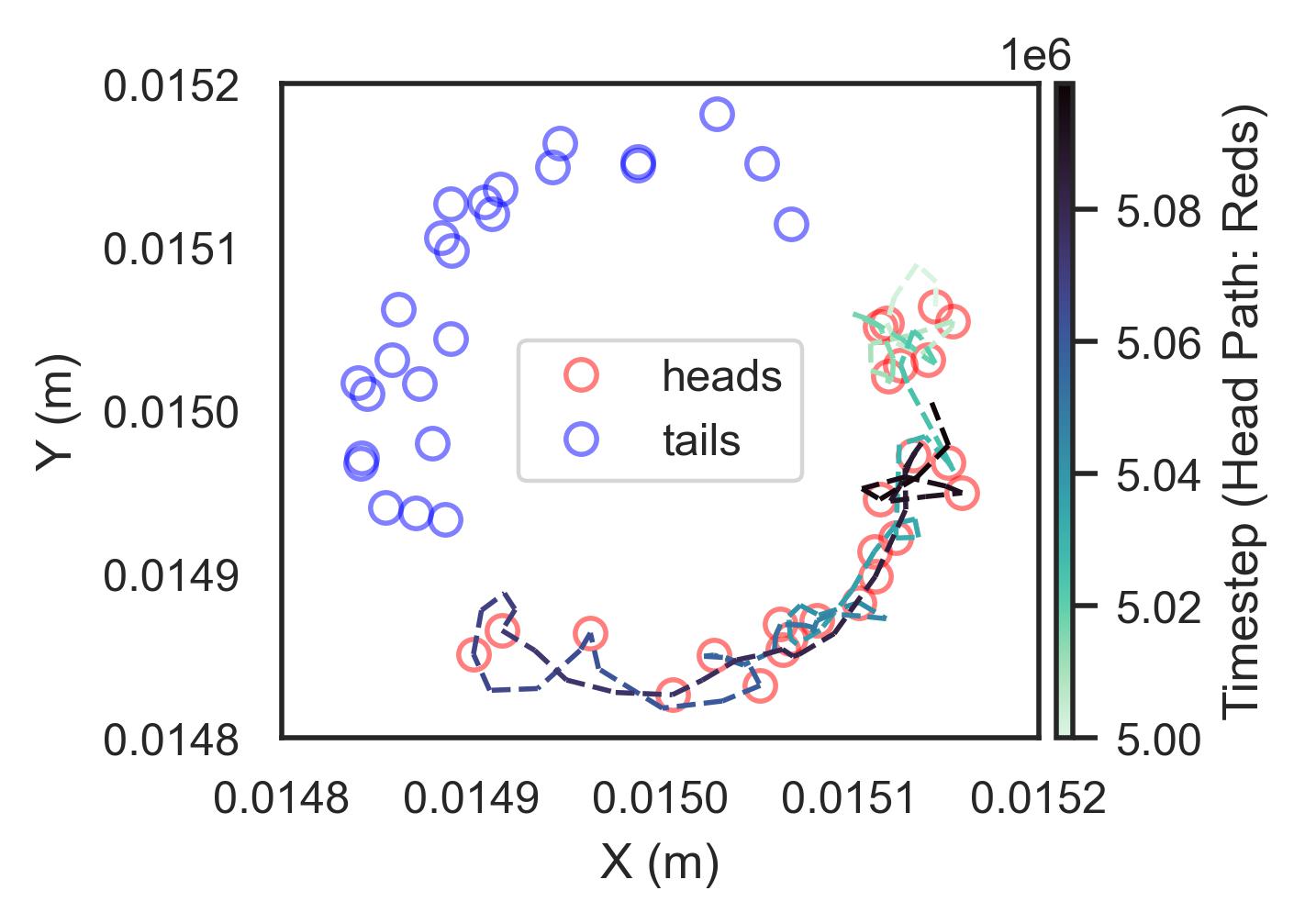}
    \caption{Dynamical evolution of the dimer body. The head (red) of the dimer is traced with time.}
    \label{fig:1dimersub2}
  \end{subfigure}
  \begin{subfigure}[h!]{0.45\textwidth}
    \centering
    \includegraphics[width=\textwidth]{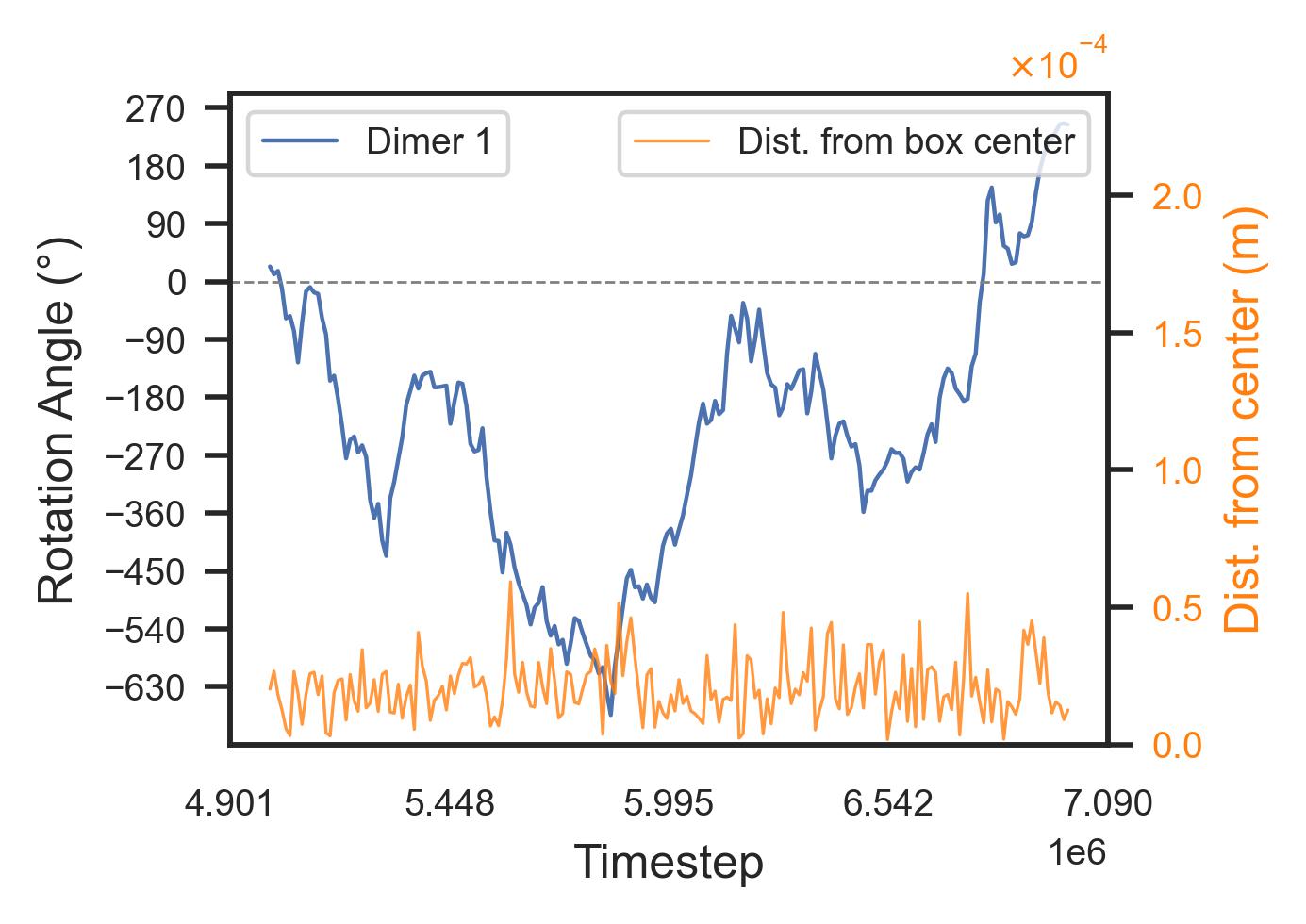}
    \caption{Angle of rotation is plotted for the dimer vector with time on the left. On the right axis, the distance of dimer body center from the center of the radial field is plotted.}
    \label{fig:1dimersub3}
  \end{subfigure}
  \caption{Dynamics of a single dimer confined using a radial electric field.}
  \label{fig:1dimerdynamics}
\end{figure}
We begin by considering the simplest case of a single dimer and investigate its relaxation dynamics within the confining potential. In this configuration, the only external force acting on the constituent particles arises from the radially directed confining electric field. As expected, the center of mass of the dimer migrates toward and eventually settles at the center of the potential well, corresponding to the point of minimum potential energy. However, while the translational degree of freedom is frozen out, the dimer retains complete rotational freedom because its energy is independent of orientation. Consequently, all angular configurations are energetically equivalent. This orientational degeneracy prevents the dimer from reaching a truly stationary equilibrium state. Instead, the dimer continues to undergo rotational motion, as illustrated in Fig.~\ref{fig:1dimerdynamics}. The two ends of the dimer, represented by the blue and red circles, clearly trace out this rotational dynamics. The temporal evolution of the motion is highlighted by the color gradient along the dashed trajectory connecting successive positions of the red particle. The corresponding time evolution of the orientation angle is presented in Fig.~\ref{fig:1dimerdynamics}(b). It is evident that the angle changes sign repeatedly, indicating the occurrence of both clockwise and anticlockwise rotations. Furthermore, the angle frequently exceeds $360^{\circ}$, demonstrating that the dimer often executes complete revolutions rather than merely oscillating about a preferred orientation. In addition to the rotational motion, small radial fluctuations of the center of mass are also observed. The temporal evolution of the radial position is plotted in Fig.~\ref{fig:1dimerdynamics}(b) and is referenced to the right-hand axis. These fluctuations reflect the residual thermal motion of the dimer within the confining potential, superimposed on its persistent rotational dynamics.

\begin{figure}[h!]
  \centering
  \begin{subfigure}[h!]{0.45\textwidth}
    \centering
    \includegraphics[width=\textwidth]{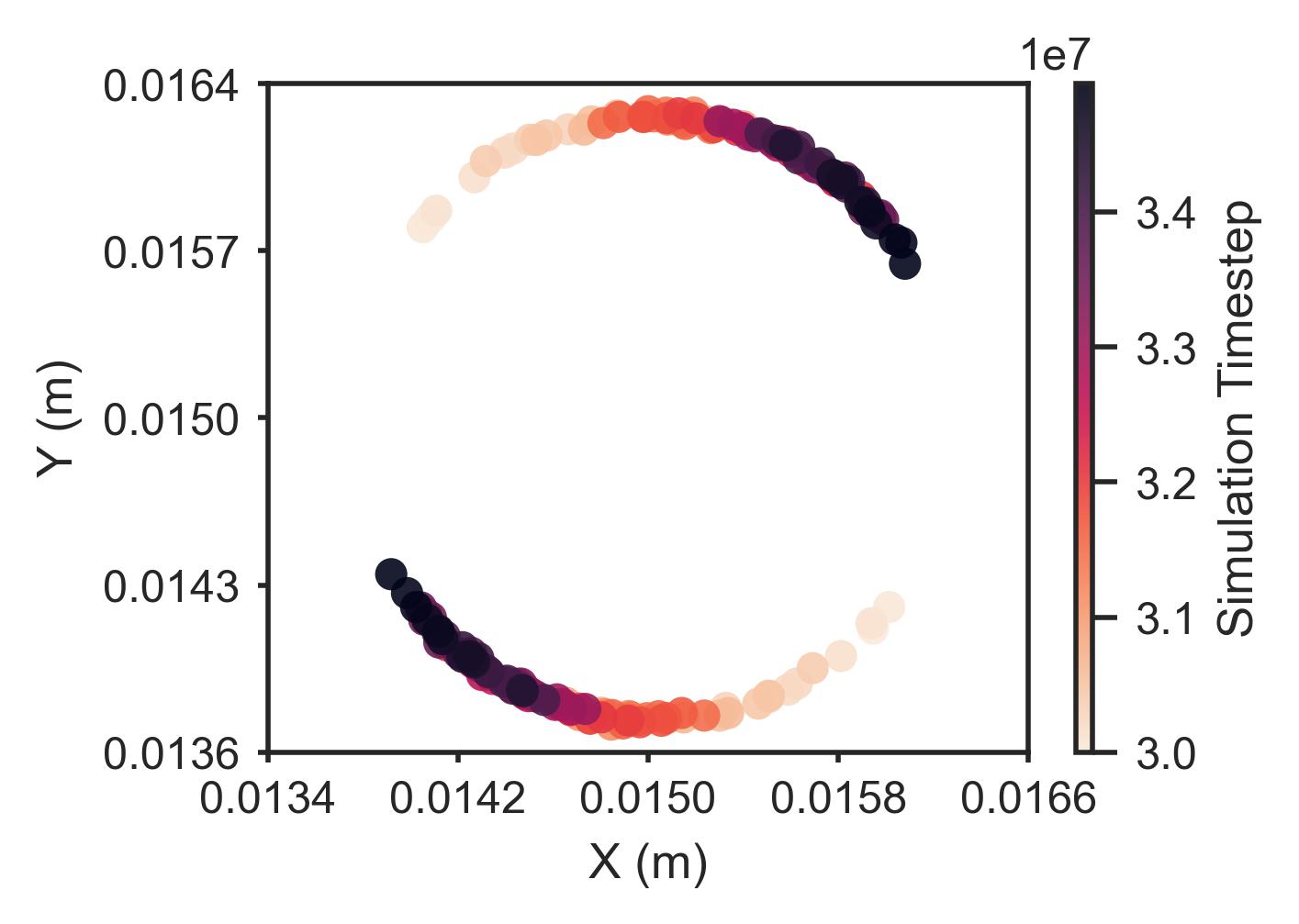}
    \caption{The two body centers of the dimers plotted with evolving timestep. Each dimer travels on a circular path.}
    \label{fig:2dimersub1}
  \end{subfigure}
  \begin{subfigure}[h!]{0.45\textwidth}
    \centering
    \includegraphics[width=\textwidth]{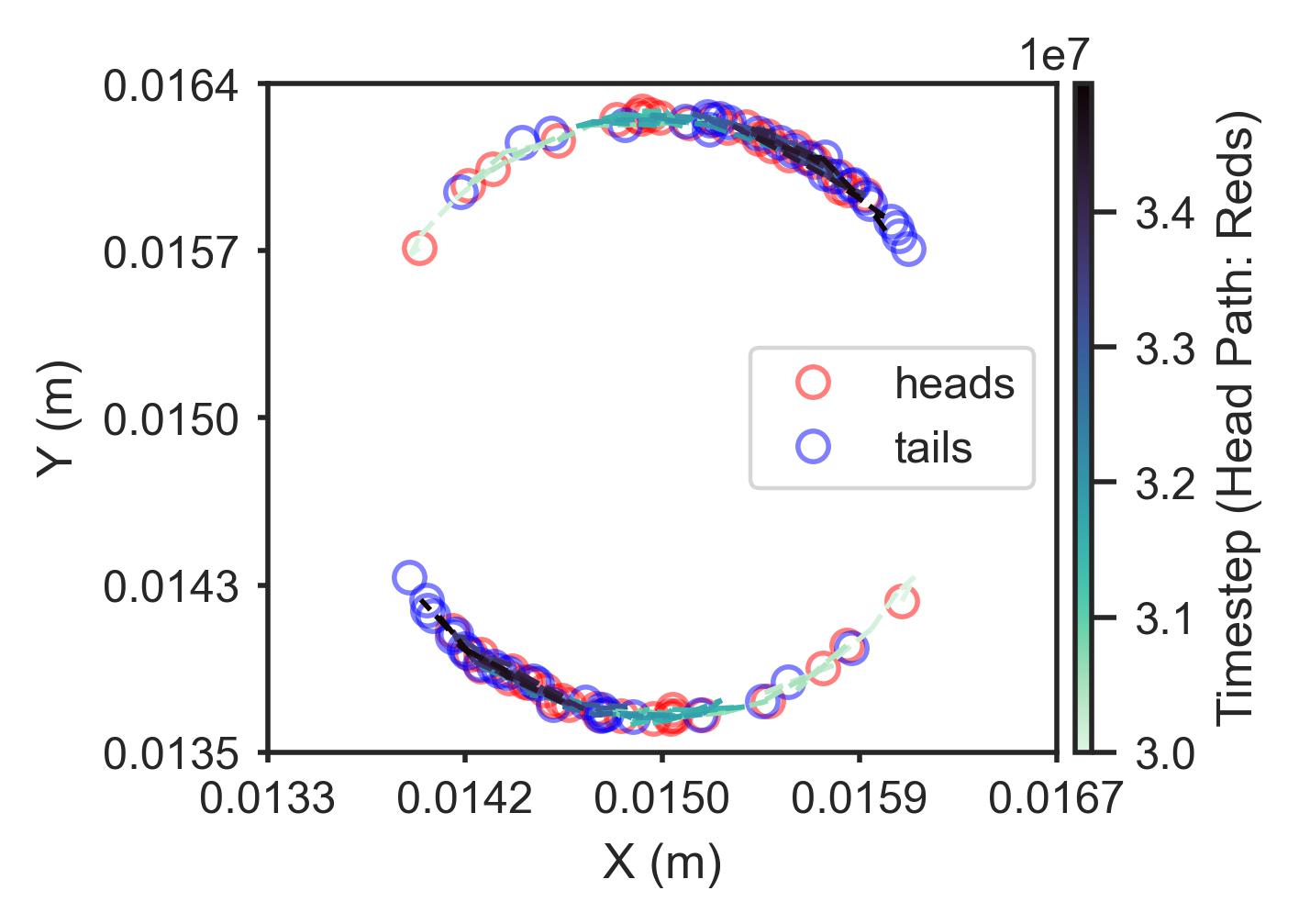}
    \caption{Dynamical evolution of the dimer bodies with increasing timestep. The head (red) of the dimer is traced with time.}
    \label{fig:2dimersub2}
  \end{subfigure}
  \begin{subfigure}[h!]{0.45\textwidth}
    \centering
    \includegraphics[width=\textwidth]{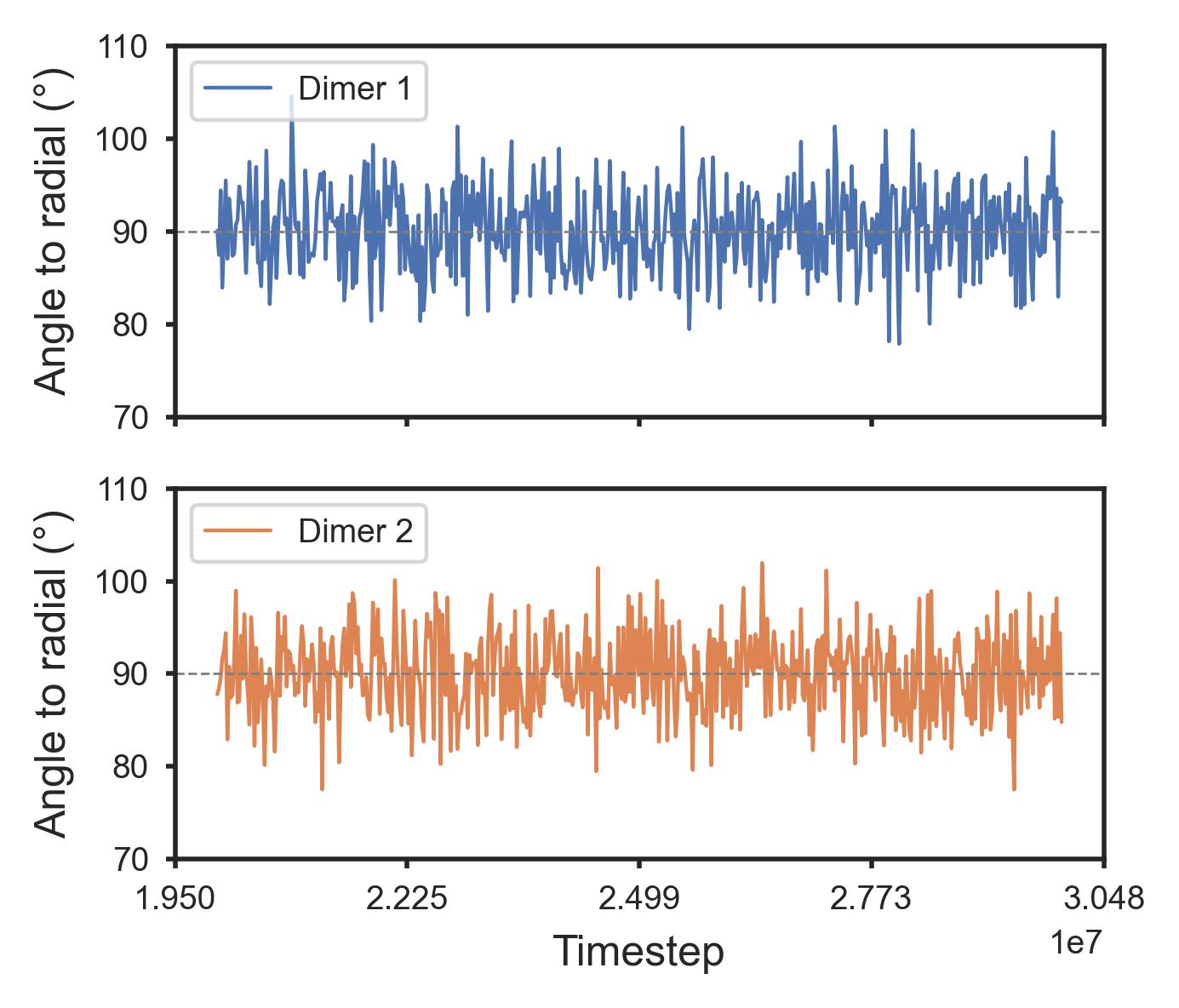}
    \caption{Angle of the dimer vector with respect to the radial direction plotted with time.}
    \label{fig:2dimersub3}
  \end{subfigure}
  \caption{Dynamics of two dimers confined using a radial electric field.}
  \label{fig:2dimerdynamics}
\end{figure}

We next consider the case of two dimers. Unlike the single-dimer configuration, where the dynamics are governed solely by the confining potential, the equilibrium structure and dynamics now result from the interplay between the radial confinement and the mutual interactions between the dimers. The simulations reveal that the centers of mass of the two dimers execute a collective rotational motion along a circular orbit while remaining approximately diametrically opposite to one another, as illustrated in Fig.~\ref{fig:2dimerdynamics}. This arrangement minimizes the mutual repulsion between the dimers while maintaining confinement within the potential well. An additional feature of this configuration is the preferred orientation of the dimer axes. The dimers are found to align predominantly along the tangential direction of the ring rather than along the radial direction. This tendency is quantified by evaluating the angle between the dimer axis and the local radial vector, which remains close to $90^\circ$ throughout the evolution, as shown in Fig.~\ref{fig:2dimersub3}. Thus, while the dimers collectively rotate around the center, their orientations remain largely locked to the tangential direction of the orbit. As the number of dimers is increased to three and four, the qualitative characteristics of the system remain largely unchanged. The dimers self-organize into nearly equispaced angular positions on a single-ring structure, reflecting the balance between confinement and mutual repulsion. At the same time, the orientational ordering persists, with the dimer axes remaining predominantly tangential to the ring. These observations suggest the emergence of a robust collective state characterized by simultaneous positional ordering and orientational alignment, which is maintained even as the number of dimers increases.

\begin{figure}[h!]
  \centering
    \centering
    \includegraphics[width=\linewidth]{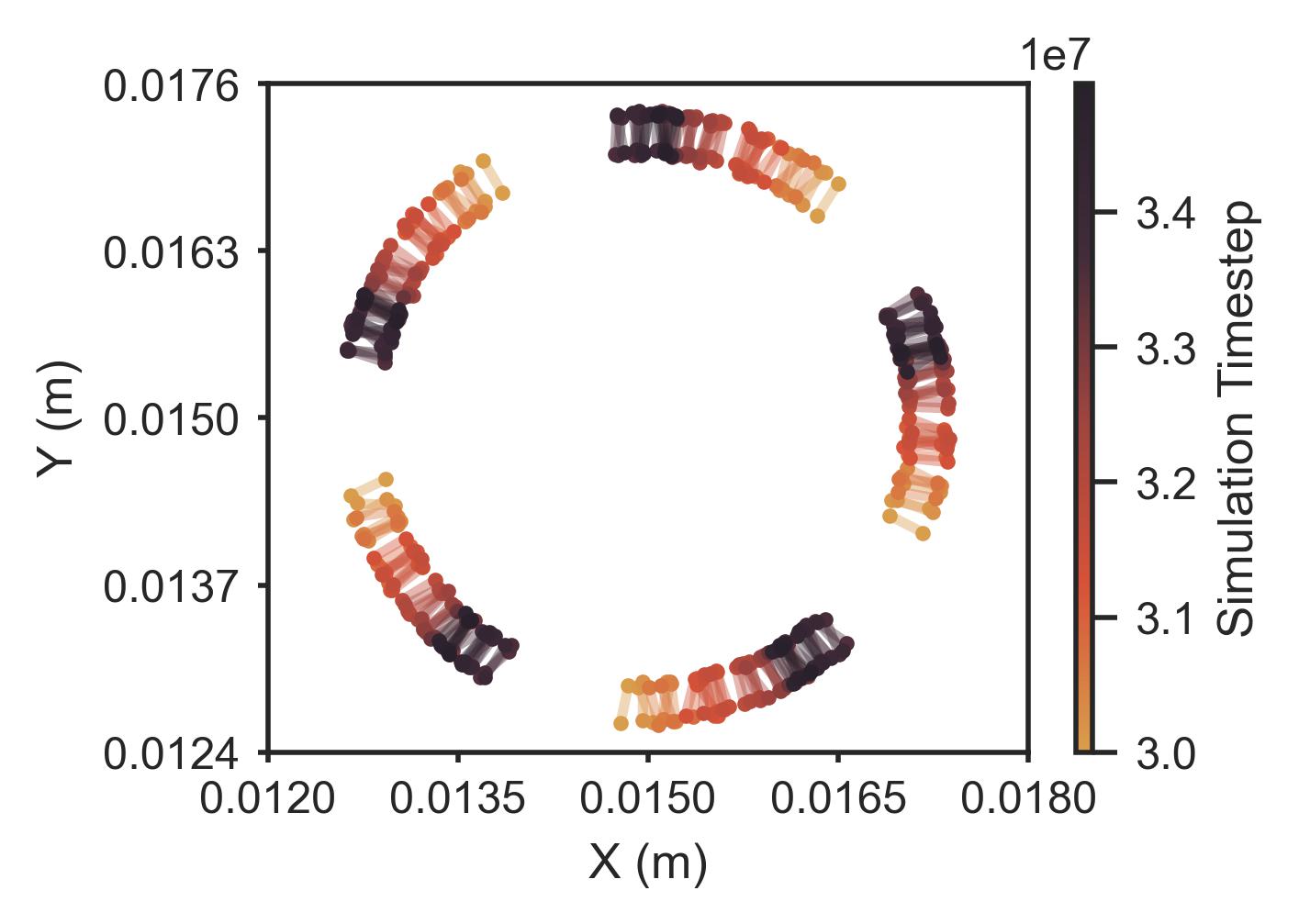}
    \caption{Dynamics of dimer clusters having 5 dimers confined using a radial electric field.  The dimers move in a circular manner while remaining radial with the electric field lines.}
    \label{fig:5dimersub1}
  
\end{figure}

However, when the number of dimers is increased to five, a unique structural transition is observed. Although the dimers continue to organize themselves on a dynamically evolving ring-like configuration, their orientational ordering undergoes a marked change. In contrast to the predominantly tangential alignment seen for smaller numbers of dimers, the dimer axes now preferentially orient along the radial direction, pointing outward from the center of the trap. This transition is clearly visible in Fig.~\ref{fig:5dimersub1}, where the radial alignment of the dimer axes is readily apparent. The color gradient employed in the figure highlights the temporal evolution of the configuration and illustrates the collective rotational motion of the ring structure.

A further intriguing reorganization occurs when a sixth dimer is introduced into the system. Rather than accommodating all six dimers on a single ring, the system lowers its energy by placing one dimer at the center of the confinement region, while the remaining five occupy positions along the periphery of an outer ring. Accompanying this structural rearrangement is another change in orientational order: the dimers located on the outer ring revert to a predominantly tangential alignment. This behavior, summarized in Table~I, demonstrates that both the positional and orientational ordering of the dimers are highly sensitive to the particle number, giving rise to a sequence of distinct collective states as the system size increases.

A qualitatively similar structural transition has been reported previously for monomer systems \cite{deshwal2022chaotic}. In that case, the addition of a sixth particle marked the onset of a shell reorganization, with one monomer occupying the center of the trap while the remaining five formed an outer ring. The dimer system exhibits the same positional transition; however, it is accompanied by an additional degree of complexity arising from the orientational freedom of the dimers. Before the emergence of the next shell structure, the system undergoes a distinct orientational transition in which the dimer axes switch from a predominantly tangential alignment to a radial one. Upon the addition of a sixth dimer and the consequent formation of the central-particle configuration, the orientational order changes once again, reverting to a predominantly tangential alignment. Thus, unlike the monomer case, where only the positional arrangement evolves with particle number, the dimer system displays coupled positional and orientational transitions, leading to a richer sequence of equilibrium states.

In fact, this behavior appears to be a recurring feature of the system whenever the addition of an extra dimer triggers the formation of a new shell. A clear example is provided by the transition from the 16-dimer to the 17-dimer configuration. For 16 dimers, the equilibrium structure consists of two concentric rings containing $(5,11)$ dimers, respectively, with the dimer axes in both rings predominantly aligned along the radial direction. However, the addition of a single dimer induces a substantial reorganization of the system into a three-shell configuration with a $(1,6,10)$ arrangement. Remarkably, this structural transition is accompanied by a simultaneous change in orientational order, with the dimers in the outer shells reverting to a predominantly tangential alignment. Thus, the emergence of a new shell is not merely a positional rearrangement; it is closely linked to a collective reorientation of the dimer axes. This coupling between shell formation and orientational ordering appears to be a robust characteristic of the system and highlights the rich interplay between translational and rotational degrees of freedom in confined dimer assemblies. 

Another intriguing feature emerges for intermediate cluster sizes, where the system does not exhibit a uniform orientational order. Instead, a mixed state develops in which some dimers preferentially align along the radial direction, while others adopt tangential orientations. Although most dimers fluctuate only weakly about their preferred orientations, exhibiting angular excursions of a few degrees, certain dimers undergo intermittent transitions between nearly radial and nearly tangential configurations. These orientation-switching events indicate the presence of competing local energy minima and highlight the rich orientational dynamics of the system.

\begin{figure*}[htbp!]
    \centering
  \begin{subfigure}[h!]{0.45\textwidth}
    \centering
    \includegraphics[width=0.95\textwidth]{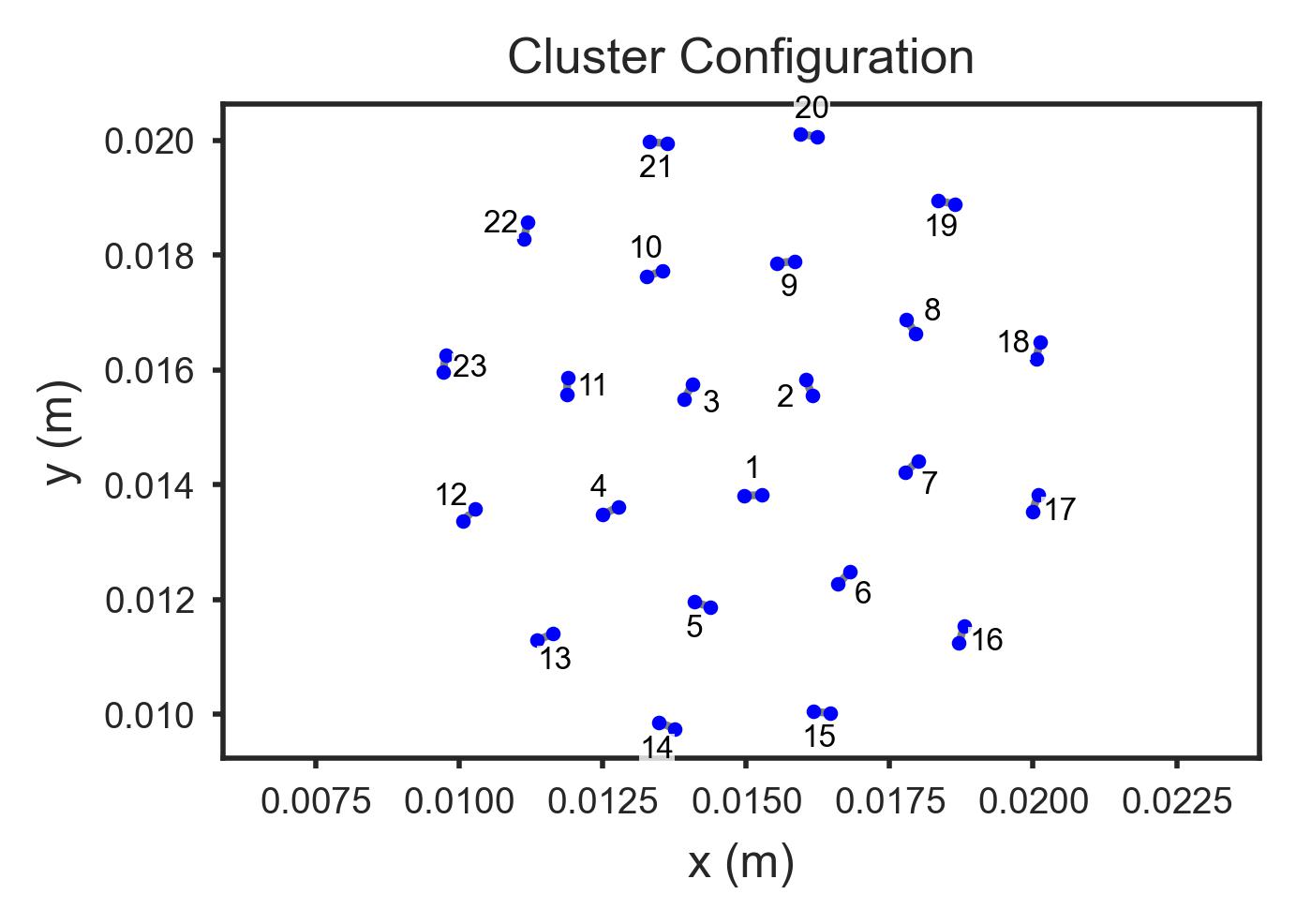}
    \caption{Structural configuration obtained after the simulation is completed.}
    \label{fig:23dimersub1}
  \end{subfigure}
  \begin{subfigure}[h!]{0.45\textwidth}
    \centering
    \includegraphics[width=0.95\textwidth]{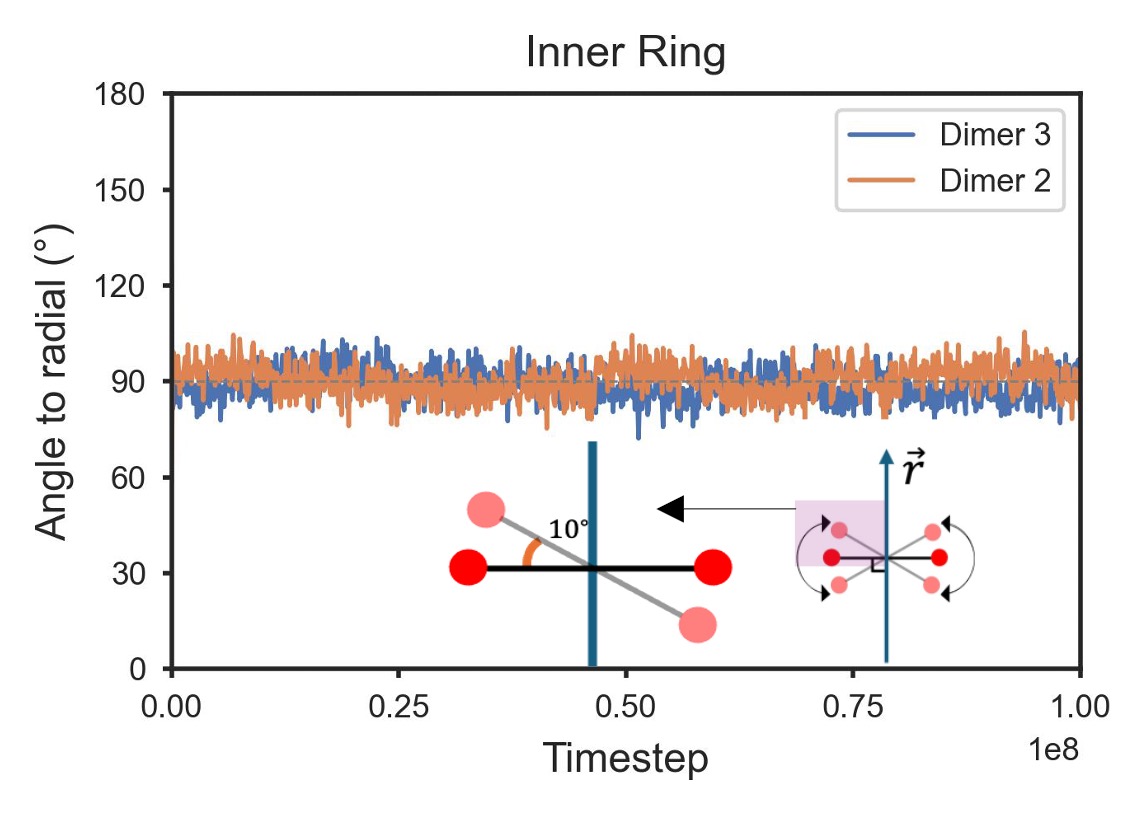}
    \caption{Variation of angle between the dimer axis and radial axis, for a dimer placed in the inner ring/shell. It can be seen that no significant variation is observed in the angle during the simulation.}
    \label{fig:23dimersub2}
  \end{subfigure}
  \begin{subfigure}[h!]{0.45\textwidth}
    \centering
    \includegraphics[width=0.95\textwidth]{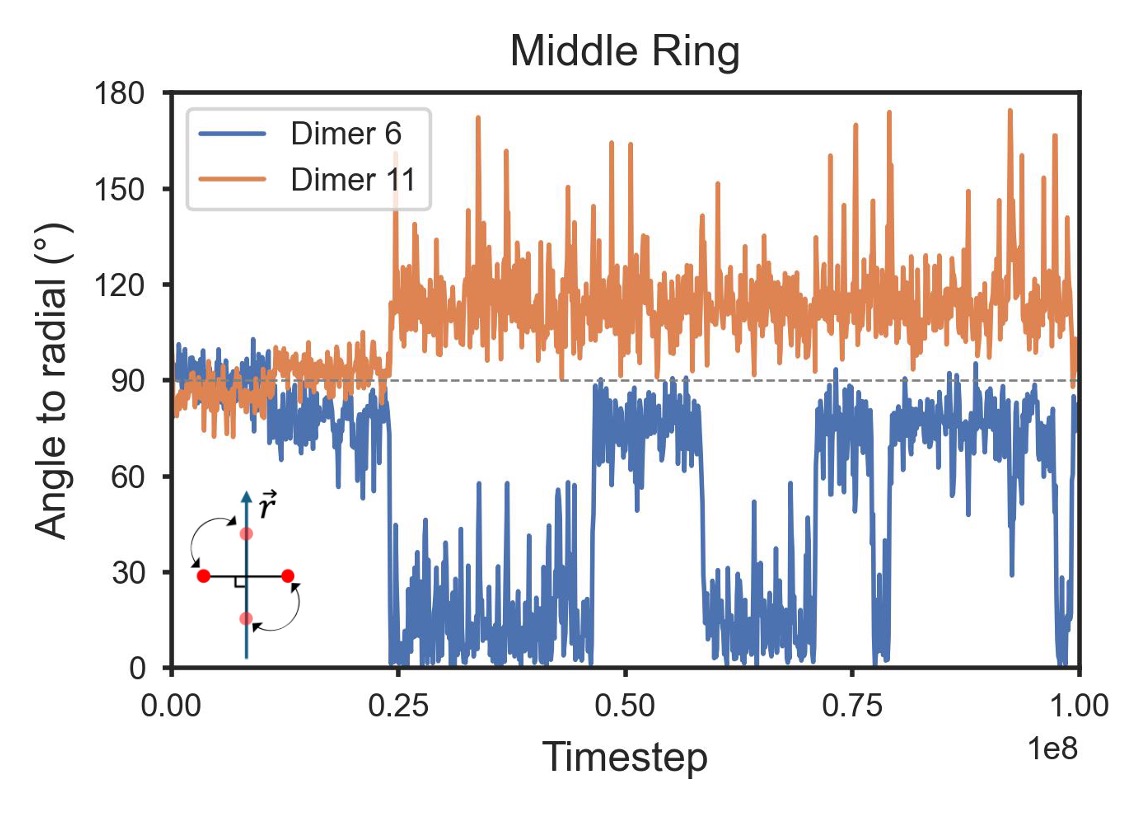}
    \caption{Variation of angle between the dimer axis and radial axis, for a dimer placed in the middle ring/shell. It can be seen that the dimer many times goes from tangential ($90^\circ$) to radial ($0^\circ$ or $180^\circ$) orientation and vice-versa.}
    \label{fig:23dimersub3}
  \end{subfigure}
  \begin{subfigure}[h!]{0.45\textwidth}
    \centering
    \includegraphics[width=0.95\textwidth]{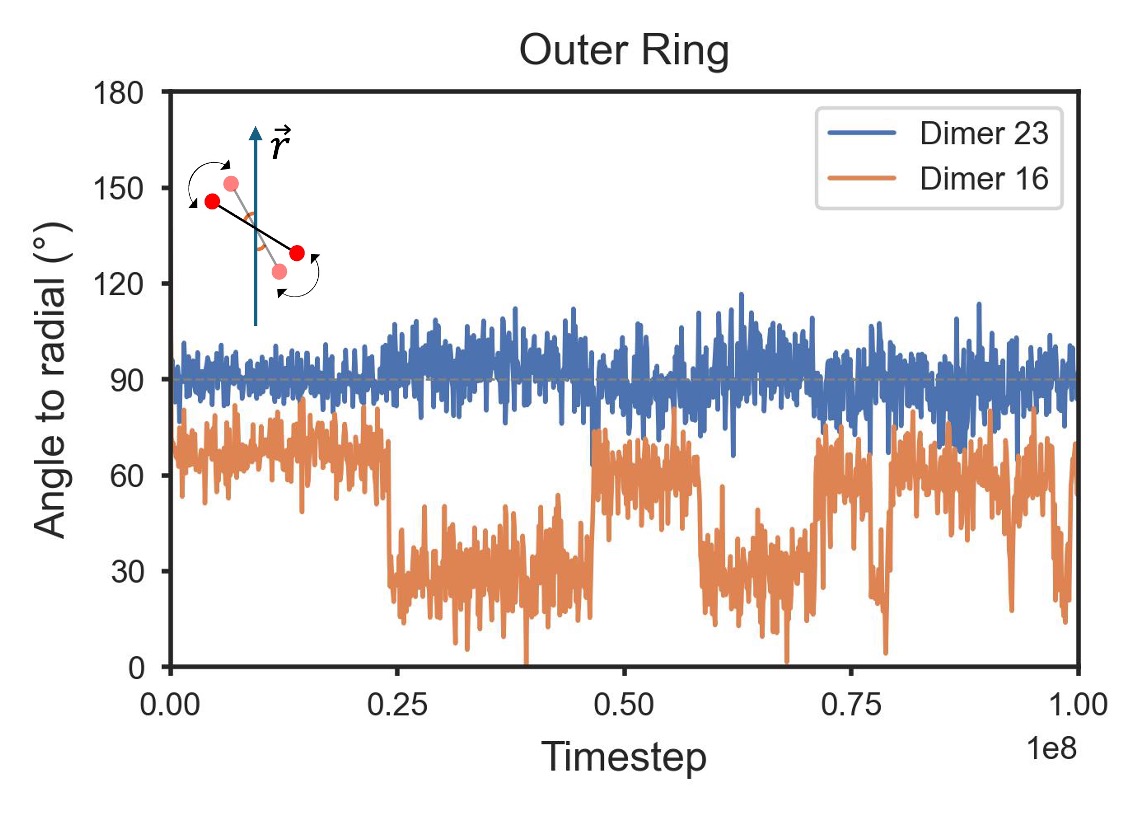}
    \caption{Variation of angle between the dimer axis and radial axis, for a dimer placed in the outer ring/shell. Again the dimer transitions from almost tangential ($90^\circ$) to radial ($0^\circ$ or $180^\circ$) orientation and vice-versa.}
    \label{fig:23dimersub4}
  \end{subfigure}
  \caption{Dynamics of dimer cluster having 23 dimers confined using a radial electric field. Different orientations are observed in the cluster and these orientations are dynamic in nature.
  }
  \label{fig:23dimerdynamics}
\end{figure*}

A representative example is provided by the three-ring cluster containing 23 dimers, shown in Fig.~\ref{fig:23dimerdynamics}. To facilitate tracking of the orientational dynamics, the dimers are numbered in Fig.~\ref{fig:23dimersub1}. The temporal evolution of the orientations of selected dimers is presented in the remaining subplots. Figure~\ref{fig:23dimersub2} shows the orientations of two dimers belonging to the innermost ring. All dimers in this ring remain predominantly tangentially aligned throughout the simulation, displaying only small fluctuations of approximately $10^\circ$ about their preferred orientation.

The orientational dynamics of dimers belonging to the middle and outermost rings are shown in Fig.~\ref{fig:23dimersub3} and \ref{fig:23dimersub4}, respectively. In contrast to the inner ring, the dimers in the middle ring exhibit repeated transitions between radial and tangential states, spending appreciable periods of time in both configurations. The outermost ring displays even richer behavior. While many dimers predominantly occupy either radial or tangential orientations, some remain trapped for extended durations in intermediate configurations, with their axes oriented at angles close to $60^\circ$. The existence of these long-lived intermediate states, together with the observed switching events, suggests a complex orientational energy landscape in which multiple nearly degenerate configurations coexist. Such behavior underscores the intricate coupling between positional ordering and orientational degrees of freedom in confined dimer clusters.

The structural and dynamical properties of the dimer assemblies are summarized in Tables~\ref{tab:1-1} and \ref{tab:1-2}, covering systems containing 2--13 and 14--25 dimers, respectively. As the number of dimers increases, the system undergoes a sequence of structural reorganizations characterized by the emergence of multiple concentric shells. These shell transitions are often accompanied by changes in the preferred orientational ordering of the dimers, revealing a strong coupling between positional and orientational degrees of freedom.

In an earlier study of confined dust monomers \cite{deshwal2022chaotic}, both static and dynamic equilibrium configurations were observed depending on the cluster size and structural arrangement. In contrast, all dimer clusters investigated here exhibit dynamic equilibrium. Even after attaining their equilibrium spatial organization, the dimers continue to display persistent collective motion in the form of rotations, oscillations, or a combination of both. The origin of this behavior lies in the anisotropic nature of the dimer. The two constituent particles, located at opposite ends of the rigid bond, generally experience unequal forces arising from the confining field and interactions with neighboring dimers. Since the particles are rigidly connected, these force imbalances generate torques that continuously drive rotational and oscillatory motion, preventing the system from settling into a completely static configuration.

The results summarized in Tables~\ref{tab:1-1} and \ref{tab:1-2} therefore reveal a rich hierarchy of equilibrium states characterized not only by shell formation and positional ordering, but also by persistent collective dynamics and nontrivial orientational organization. These features distinguish dimer clusters from their monomer counterparts and highlight the important role played by internal rotational degrees of freedom in determining the behavior of strongly coupled confined systems.

\begin{figure}[htbp!]
     \begin{subfigure}[h!]{0.45\textwidth}
    \centering
    \includegraphics[width=\textwidth]{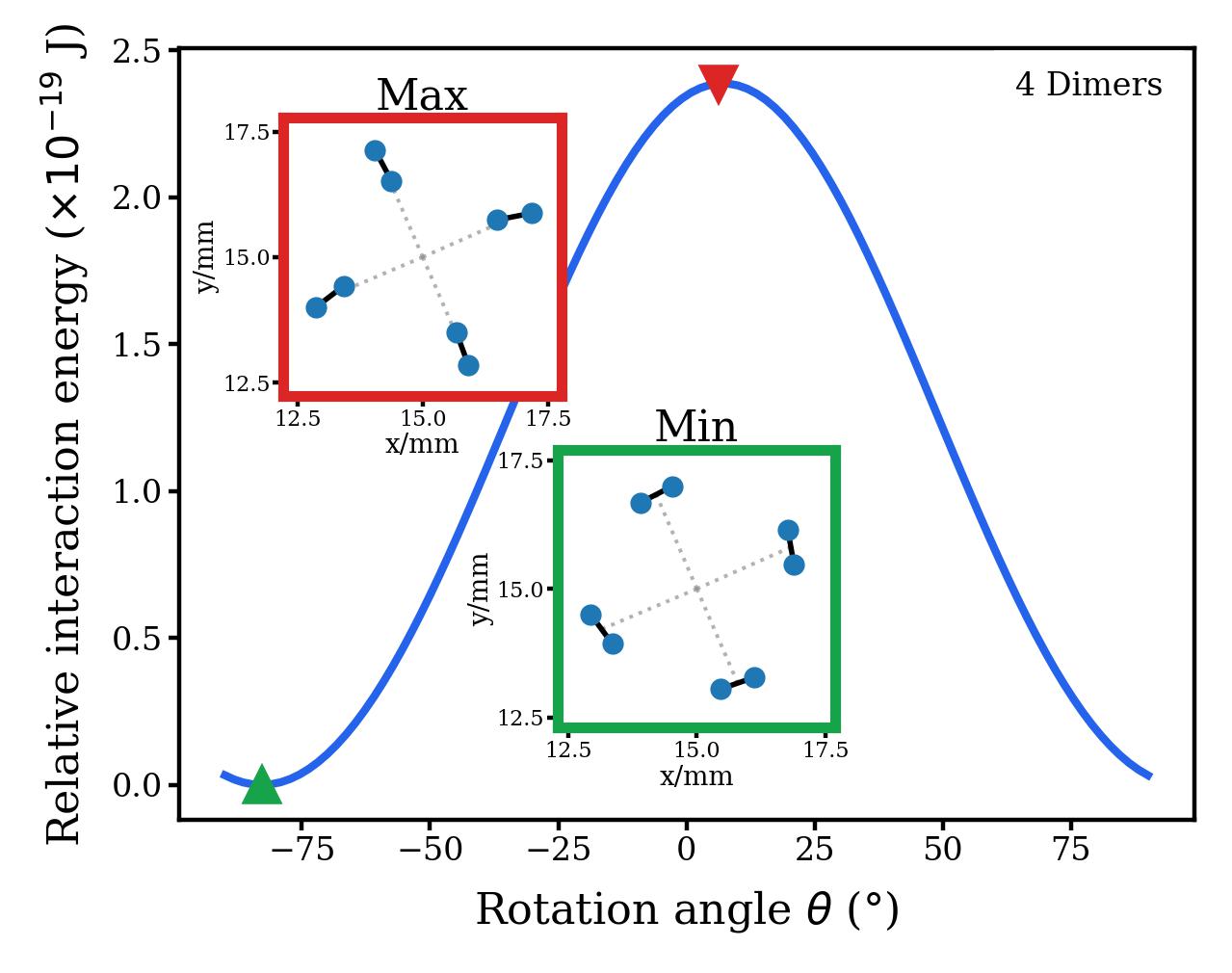}
    \caption{Potential energy variation of 4 dimer cluster  as the dimers are rotated around its center making angle $\theta$ with the radial direction.}
    \label{fig:4dimerPE}
  \end{subfigure}
  \begin{subfigure}[h!]{0.45\textwidth}
    \centering
    \includegraphics[width=\textwidth]{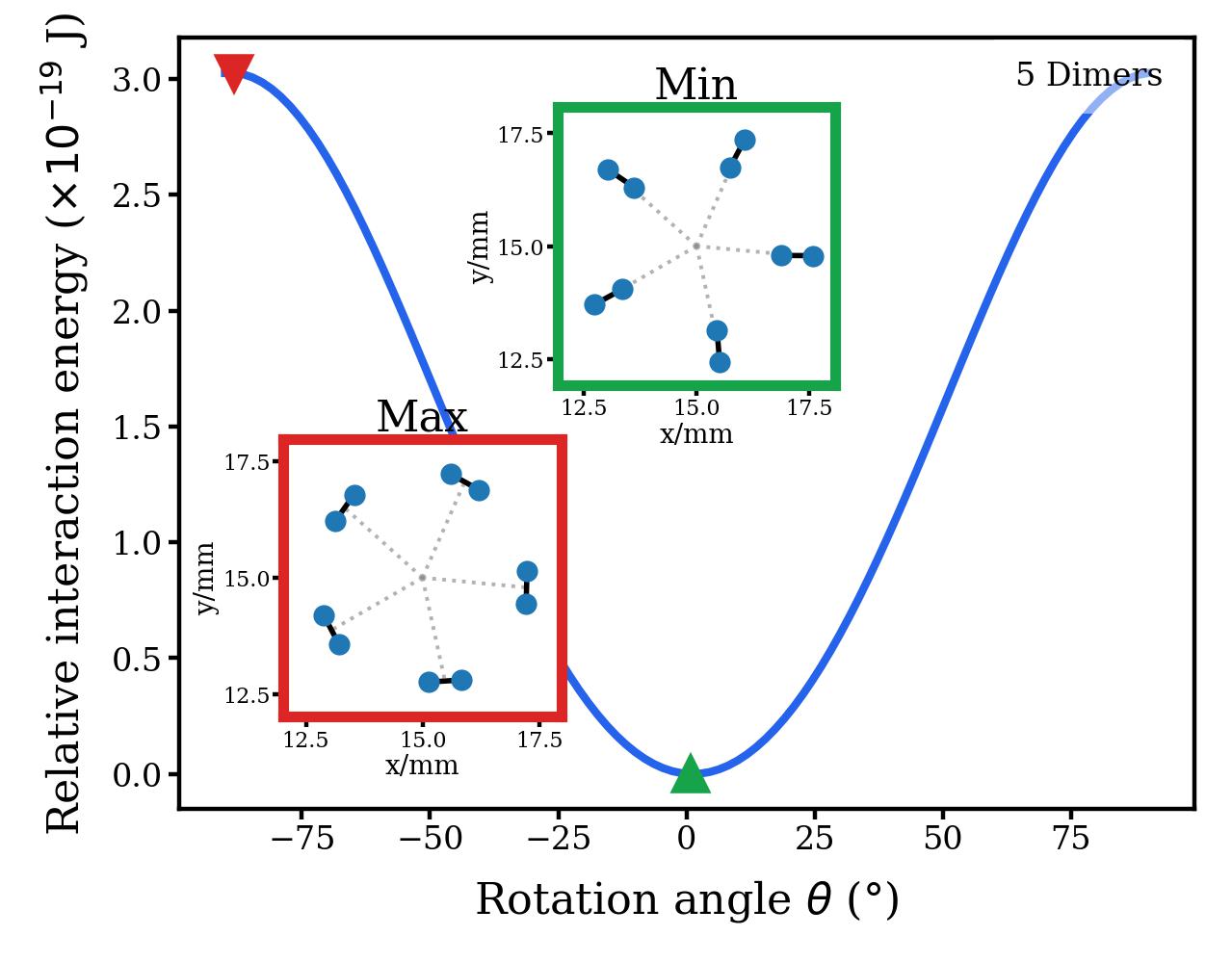}
    \caption{Potential energy variation of 4 dimer cluster  as the dimers are rotated around its center making angle $\theta$ with the radial direction.}
    \label{fig:5dimerPE}
  \end{subfigure}
     \caption{Plot of the potential energy of dimers as the orientation angles are changed from tangential to radial. The potential energy has been scaled such that the minimum is at 0. The minimum and maximum interaction energy configurations are shown as insets.}
     \label{fig:PE}
\end{figure}
To gain insight into the observed orientational ordering, we now examine the dependence of the cluster potential energy on the dimer orientation. Figure~\ref{fig:PE} shows the variation of the total potential energy for clusters containing four and five dimers as a function of their orientation angle. For the four-dimer cluster [Fig.~\ref{fig:PE}(a)], the potential energy attains its minimum value when the dimers are aligned tangentially to the ring and increases monotonically as the orientation is rotated toward the radial direction. The energy reaches a maximum when the dimers are oriented radially outward. This energetic preference naturally explains the tangential ordering observed in the equilibrium configuration of the four-dimer cluster.

In contrast, the behavior changes qualitatively for the five-dimer cluster. As shown in Fig.~\ref{fig:PE}(b), the minimum of the potential energy now occurs for a predominantly radial orientation of the dimers. Consequently, the equilibrium state favors radial alignment rather than the tangential arrangement observed for smaller cluster sizes. Thus, the orientational transitions identified in the simulations can be understood as a direct consequence of changes in the underlying potential-energy landscape with increasing particle number. The competition between confinement and inter-dimer interactions modifies the energetically preferred configuration, leading to the observed alternation between tangential and radial ordering as the cluster structure evolves.

\begin{figure}[htbp!]
    \centering
    \includegraphics[width=\linewidth]{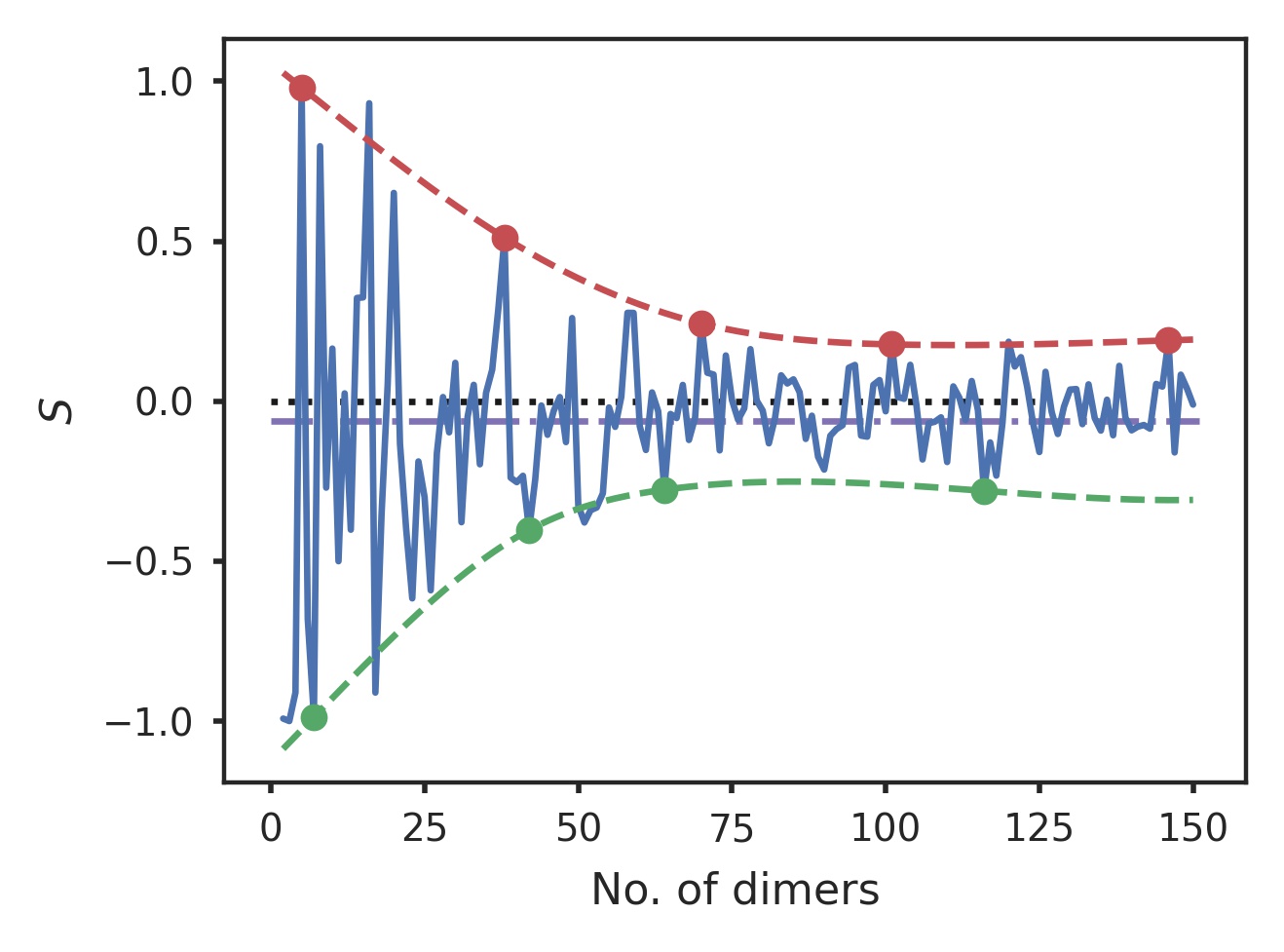}
    \caption{Order parameter, $S$, plotted for increasing number of dimers. It can be seen that for low number of dimers the order parameter reaches extreme values. As the numbers in cluster increase, it fluctuates within a certain range of -0.2 o 0.2.}
    \label{fig:order_parameter}
\end{figure}
As the number of dimers in the cluster increases, maintaining a globally ordered radial or tangential arrangement becomes progressively more difficult. Consequently, the system evolves toward configurations exhibiting mixed orientational order, in which dimers with different preferred orientations coexist. Although the degree of ordering decreases with increasing cluster size, the system does not become completely disordered. Instead, a finite level of orientational correlation persists even in the largest clusters considered here.
\begin{equation}
S = \left(\left<2\cos^2\theta - 1\right>\right)
\end{equation}
where ($\theta$) is the angle between the dimer axis and the radially inward confining electric field. An order parameter value of (S=1) corresponds to perfect radial alignment, with the dimers oriented parallel or antiparallel to the electric field, whereas (S=-1) indicates perfect tangential alignment, with the dimer axes perpendicular to the field.

The variation of the order parameter with cluster size is shown in Fig.~\ref{fig:order_parameter}. For the simulation parameters considered here, (S) approaches an asymptotic value of approximately (-0.06) as the number of dimers increases. It is evident that the order parameter remains bounded within an envelope. For small clusters, (S) often attains values close to its extrema, reflecting the highly ordered radial or tangential configurations observed in these systems. However, as the cluster size increases and multiple concentric shells emerge, the magnitude of the order parameter decreases, indicating a gradual reduction in global orientational order and the appearance of mixed configurations.

This trend can be understood in terms of the increasingly complex shell structure of larger clusters. As additional rings form, dimers located in the intermediate shells experience competing interactions from neighboring inner and outer shells. These interactions partially screen the influence of the confining field and reduce the ability of the dimers to adopt a single globally preferred orientation. As a result, the orientational ordering is suppressed by competing interactions, leading to the coexistence of radial, tangential, and intermediate alignments. Nevertheless, the fact that (S) does not approach zero indicates that complete orientational disorder is never achieved. Even in large clusters, a residual degree of collective ordering survives, reflecting the persistent influence of confinement and inter-dimer interactions on the equilibrium structure.


\begin{figure*}[!htbp]
    \centering
    \includegraphics[width=0.9\linewidth]{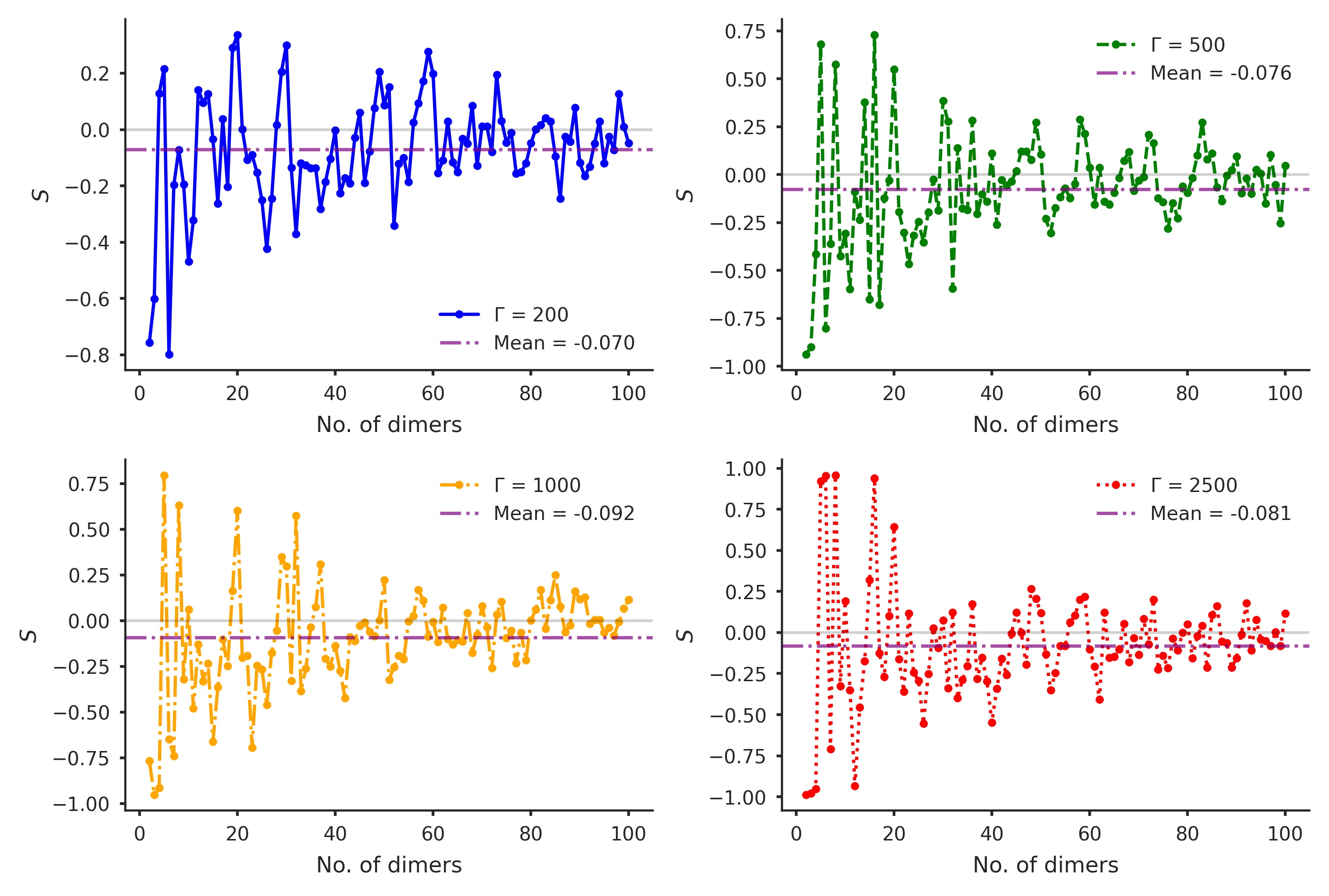}
    \caption{A comparison of the order parameter calculated for dimers at different $\Gamma$ values.}
    \label{fig:gammacompare}
\end{figure*}

The influence of the coupling parameter on the equilibrium dimer configurations has also been investigated. The resulting structures for ($\Gamma = 200, 500, 1000$)  are summarized in Table~III. Across this range of coupling strengths, the overall structural characteristics show some variations.  This behavior can be understood by noting that all these high values of ($\Gamma$) being in a very strongly coupled regime,  the interaction energy significantly exceeds the thermal energy.

As ($\Gamma$) increases, thermal fluctuations become progressively weaker, causing the particles to remain more tightly localized around their equilibrium positions. Consequently, the system explores the configurational phase space less efficiently, and relaxation toward the true minimum-energy state becomes increasingly slow. In finite-duration simulations, this can lead to small differences in the observed configurations, particularly at the highest values of ($\Gamma$), where the dynamics become sluggish and metastable states can persist for extended periods. 

We have also investigated the dependence of the orientational order parameter on the coupling strength ($\Gamma$). Care has been taken to ensure that the order parameter is evaluated only after the system has evolved for a sufficiently long duration and attained a stable equilibrium configuration, although, particularly at very high values of ($\Gamma$), the system may still remain trapped in a metastable state rather than reaching the absolute minimum-energy configuration. The variation of the order parameter (S) with the number of dimers is shown in Fig.~\ref{fig:gammacompare} for several values of ($\Gamma$).

A key observation from the figure is that, irrespective of the coupling strength, the magnitude of the order parameter decreases as the cluster size increases. For sufficiently large clusters, (S) remains close to zero for all values of ($\Gamma$), indicating the absence of a globally preferred orientation and the emergence of mixed orientational states. Thus, while local orientational correlations persist, the overall directional ordering becomes increasingly weak as additional shells form within the cluster.

A comparison of the different subplots in Fig.~\ref{fig:gammacompare} further reveals that the orientational ordering of large clusters is relatively insensitive to the value of ($\Gamma$). In contrast, the ordering of smaller clusters, particularly those consisting of only one or two shells, exhibits a stronger dependence on the coupling strength. For example, at ($\Gamma = 200$), clusters containing up to approximately twenty dimers exhibit order-parameter values below 0.5, indicating only partial alignment with the confining electric field. As the coupling strength is increased to ($\Gamma = 500$), the maximum value of the order parameter rises to about 0.75, reflecting a stronger degree of orientational ordering. Only at very high coupling strengths, such as ($\Gamma = 2500$), does the order parameter approach unity for some cluster sizes, corresponding to an almost perfect alignment of the dimer axes with the radial electric field.

These trends suggest a gradual shift in the dominant mechanism governing orientational ordering. At lower values of ($\Gamma$), thermal fluctuations are sufficiently strong to compete with the torque exerted by the confining electric field, thereby reducing the degree of alignment. As ($\Gamma$) increases, thermal agitation becomes less important and the orientational dynamics are increasingly controlled by the confining field and inter-dimer interactions. Consequently, the dimers are able to adopt configurations closer to their energetically preferred orientations, resulting in a higher degree of order. For large clusters, however, the competing interactions arising from multiple neighboring shells introduce orientational frustration, causing the global order parameter to remain small even in the strongly coupled limit.

Another limitation affecting clusters with a large number of dimers arises from the finite-size nature of the confinement itself. As the cluster grows, an increasing fraction of the dimers occupy outer shells, where their arrangement and orientation are influenced not only by inter-dimer interactions but also by the geometry and boundaries of the confining potential. Consequently, some of the observed structural and orientational features may reflect the constraints imposed by the finite confinement rather than the intrinsic behavior of an extended dimer assembly.

To disentangle these effects, we now turn to a complementary system employing periodic boundary conditions. By eliminating the influence of confining boundaries, periodic simulations provide an opportunity to investigate the positional and orientational ordering that emerges solely from the interactions between the dimers. In the next section, we examine the equilibrium structures formed in a periodic simulation box and analyze the corresponding translational and orientational order exhibited by the dimer ensemble.

\begin{table*}[htbp!]
    \caption{Dimer configuration observed for 2 - 13 dimers. Distinct radial and tangential configurations are seen. }
     \label{tab:1-1}
      \centering
     \begin{tabular}{|p{3em} | c | c | p{2cm}||p{3em} | c | c | p{2cm}|} \toprule
        No. of dimers & Configuration & Structure & Orientation & No. of dimers & Configuration & Structure & Orientation\\ \hline
        2 & (0,2) & \includegraphics[width=0.2\linewidth]{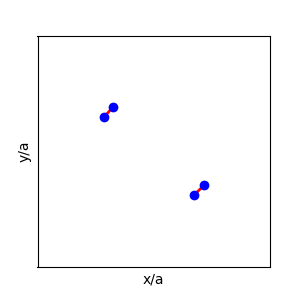} & tangential & 8 & (1,7) & \includegraphics[width=0.2\linewidth]{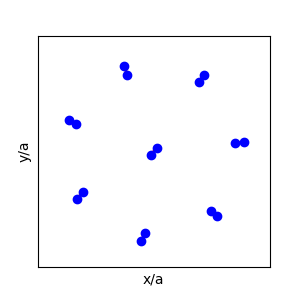} & radial\\ \hline
        3 & (0,3) & \includegraphics[width=0.2\linewidth]{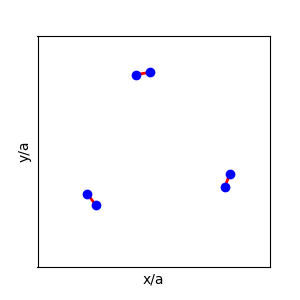} & tangential & 9 & (2,7) & \includegraphics[width=0.2\linewidth]{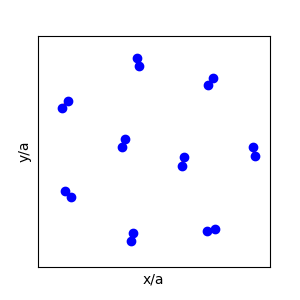} & mix\\ \hline
        4 & (0,4) & \includegraphics[width=0.2\linewidth]{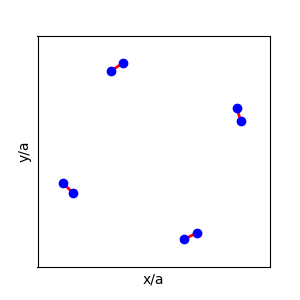} & tangential & 10 & (2,8) & \includegraphics[width=0.2\linewidth]{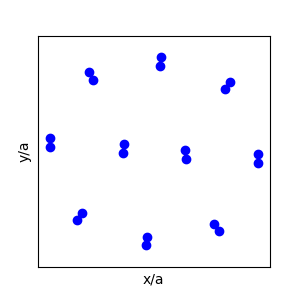} & mix\\ \hline
        5 & (0,5) & \includegraphics[width=0.2\linewidth]{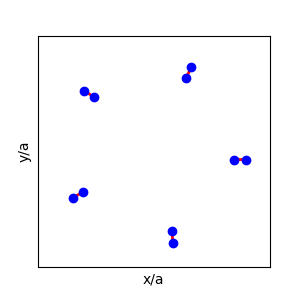} & radial & 11 & (3,8) & \includegraphics[width=0.2\linewidth]{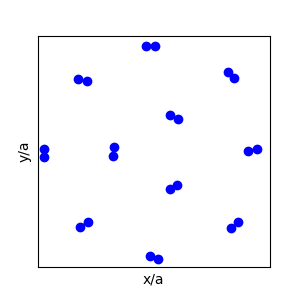} & mix\\ \hline
        6 & (1,5) & \includegraphics[width=0.2\linewidth]{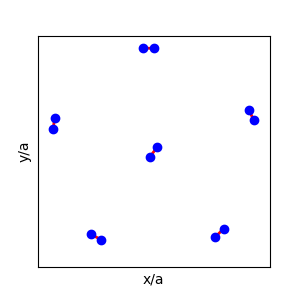} & tangential & 12 & (3,9) & \includegraphics[width=0.2\linewidth]{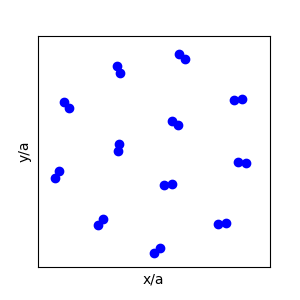} & mix\\ \hline
        7 & (1,6) & \includegraphics[width=0.2\linewidth]{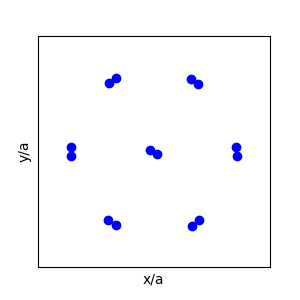} & tangential & 13 & (4,9) & \includegraphics[width=0.2\linewidth]{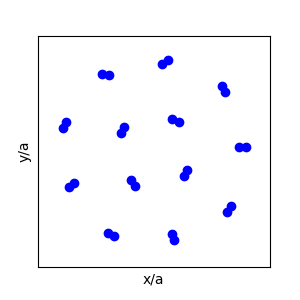} & tangential\\ \hline
     \end{tabular}
\end{table*}

\begin{table*}[htbp!]
    \caption{Dimer configuration observed for 14 - 25 dimers. The distinct radial and tangential configurations are now much harder to observe as we go beyond 20 dimers in a single configuration. }
    \label{tab:1-2}
     \centering
     \begin{tabular}{|m{3em} | c | c | b{1.8cm}||m{3em} | c | c | b{1.8cm}|} \toprule
        No. of dimers & Configuration & Structure & Orientation & No. of dimers & Configuration & Structure & Orientation \\ \hline
        14 & (4,10) & \includegraphics[width=0.2\linewidth]{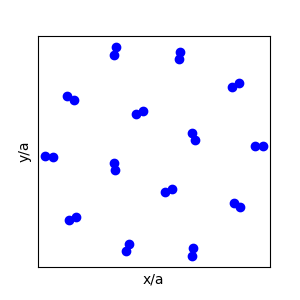} & tangential, radial & 20 & (1,7,12) & \includegraphics[width=0.2\linewidth]{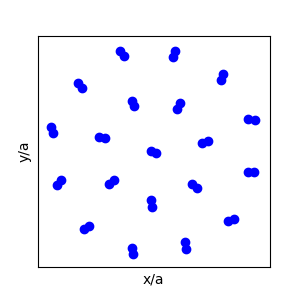} & radial, radial \\ \hline
        15 & (5,10) & \includegraphics[width=0.2\linewidth]{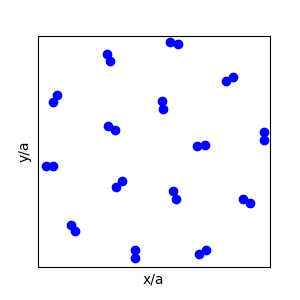} & radial, mix & 21 & (2,7,12) & \includegraphics[width=0.2\linewidth]{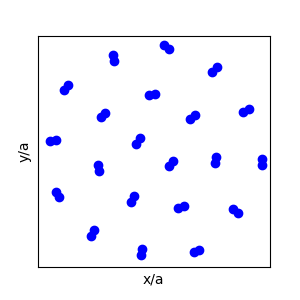} & mix\\ \hline
        16 & (5,11) & \includegraphics[width=0.2\linewidth]{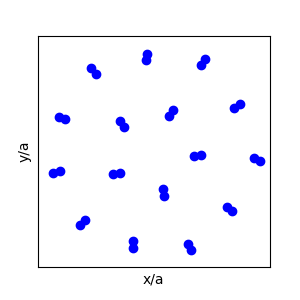} & radial & 22 & (2,8,12) & \includegraphics[width=0.2\linewidth]{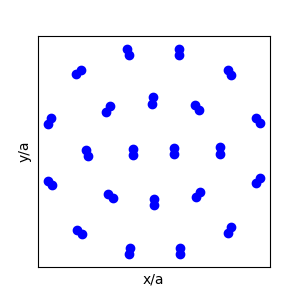} & mix\\ \hline
        17 & (1,6,10) & \includegraphics[width=0.2\linewidth]{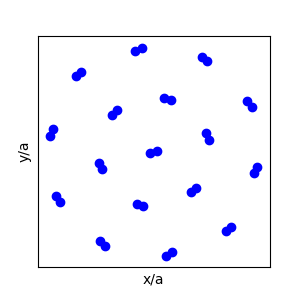} & tangential & 23 & (3,8,12) & \includegraphics[width=0.2\linewidth]{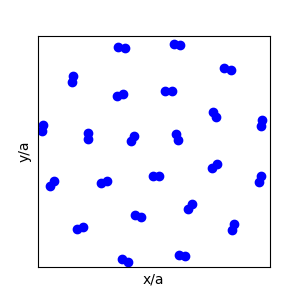} & mix\\ \hline
        18 & (1,6,11) & \includegraphics[width=0.2\linewidth]{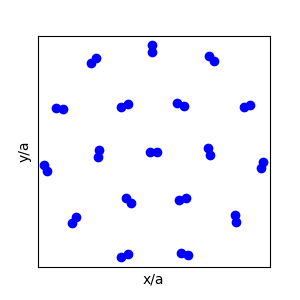} & tangential & 24 & (3,8,13) & \includegraphics[width=0.2\linewidth]{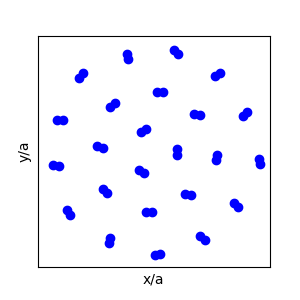} & mix\\ \hline
        19 & (1,7,11) & \includegraphics[width=0.2\linewidth]{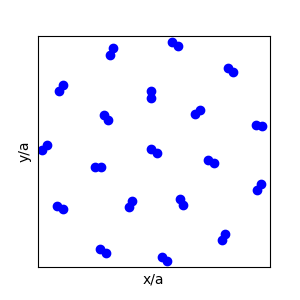} & radial, tangential & 25 & (3,9,13) & \includegraphics[width=0.2\linewidth]{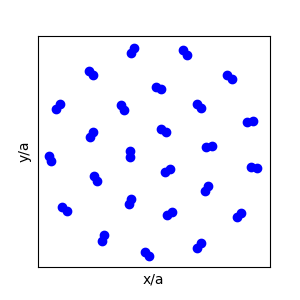} & mix\\ \hline
     \end{tabular}
\end{table*}

\begin{table*}[htbp!]
    \caption{Change in structural configuration for dimer crystals under different coupling parameter ($\Gamma$) values.}
     \label{tab:3}
     \centering
     \begin{tabular}{|m{3em} | c | c | c | c |}\toprule
        No. of dimers & $\Gamma$ = 200 & $\Gamma$ = 500 & $\Gamma$ = 1000 & $\Gamma$ = 2500 \\ \hline
        6 & \includegraphics[width=0.19\linewidth]{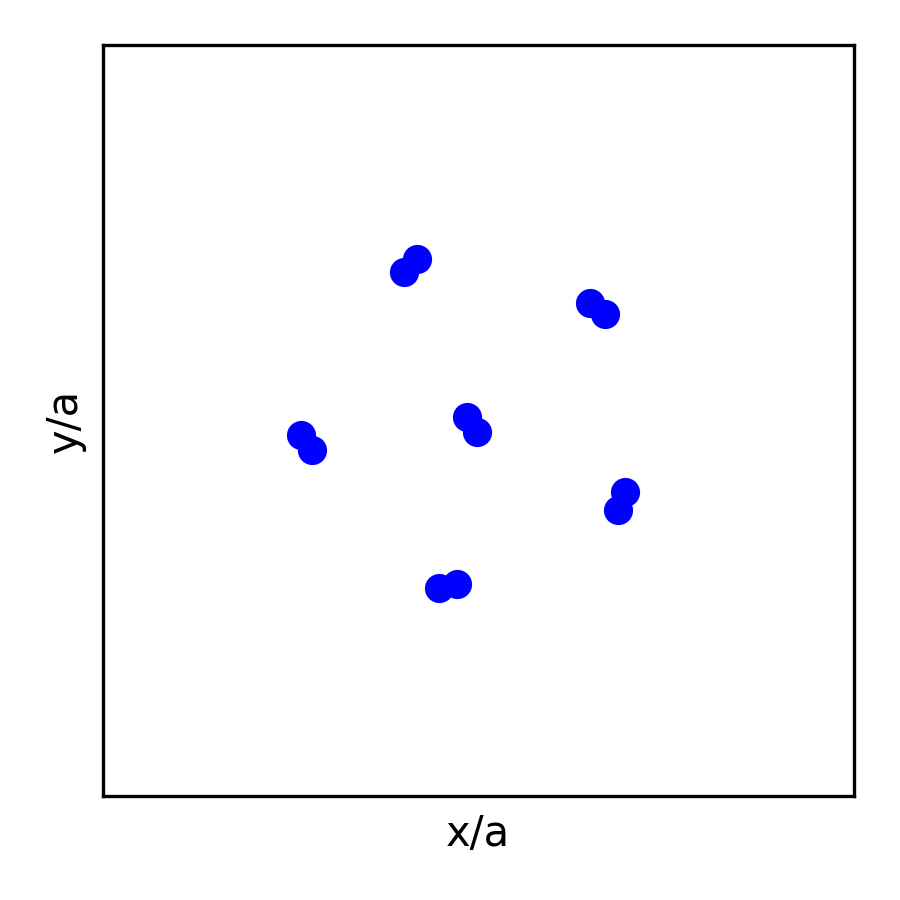} & 
            \includegraphics[width=0.19\linewidth]{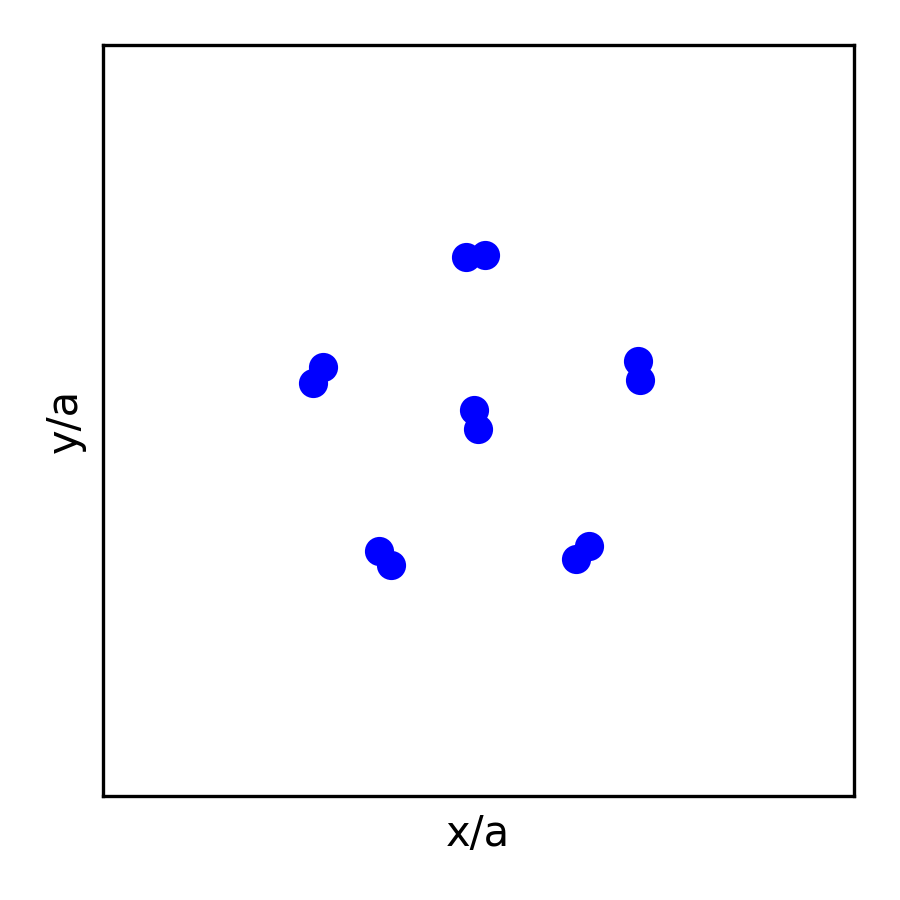} &
            \includegraphics[width=0.19\linewidth]{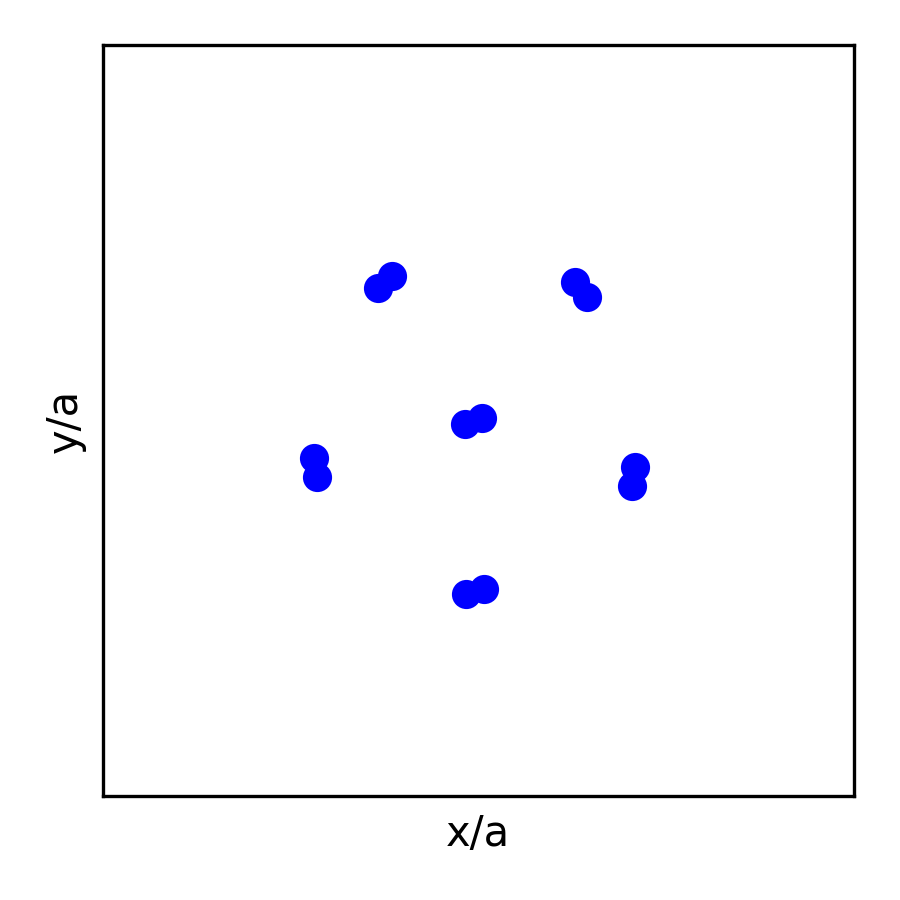} &
            \includegraphics[width=0.19\linewidth]{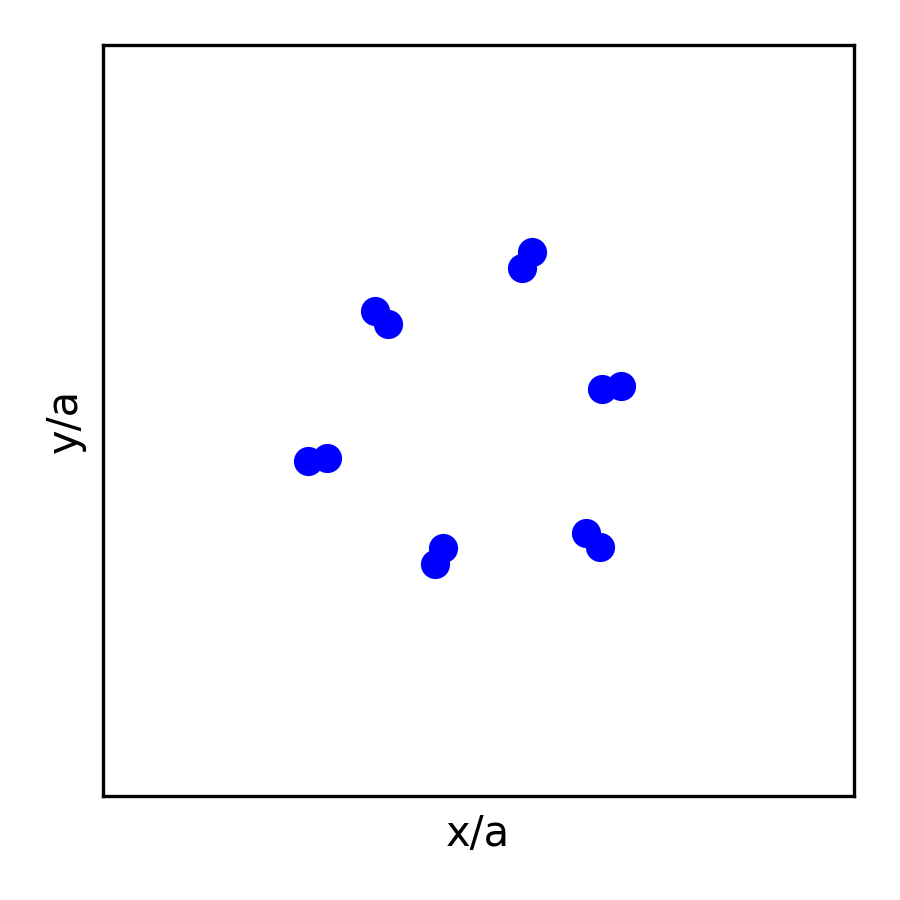} \\ \hline
        10 & \includegraphics[width=0.19\linewidth]{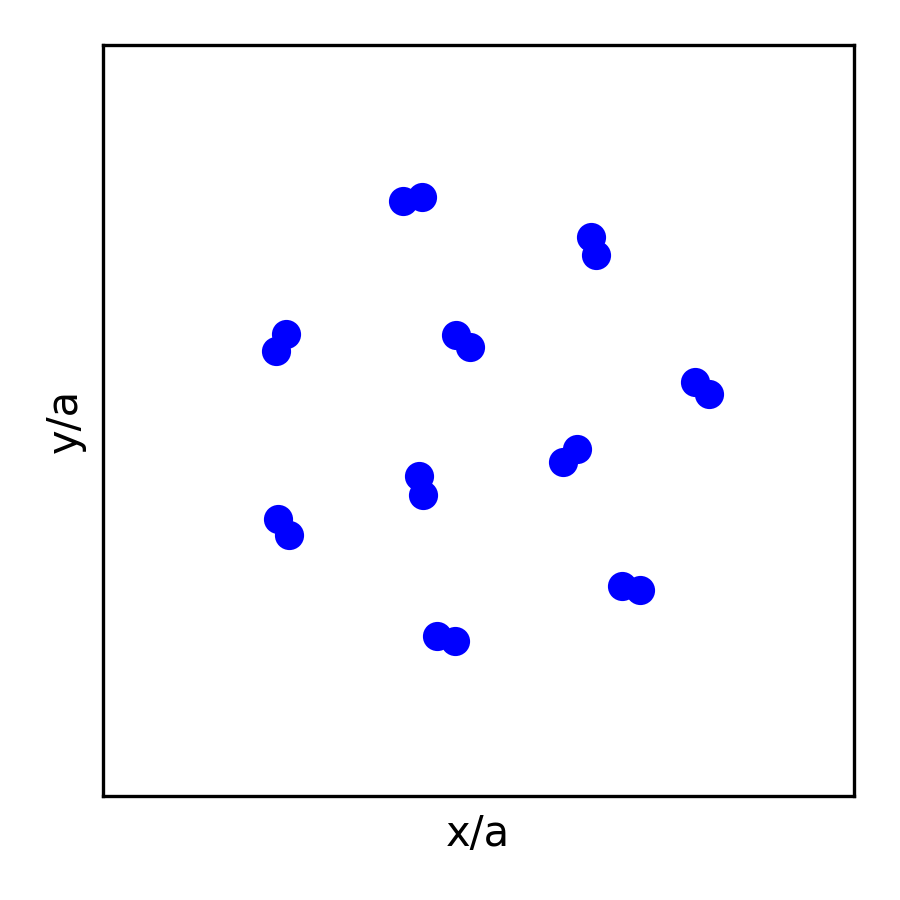} & 
            \includegraphics[width=0.19\linewidth]{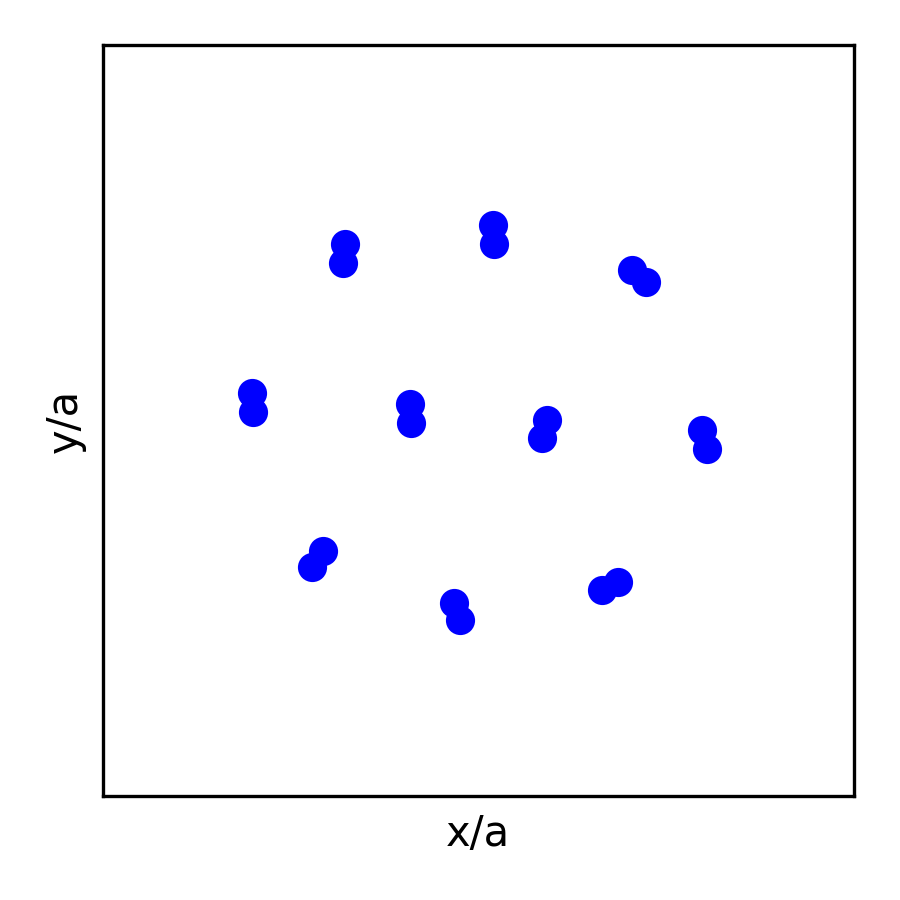} &
            \includegraphics[width=0.19\linewidth]{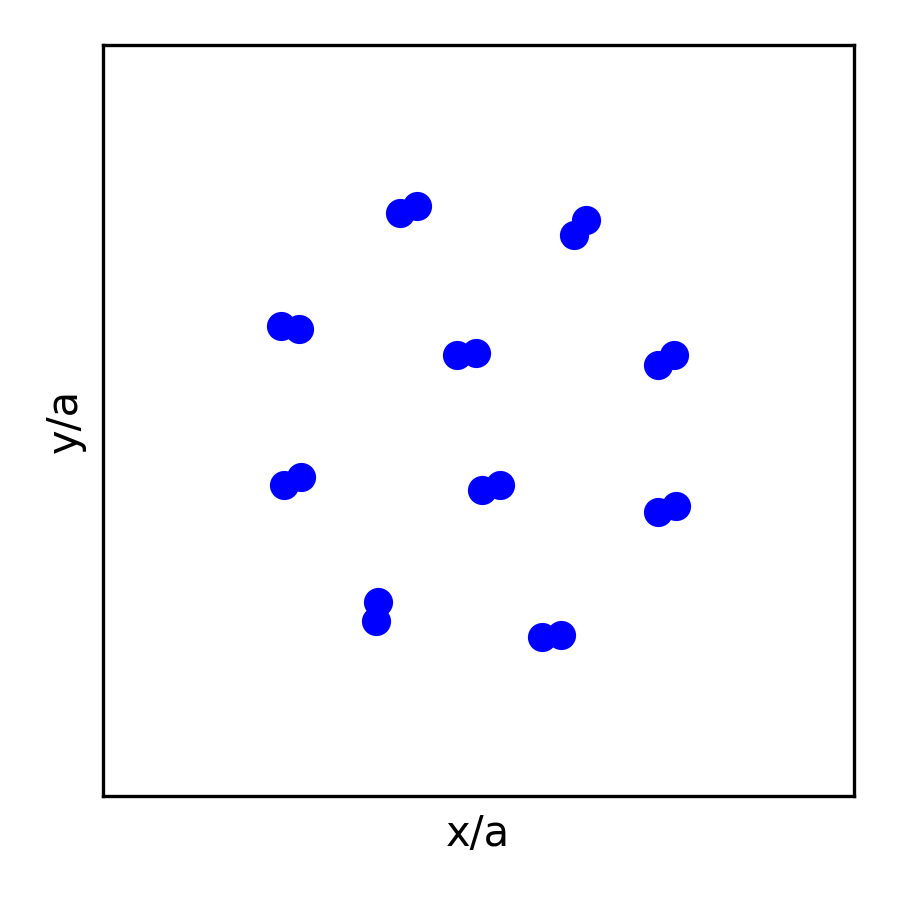} &
            \includegraphics[width=0.19\linewidth]{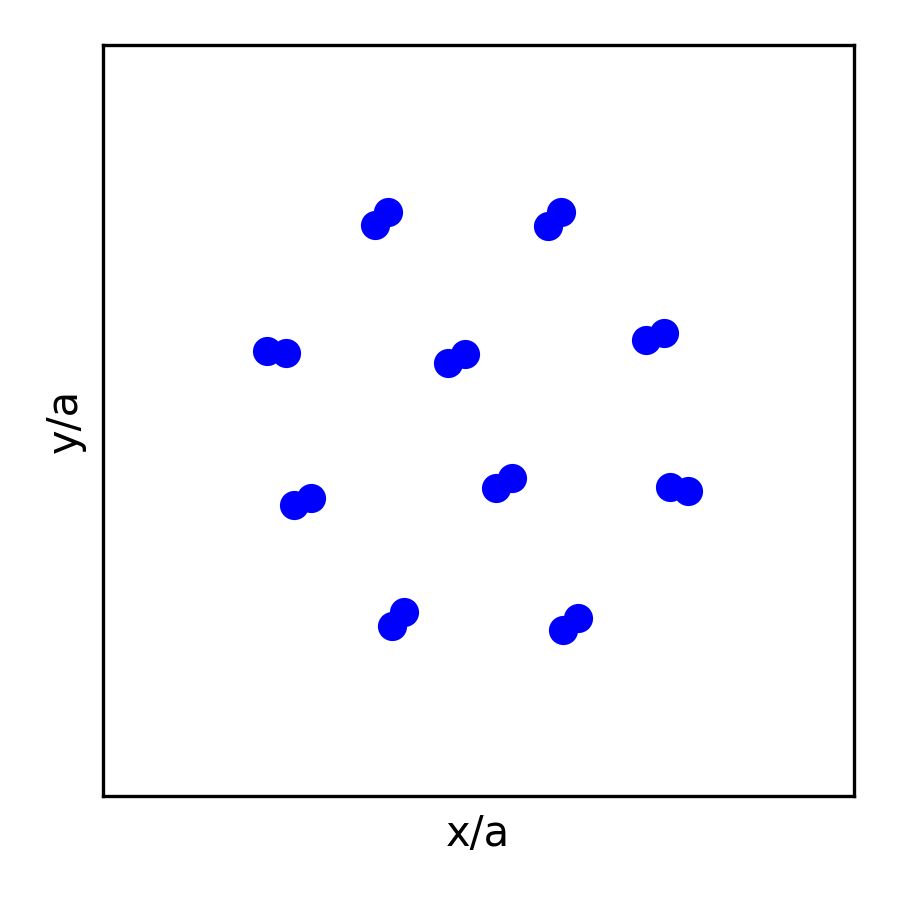} \\ \hline
        12 & \includegraphics[width=0.19\linewidth]{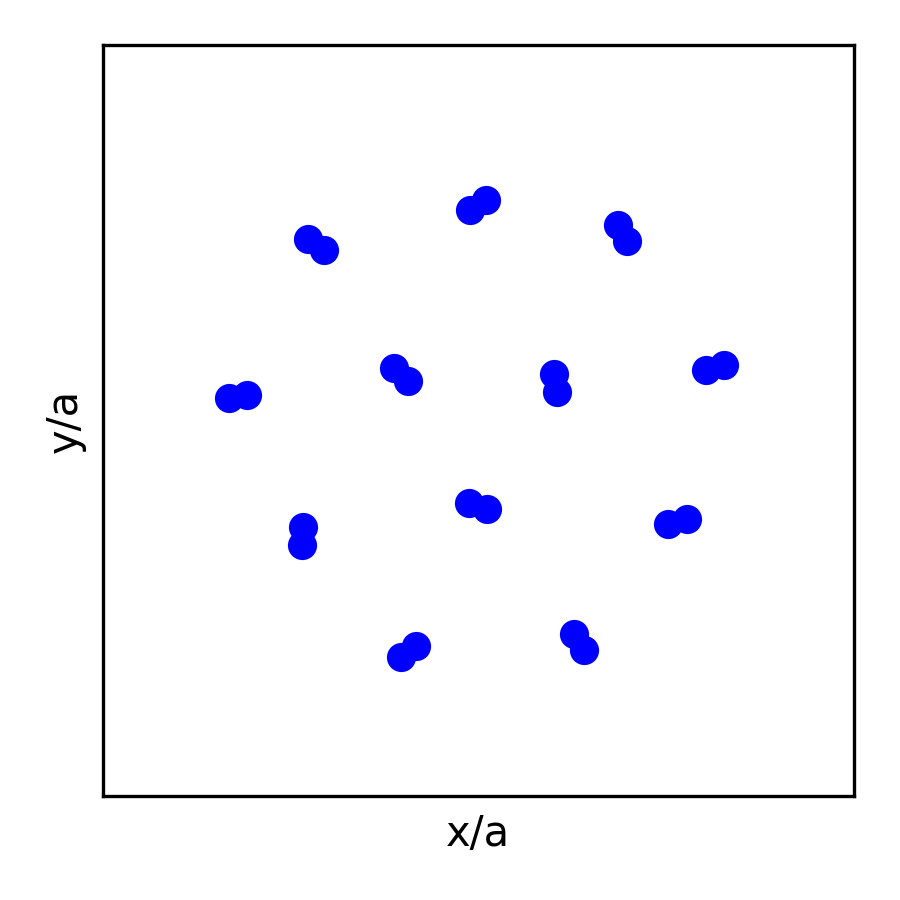} & 
            \includegraphics[width=0.19\linewidth]{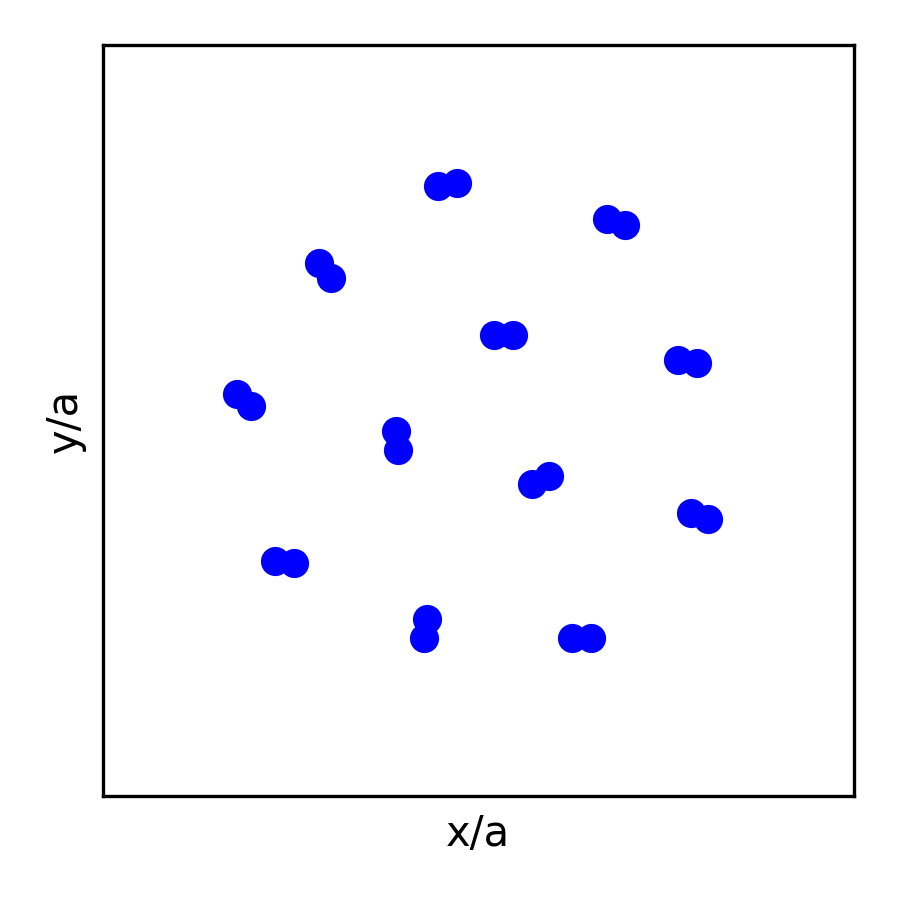} &
            \includegraphics[width=0.19\linewidth]{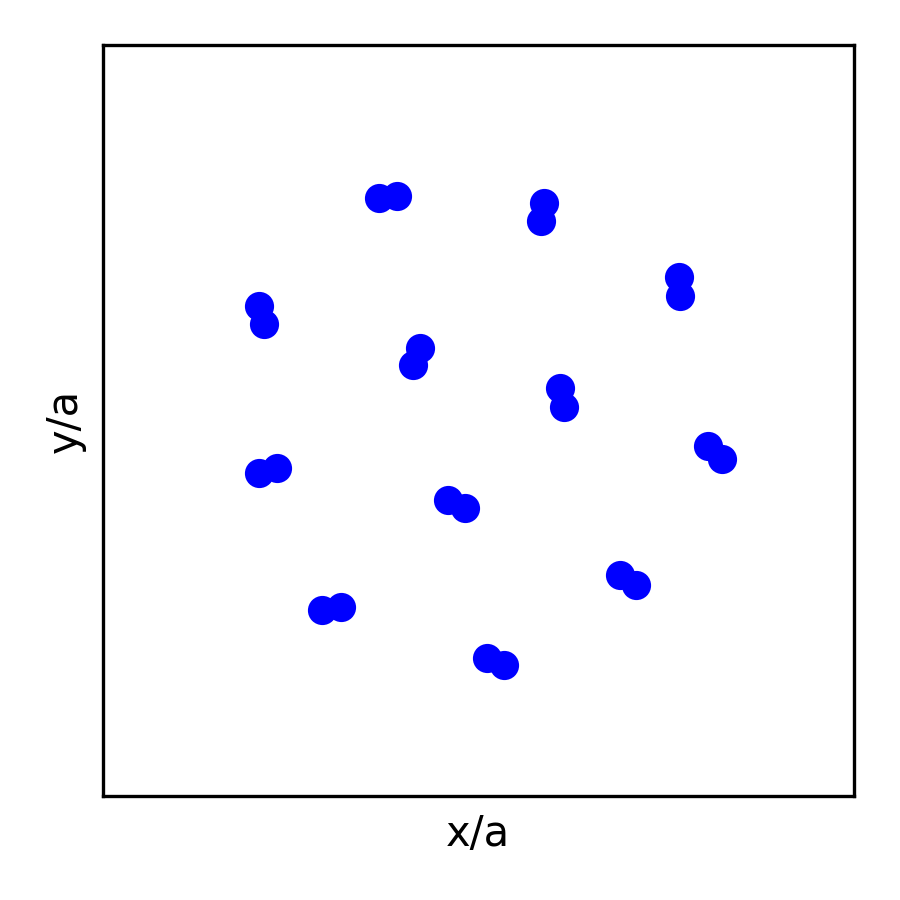} &
            \includegraphics[width=0.19\linewidth]{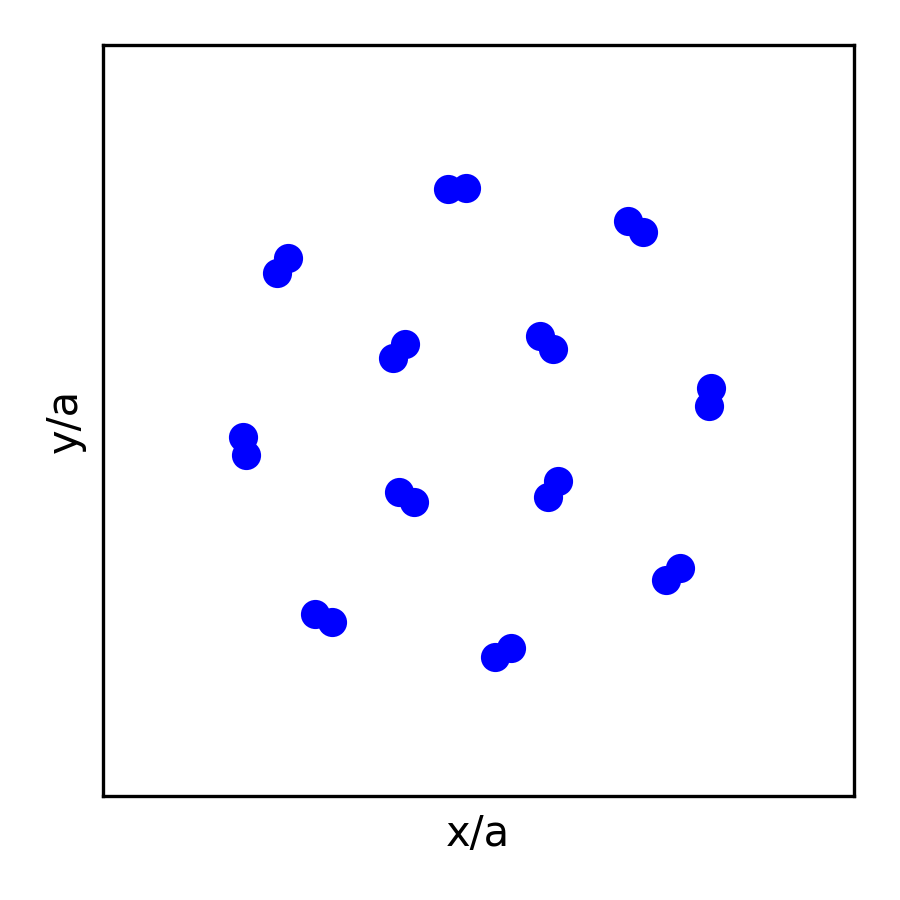} \\ \hline
        14 & \includegraphics[width=0.19\linewidth]{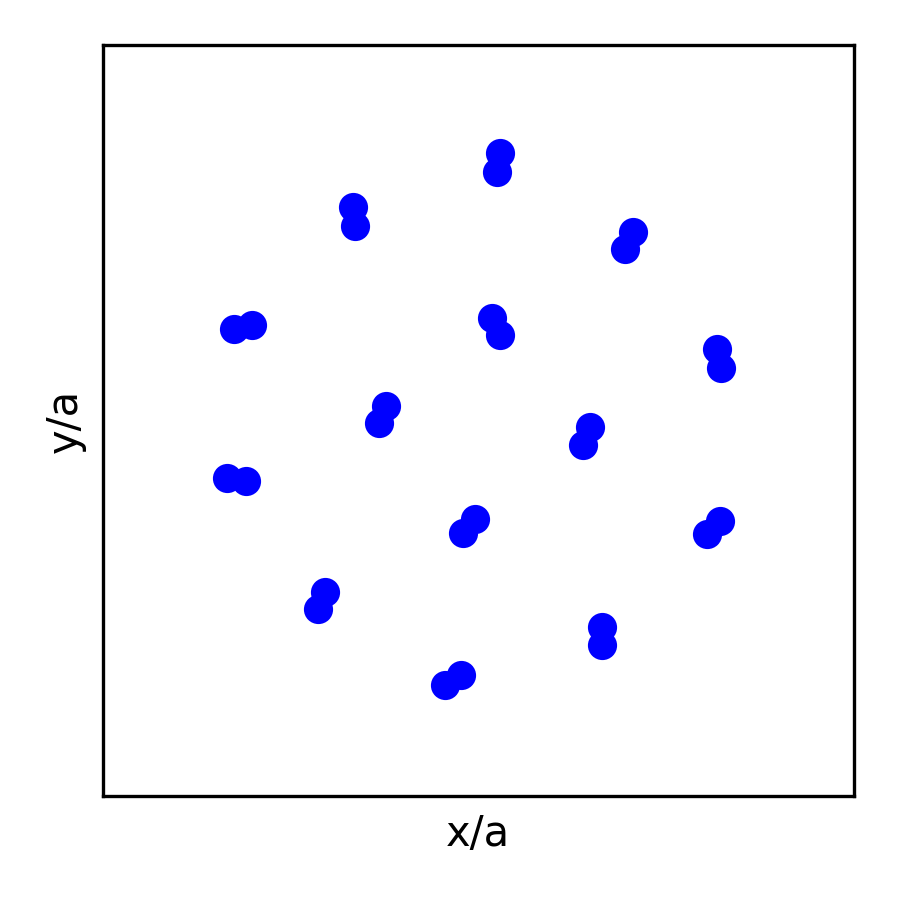} & 
            \includegraphics[width=0.19\linewidth]{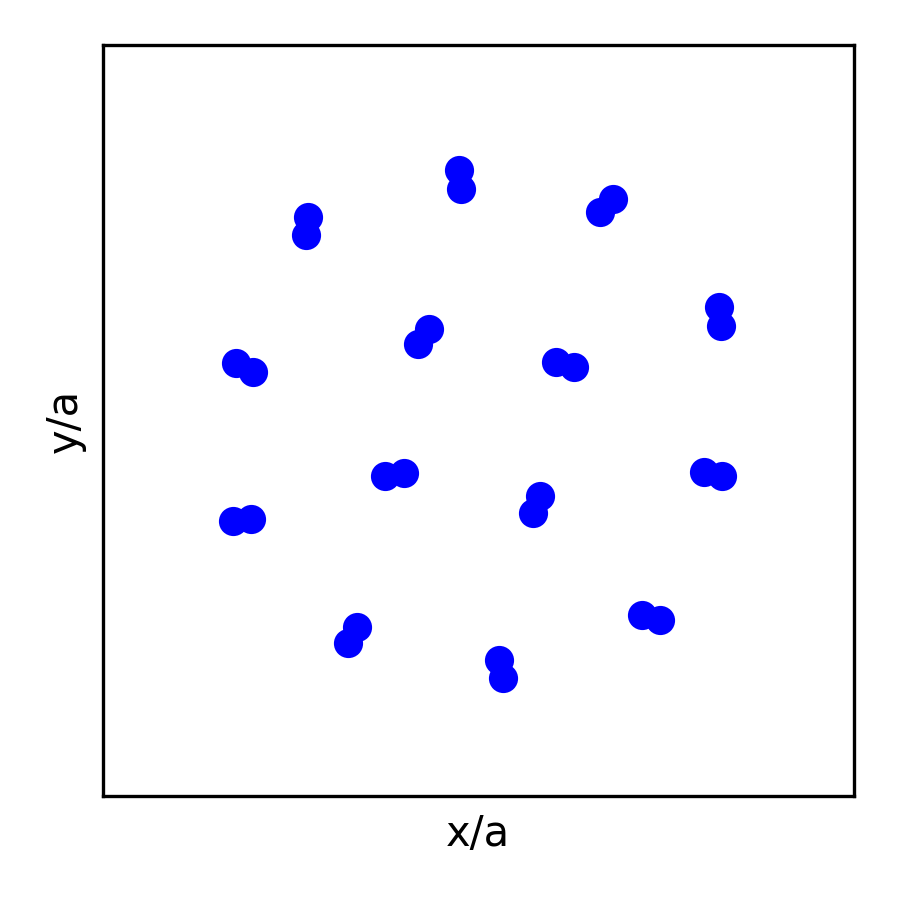} &
            \includegraphics[width=0.19\linewidth]{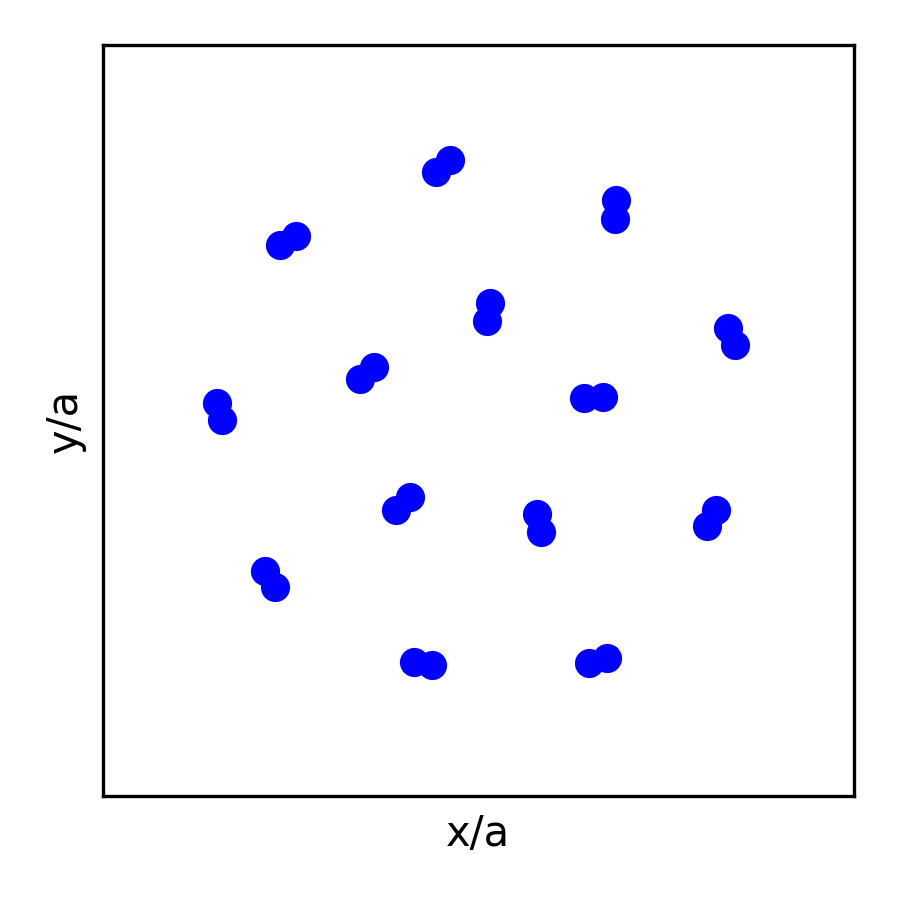} &
            \includegraphics[width=0.19\linewidth]{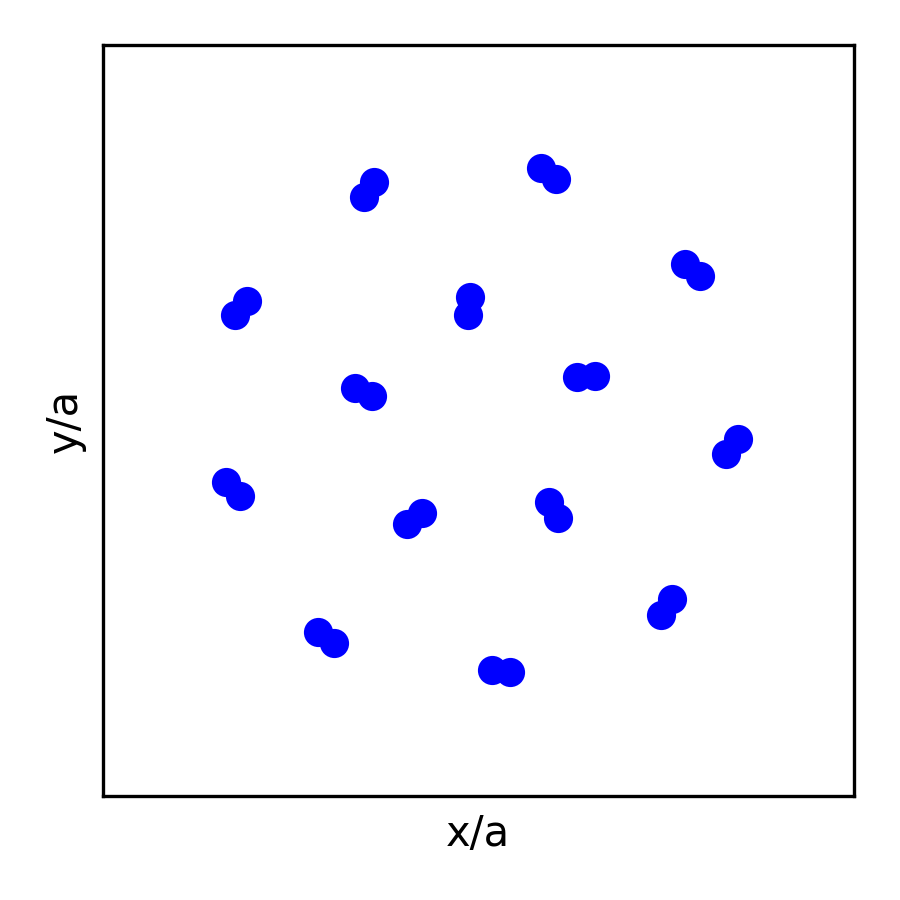} \\ \hline
        15 & \includegraphics[width=0.19\linewidth]{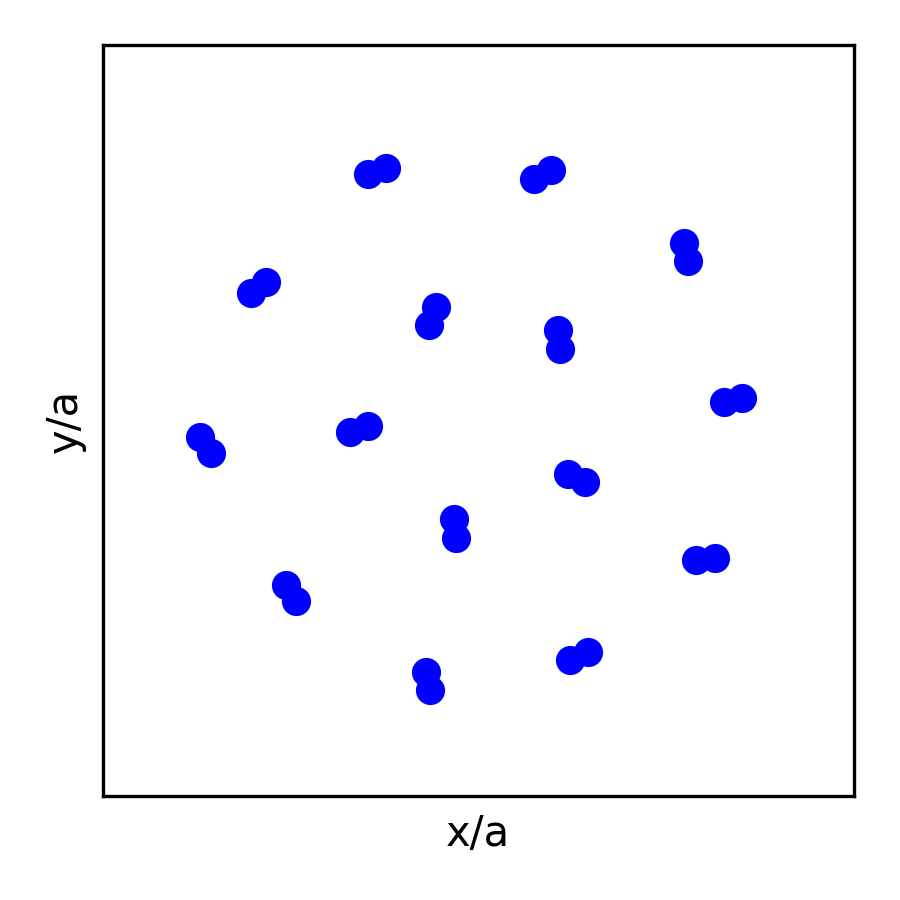} & 
            \includegraphics[width=0.19\linewidth]{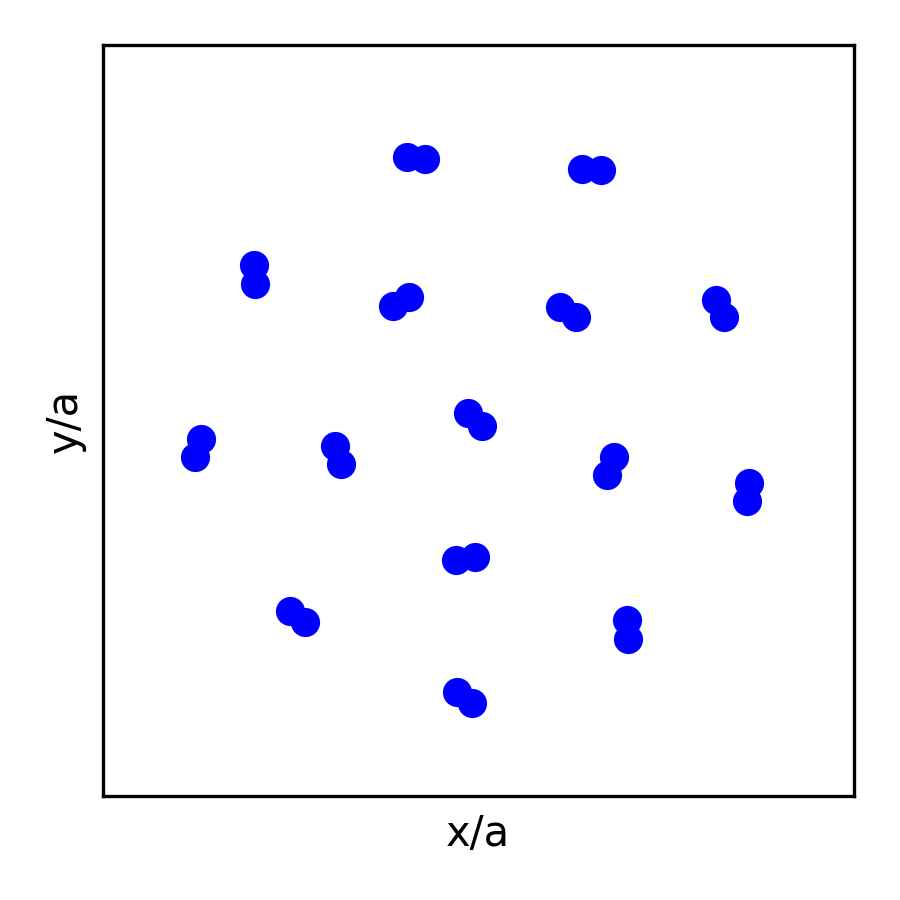} &
            \includegraphics[width=0.19\linewidth]{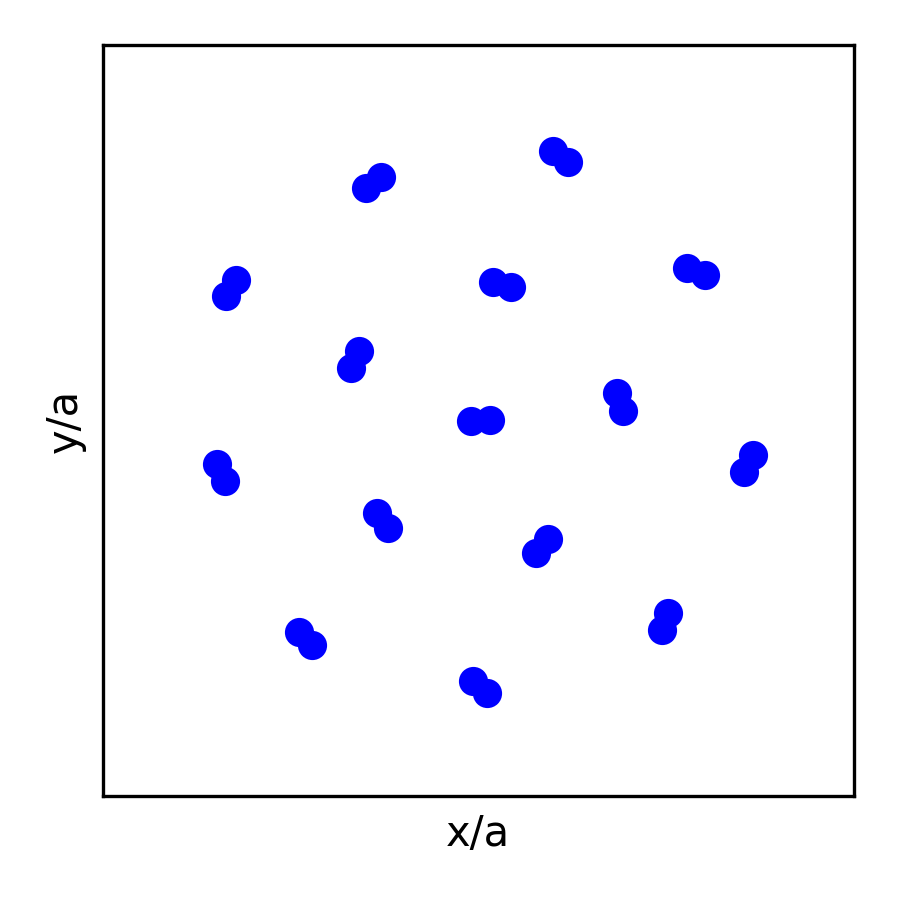} &
            \includegraphics[width=0.19\linewidth]{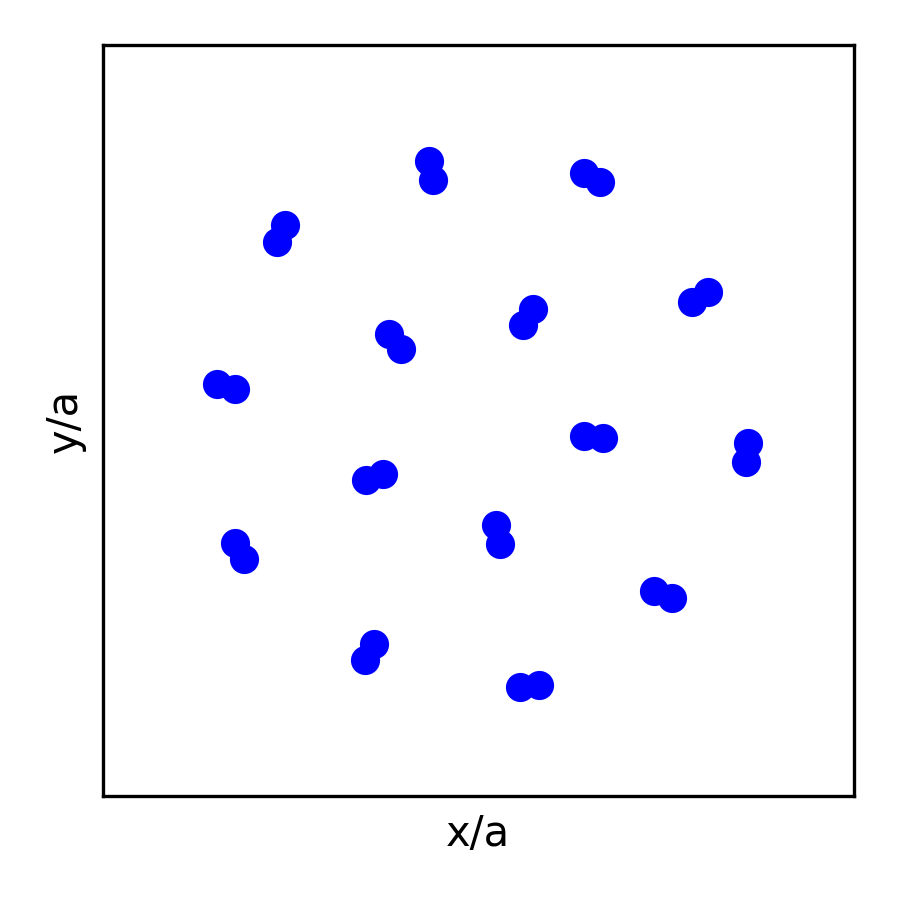} \\ \hline
        16 & \includegraphics[width=0.19\linewidth]{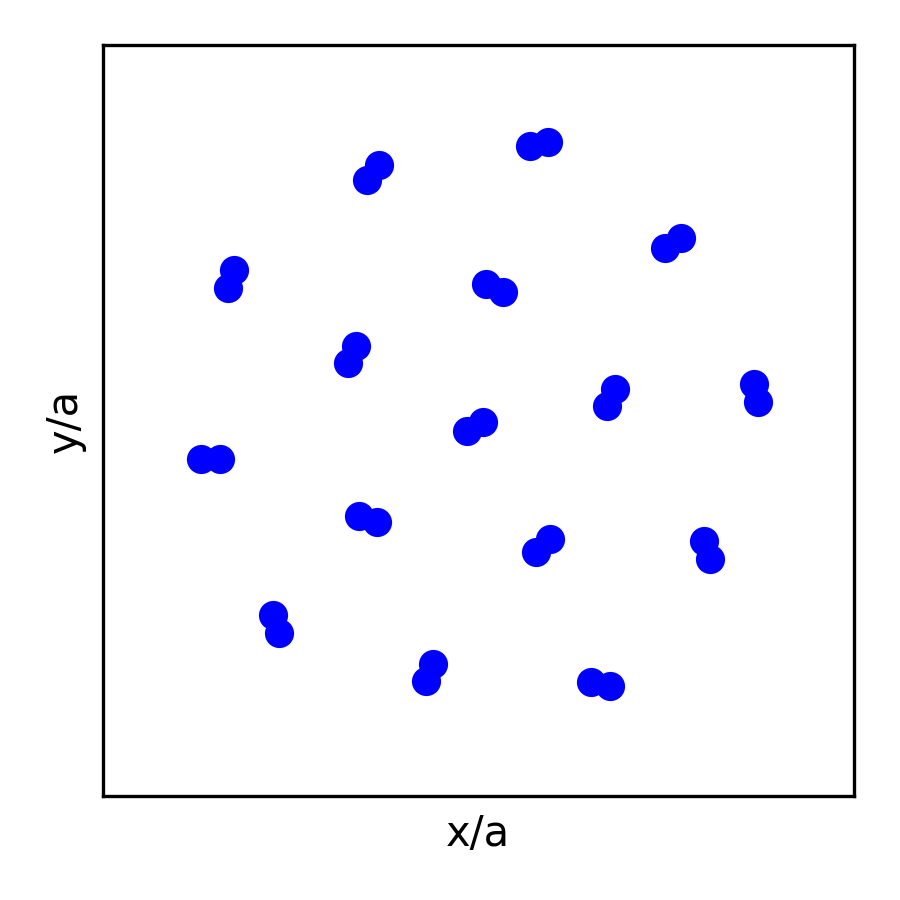} & 
            \includegraphics[width=0.19\linewidth]{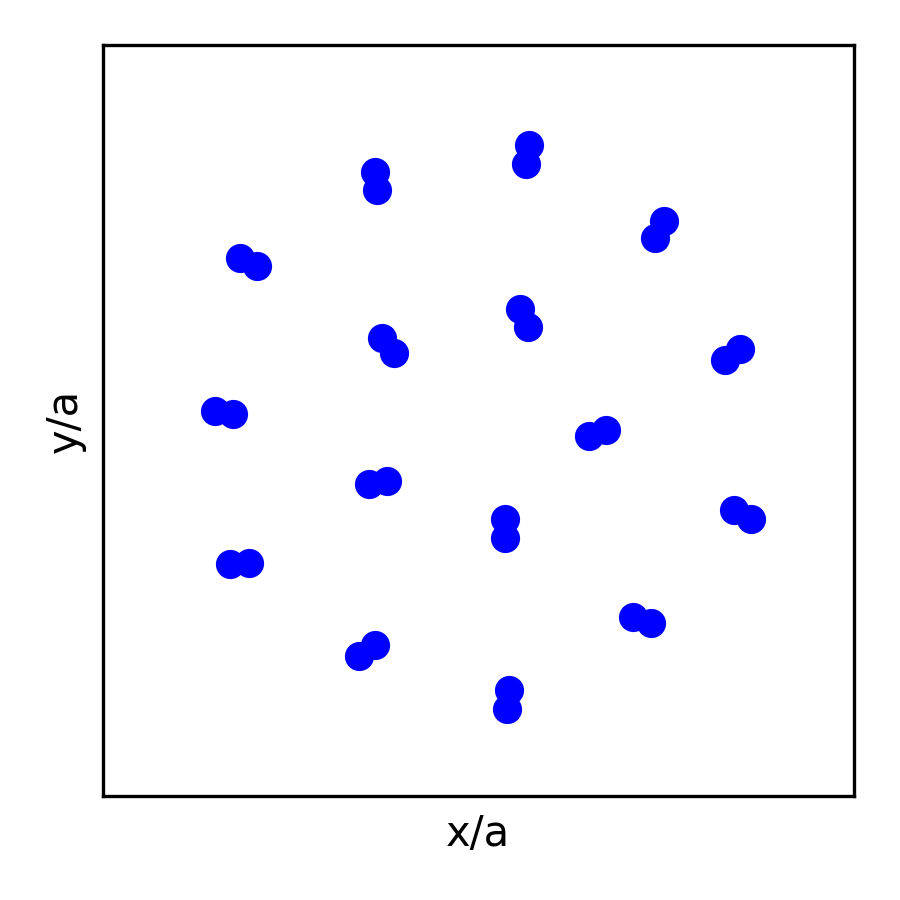} &
            \includegraphics[width=0.19\linewidth]{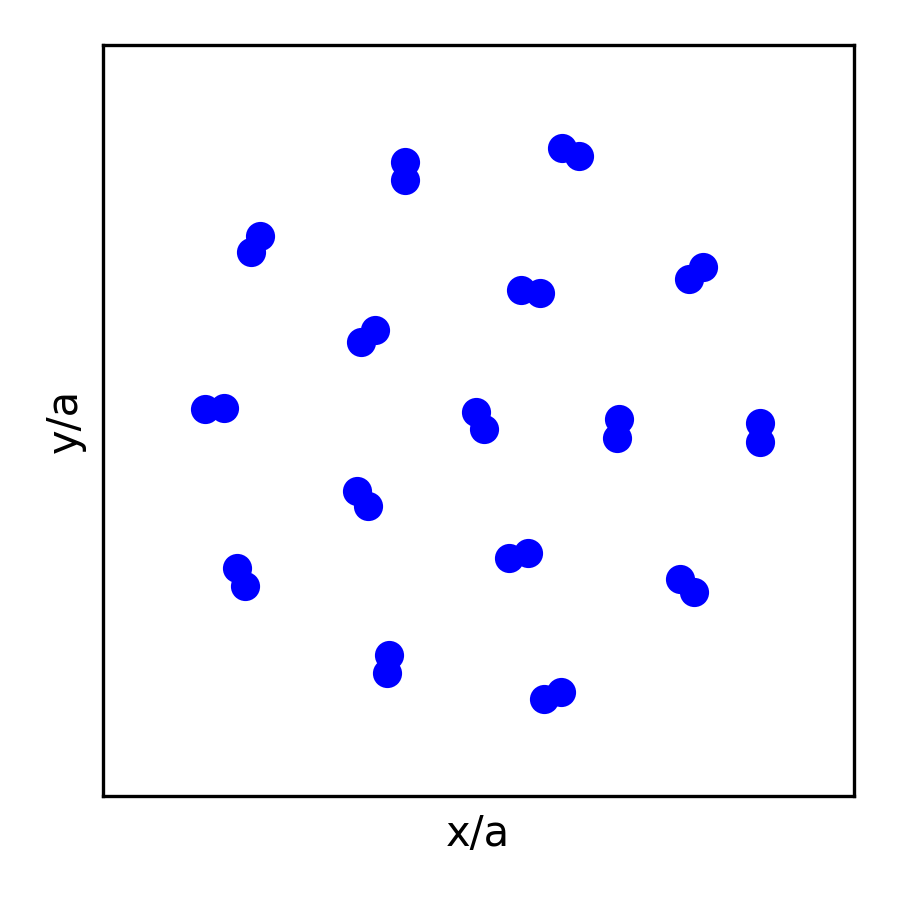} &
            \includegraphics[width=0.19\linewidth]{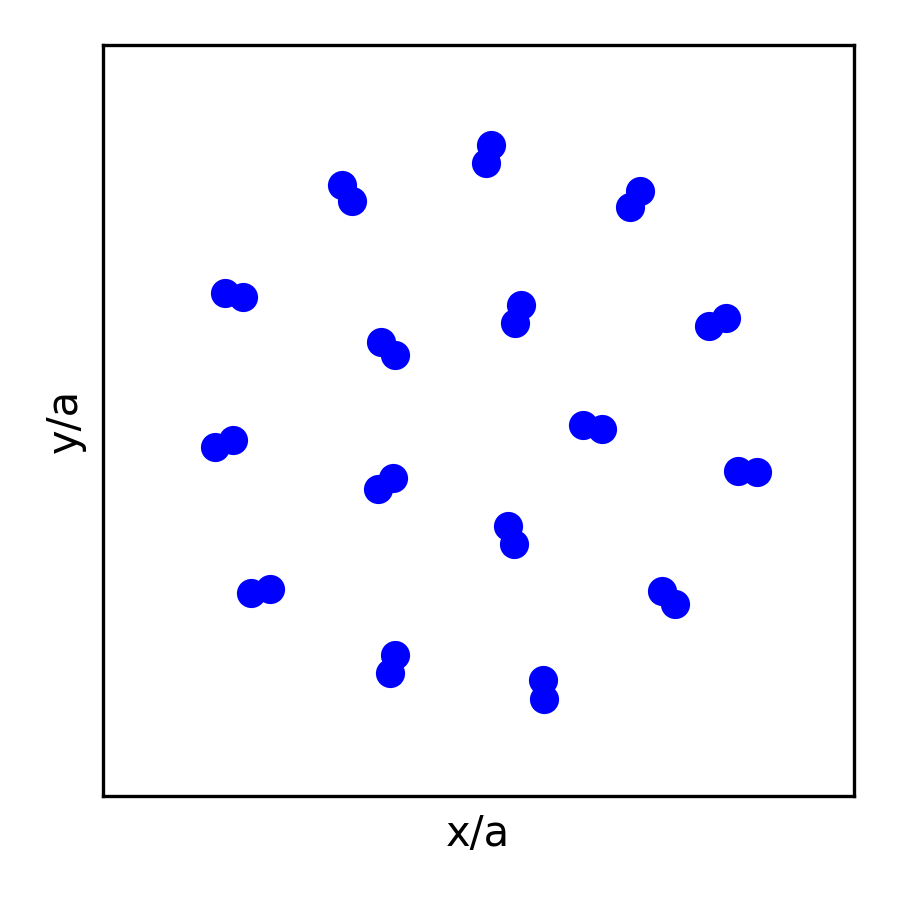} \\ \hline
     \end{tabular}
 \end{table*}

\section{Self-organization in periodic simulation box}

\begin{figure*}[htbp!]
    \centering
    \includegraphics[width=0.8\linewidth]{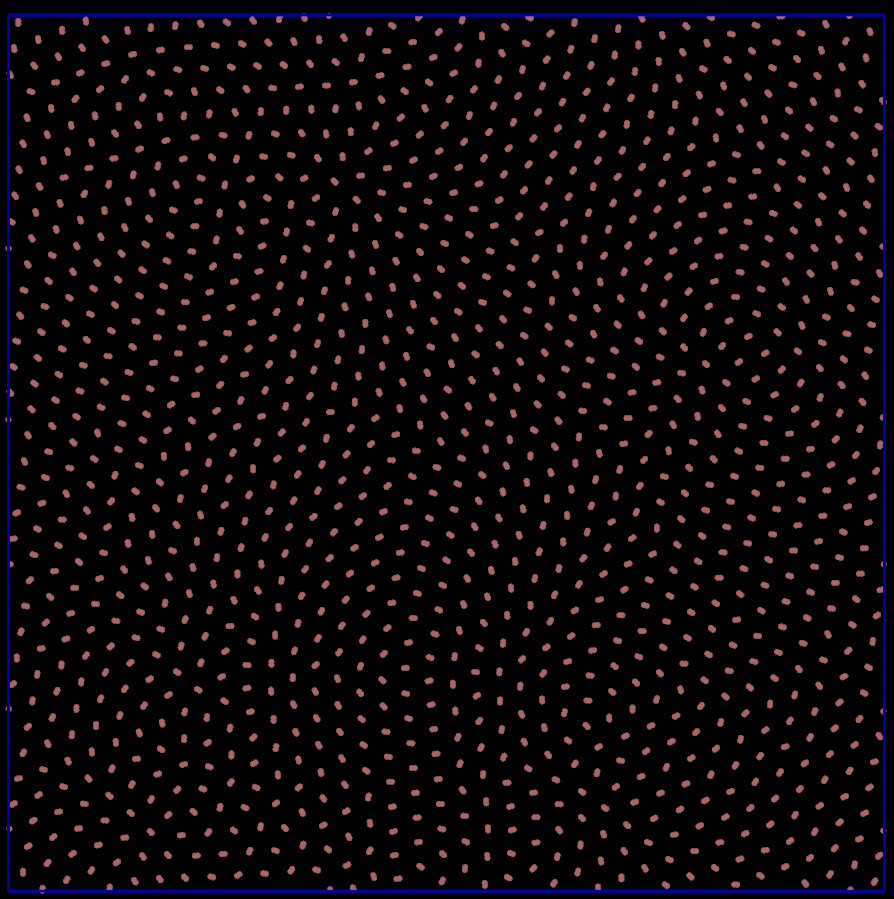}
    \caption{Structural arrangement of dimers in a periodic boundary condition where blue line defines the boundary. It can be seen that a periodic lattice like picture emerges from the dimer positions. However, their orientation seems to be random overall, with some directionality in some local regions.}
    \label{figr:1288}
\end{figure*}

\begin{figure}[htbp!]
    \centering
    \includegraphics[width=0.9 \linewidth]{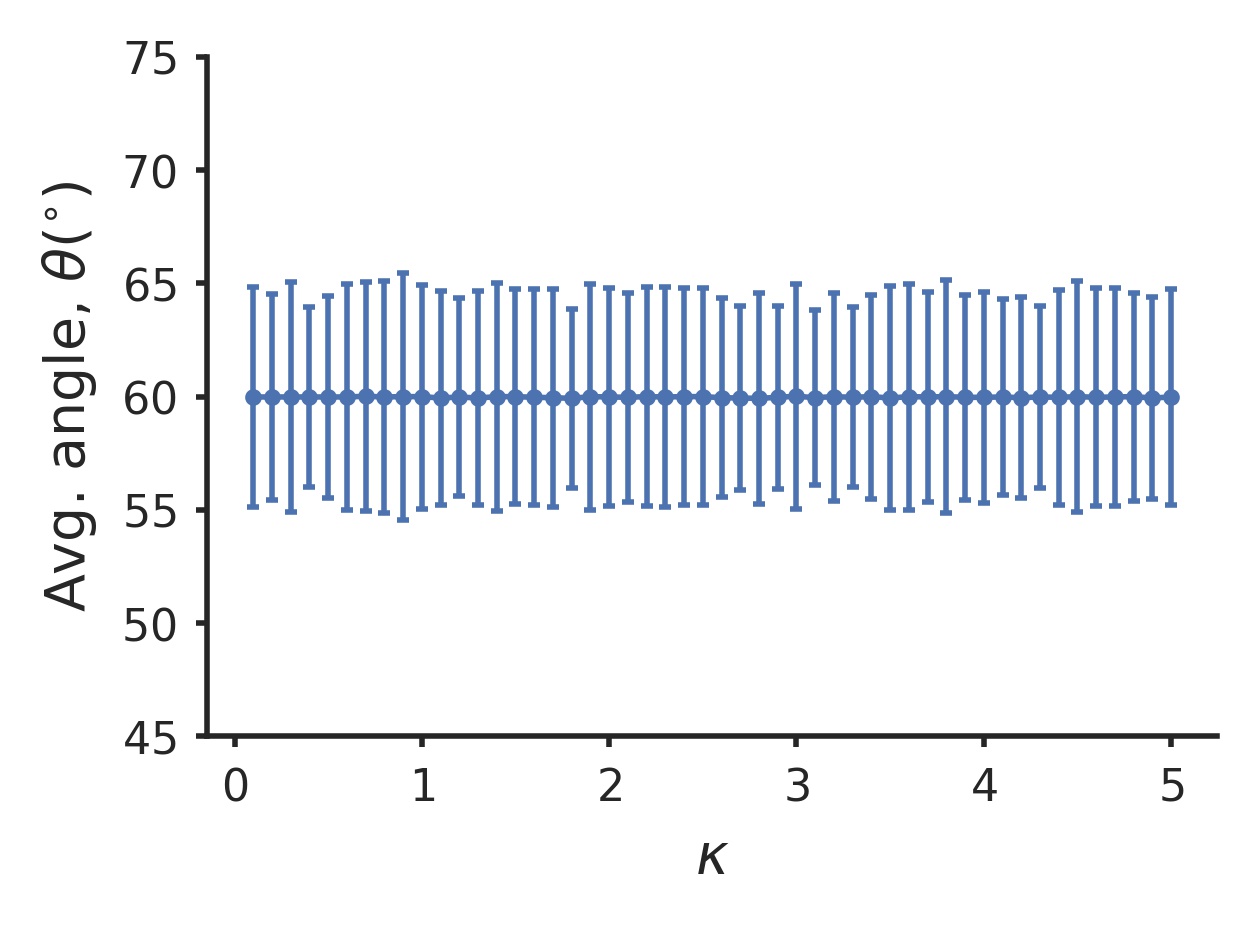}
    \caption{Average angle between the dimer centers arranged in periodic boundary conditions}
    \label{fig:avgangle}
\end{figure}
As the number of dimers in the system increases and the simulation box becomes progressively filled, ordered lattice-like structures begin to emerge even in the absence of an external confining field. In this regime, the periodic boundaries effectively impose a global constraint on the system, promoting long-range ordering through the interplay of inter-dimer interactions and finite packing density. To investigate this behavior, we perform simulations under periodic boundary conditions and examine the resulting positional and orientational ordering of the dimers.

A system containing 1288 dimers is initialized in a two-dimensional periodic box, with all dimers initially aligned parallel to the (y)-axis. A Langevin thermostat is then employed to allow the system to relax toward equilibrium. Upon equilibration, the dimers self-organize into well-defined periodic lattice structures. The centers of mass of the dimers predominantly occupy the sites of a hexagonal lattice, similar to the ordering observed previously for monomeric particles. Figure~\ref{figr:1288} illustrates this arrangement. Although the lattice is largely hexagonal, a small number of topological defects are present, where the local coordination resembles pentagonal or heptagonal packing rather than the ideal sixfold symmetry.

While the positional ordering is strongly hexagonal, the orientational ordering of the dimer axes is considerably more complex. On a global scale, the dimer orientations appear largely disordered; however, closer inspection reveals the existence of locally correlated domains in which neighboring dimers exhibit a preferred common orientation. Thus, the system displays long-range translational order together with only short-range orientational order.

An interesting comparison can be made with the behavior of monomer systems. Previous studies have shown that increasing the screening parameter ($\kappa$) can induce a structural transition from a hexagonal lattice to a square lattice. In particular, it has been reported in \cite{maity2019molecular} that the average angle between the primitive lattice vectors changes abruptly from approximately ($60^{\circ}$) to nearly ($90^{\circ}$) as ($\kappa$) is increased, signaling a transition from hexagonal to square ordering. In contrast, no such transition is observed for dimer assemblies. As shown in Fig.~\ref{fig:avgangle}, the average angle between the lattice vectors remains close to ($60^{\circ}$) over the entire range of ($\kappa$) considered, indicating that the hexagonal lattice remains the energetically preferred structure.

This robustness of the hexagonal phase can be attributed to the additional orientational degree of freedom possessed by the dimers. Unlike monomers, which can only adjust their positions in response to changes in screening, dimers can simultaneously reorganize both their positions and orientations. By rotating their body axes, neighboring dimers can partially compensate for the effects of increased screening and reduce unfavorable interactions. This additional mechanism for energy minimization stabilizes the hexagonal packing and suppresses the hexagonal-to-square lattice transition that is otherwise observed in monomer systems.

The orientational ordering of the dimers in the periodic system is again quantified through an appropriate order parameter. Unlike the confined clusters discussed earlier, the present system does not possess a natural reference direction, such as the radially inward electric field. Consequently, the orientational order cannot be characterized relative to a predefined axis and is instead described using the nematic order tensor
 $Q$,
\begin{equation}
    Q = \left< (d \cdot u_\alpha u_\beta - \delta_{\alpha\beta}) \right>
\end{equation}
where ($d=2$) for a two-dimensional system, ($\mathbf{u}$) is the unit vector directed along the dimer axis, and ($\delta_{\alpha\beta}$ ) is the Kronecker delta. The angular brackets denote an average over all dimers in the system.
Two measures of orientational order are considered. First, we evaluate the component ($Q_{xx}$ ), denoted by ($S_x$), which quantifies the degree of alignment of the dimers with respect to the ($x$)-axis. A value of ($S_x=1$) corresponds to perfect alignment parallel to the ($x$)-direction, while ($S_x=-1$) indicates perfect alignment perpendicular to it. Intermediate values represent partial alignment.

Second, we compute the global nematic order parameter ($S_{\mathrm{global}}$  ), defined as the largest eigenvalue of the tensor ($Q$). This quantity provides a coordinate-independent measure of the overall orientational order in the system. For a perfectly ordered nematic state, in which all dimers are aligned along a common direction, ($S_{\mathrm{global}}$) approaches unity. Conversely, for a completely isotropic system with randomly oriented dimers, ($S_{\mathrm{global}}$ ) tends toward zero. The combination of ($S_x$) and ($S_{\mathrm{global}}$) therefore enables us to characterize both the preferred direction of alignment and the overall degree of orientational ordering within the periodic dimer lattice.
\begin{figure}[htbp!]
    \centering
    \includegraphics[width=1.\linewidth]{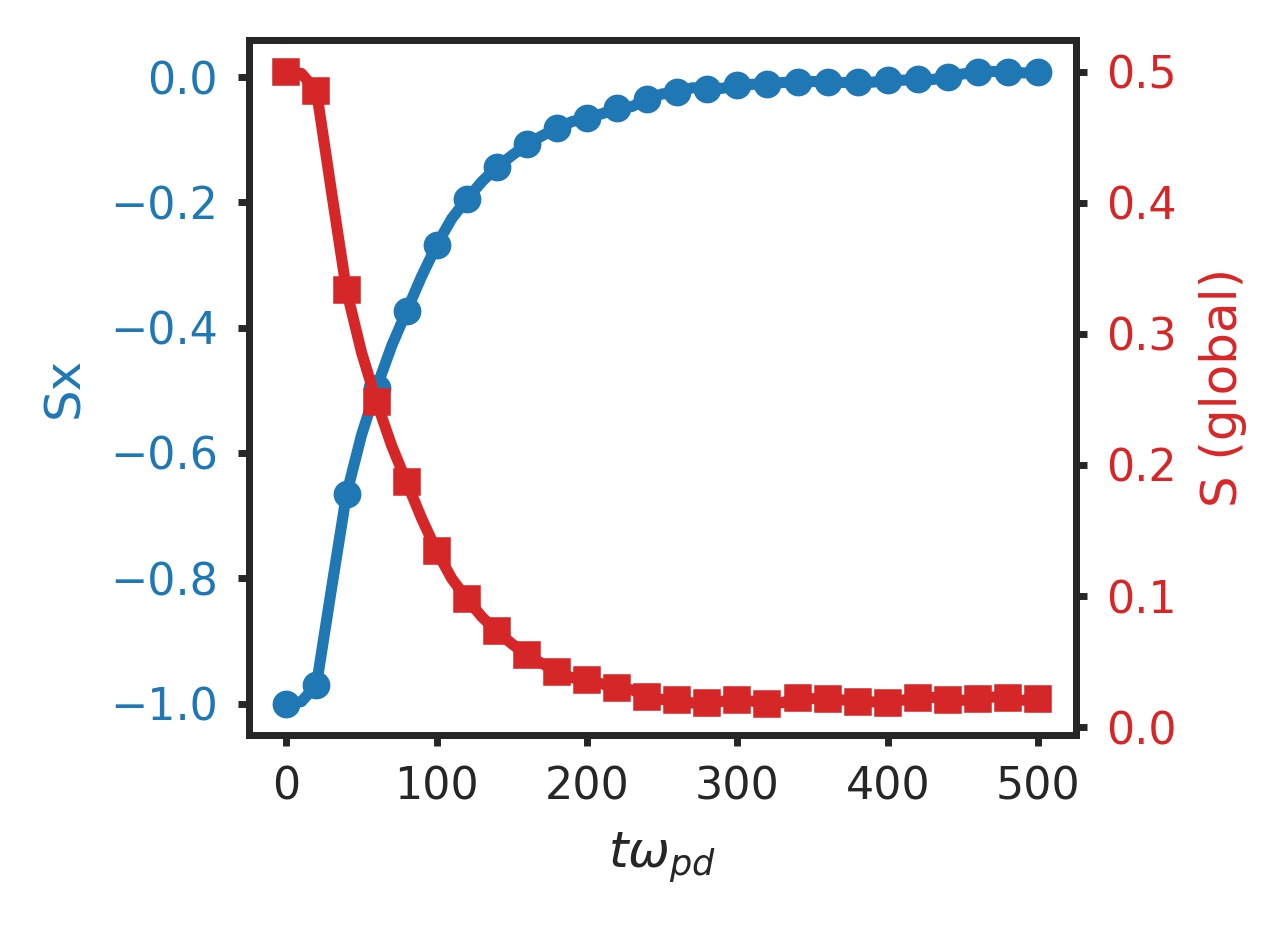}
    \caption{Evolution of the order parameter along the x-axis.}
    \label{fig:sx_vs_t}
\end{figure}
Figure~\ref{fig:sx_vs_t} shows the temporal evolution of the orientational order parameters during the relaxation process. Since all dimers are initially aligned along the (y)-axis, their orientation is perpendicular to the (x)-axis, resulting in an initial value of ($S_x=-1$). As the system evolves and approaches equilibrium, ($S_x$) gradually increases and eventually stabilizes near zero. This indicates the loss of any preferential alignment either parallel or perpendicular to the (x)-direction.

The evolution of the global nematic order parameter, ($S_{\mathrm{global}}$ ), provides further insight into the relaxation dynamics. Initially, ($S_{\mathrm{global}}\approx 0.5$), corresponding to a state with substantial, though not perfect, orientational order. As the system relaxes, ($S_{\mathrm{global}}$) steadily decreases and approaches zero, signifying a transition from a partially ordered state to a globally isotropic one. Thus, despite the emergence of a well-defined hexagonal positional lattice, the dimer orientations themselves do not exhibit long-range orientational order.

This behavior implies that, in the absence of an externally imposed directional field, the dimers possess no globally preferred orientation. Instead, each dimer tends to align locally with the net electric field generated by its neighboring particles, either parallel or antiparallel to it, resulting in short-range orientational correlations. As relaxation proceeds, these local preferences average out across the system, leading to vanishing global order parameters. The convergence of both ($S_x$) and ($S_{\mathrm{global}}$ ) toward zero therefore confirms that the equilibrium state is orientationally isotropic despite retaining strong translational order.

This observation is also consistent with the results obtained for confined dimer clusters discussed in the previous section. As the cluster size increased and multiple shells emerged, the orientational order parameter progressively decreased and eventually fluctuated within a narrow range around zero (typically between (-0.2) and (0.2)). The periodic system may therefore be viewed as the large-system limit of the confined clusters, where the orientational ordering becomes increasingly frustrated and ultimately gives way to a globally isotropic state while preserving local orientational correlations.

\begin{figure}[htbp!]
    \centering
    \includegraphics[width=\linewidth]{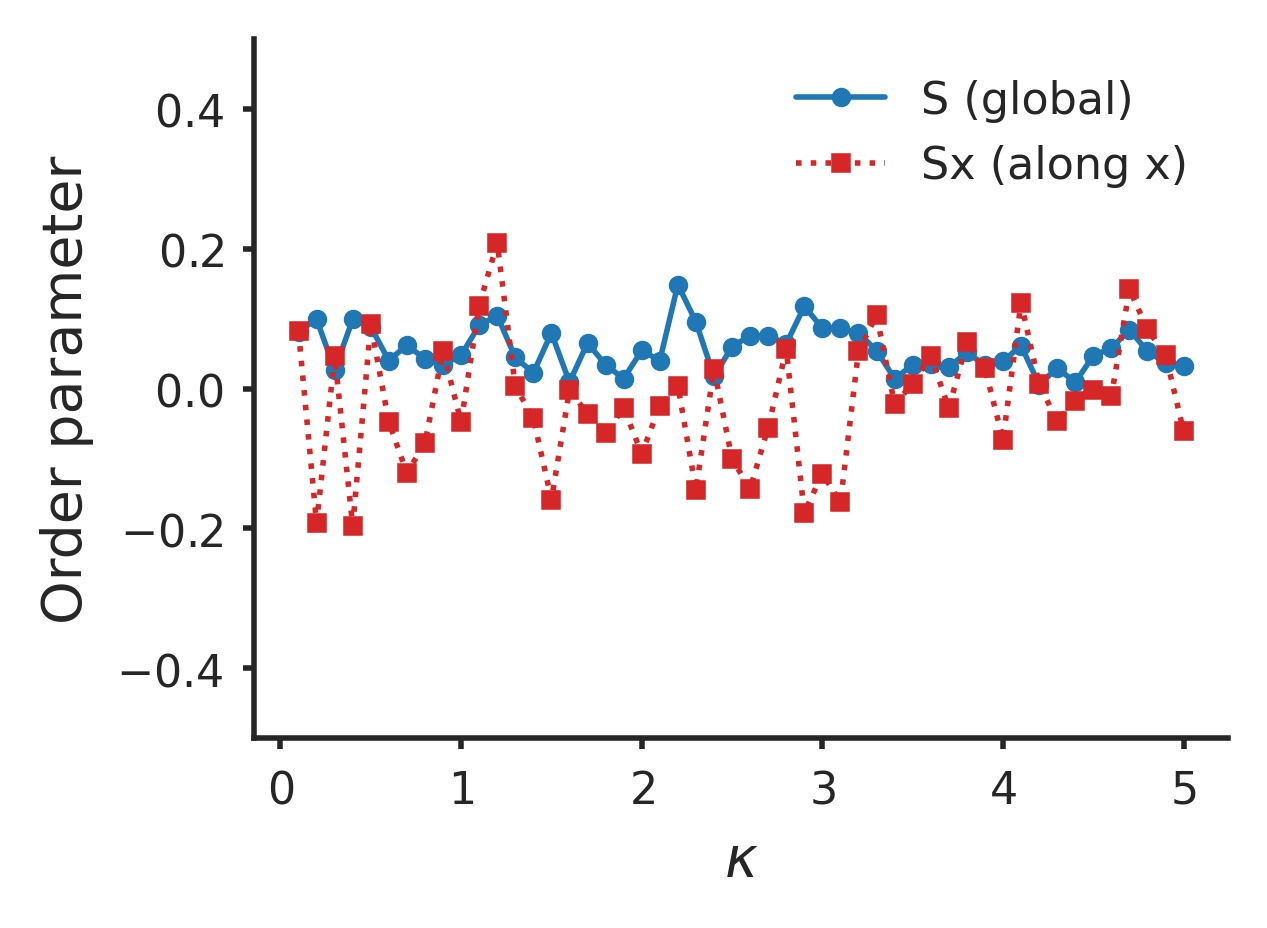}
    \caption{Order parameter v/s shielding parameter $\kappa$}
    \label{fig:svskappa}
\end{figure}
To further examine the orientational ordering in the periodic system, we study the dependence of the equilibrium order parameters ($S_x$) and ($S_{\mathrm{global}}$ ) on the normalized screening parameter ($\kappa$), varied over the range ($0 \leq \kappa \leq 5$). The results are presented in Fig.~\ref{fig:svskappa}. It is evident that ($S_x$) exhibits small fluctuations within the range (-0.2) to (0.2) over the entire range of ($\kappa$). Such small values indicate the absence of any systematic tendency for the dimers to align either parallel or perpendicular to the (x)-axis. In other words, the dimer orientations remain effectively random with respect to this reference direction. The fluctuations observed in ($S_x$) arise primarily from finite-size effects and statistical variations associated with the finite number of dimers in the simulation.

A similar conclusion is reached from the behavior of the global nematic order parameter, ($S_{\mathrm{global}}$ ). Although ($S_{\mathrm{global}}$ ) remains slightly above zero, it exhibits only a weak dependence on ($\kappa$) and never approaches values characteristic of a nematically ordered phase. The consistently small magnitude of ($S_{\mathrm{global}}$ ) indicates the absence of long-range orientational order and confirms that no preferred global alignment direction emerges in the system.

Taken together, the behaviors of ($S_x$) and ($S_{\mathrm{global}}$ ) demonstrate that the equilibrium dimer assembly remains orientationally isotropic throughout the range of screening parameters investigated. Thus, while the screening parameter can influence the positional arrangement and interaction strength between neighboring dimers, it has little effect on the global orientational ordering under periodic boundary conditions. The dimers continue to form a translationally ordered lattice while maintaining an essentially isotropic orientational distribution.

\begin{figure}
    \centering
    \includegraphics[width=\linewidth]{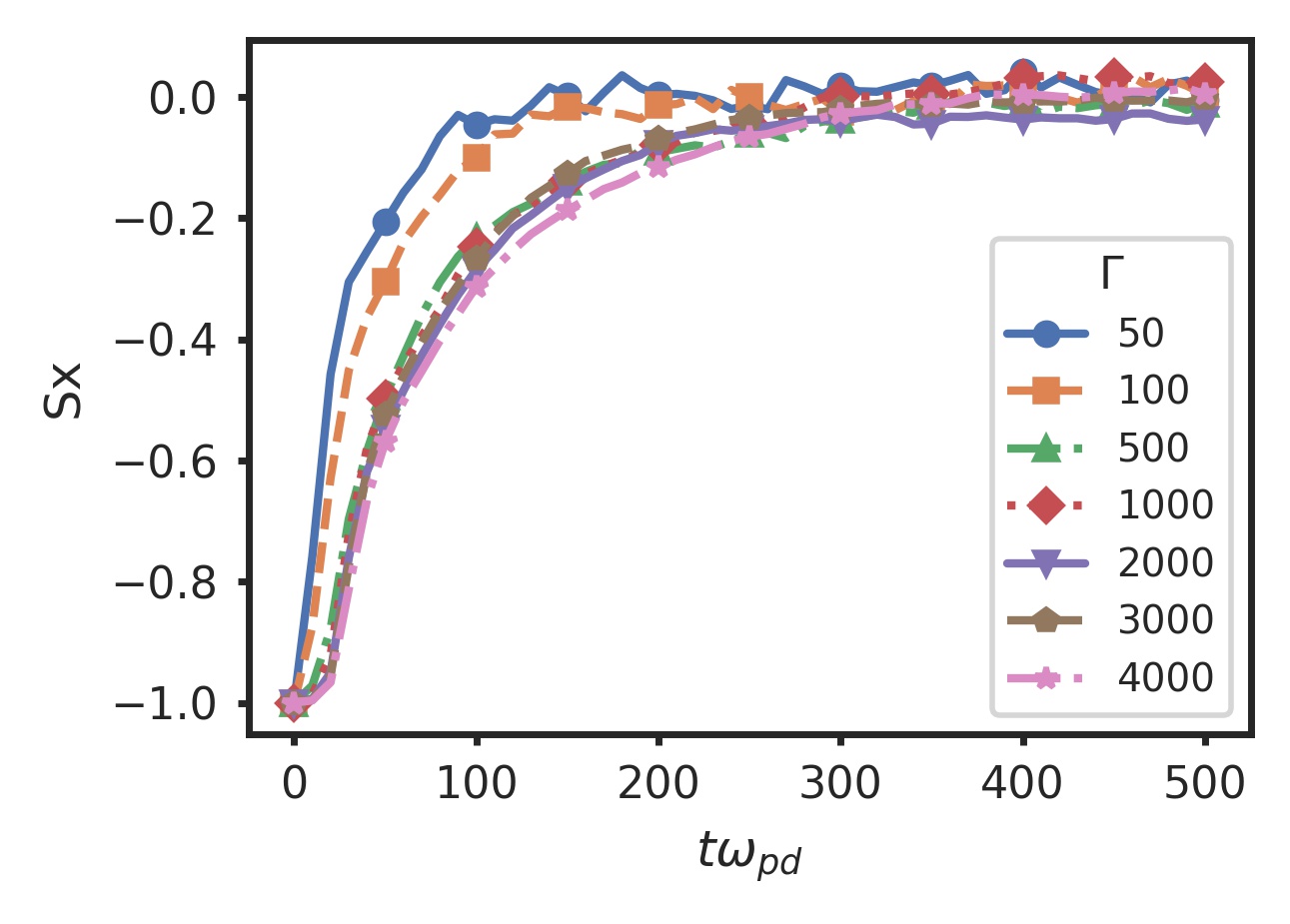}
    \caption{Variation of $S_x$ with time with increasing coupling strength $\Gamma$.}
    \label{fig:SxVsG}
\end{figure}
Finally, we investigate the influence of the coupling strength ($\Gamma$) on the orientational dynamics of large dimer assemblies under periodic boundary conditions. Since the directional order parameter ($S_x$) and the global nematic order parameter ($S_{\mathrm{global}}$  ) exhibit similar trends and provide consistent measures of orientational ordering, we restrict our discussion here to the evolution of ($S_x$). Simulations were carried out for coupling strengths spanning the range ($50 \leq \Gamma \leq 4000$).
For all cases, the dimers are initially aligned along the (y)-axis, corresponding to an initial value of ($S_x=-1$), since the dimer axes are perpendicular to the (x)-direction. Figure~\ref{fig:SxVsG} shows the temporal evolution of ($S_x$) for different values of ($\Gamma$). In every case, the order parameter evolves toward a value close to zero, indicating that the system ultimately reaches an orientationally isotropic state. However, the rate at which this relaxation occurs depends strongly on the coupling strength.

For relatively small values of ($\Gamma$), ($S_x$) rises rapidly from (-1) toward zero. In this regime, thermal kinetic energy is comparable to or larger than the interaction energy, enabling the dimers to reorient and rearrange efficiently. Consequently, the system quickly loses memory of its initial alignment and approaches the isotropic equilibrium state.
As ($\Gamma$) increases, the relaxation becomes progressively slower. In the strongly coupled regime, the interaction energy dominates over thermal fluctuations, and the particles become increasingly localized around energetically favorable configurations. Since thermal motion is suppressed, the dimers explore the configurational phase space less efficiently and require longer times to modify their orientations. As a result, the evolution of ($S_x$) becomes more gradual with increasing ($\Gamma$).
Despite these differences in the relaxation timescale, the long-time behavior remains remarkably similar across the entire range of coupling strengths considered. Even for the highest values of ($\Gamma$), the order parameter eventually approaches a value close to zero, demonstrating that no preferred global orientation develops in the equilibrium state. Thus, increasing ($\Gamma$) primarily affects the kinetics of relaxation rather than the nature of the final state. The system becomes progressively "colder" and slower to equilibrate, yet the equilibrium orientational order remains essentially unchanged, reinforcing the conclusion that large dimer assemblies under periodic boundary conditions evolve toward an orientationally isotropic state.

\section{Summary}
In this work, we have investigated the positional and orientational ordering of charged dimers in strongly coupled dusty plasmas using molecular dynamics simulations. Two complementary configurations were considered: finite dimer clusters confined by a radially symmetric electric field and extended systems subject to periodic boundary conditions. The study demonstrates that the anisotropic nature of dimers introduces a rich interplay between translational and rotational degrees of freedom, leading to phenomena absent in conventional monomer systems.

For confined clusters, the dimers self-organize into ring and shell structures whose orientational ordering evolves with cluster size. Distinct transitions between tangential, radial, and mixed alignments are observed, often accompanying the formation of new shells. Unlike monomer clusters, all dimer assemblies exhibit persistent dynamical behavior, including collective rotations and oscillations, arising from the torque generated by unequal forces acting on the two constituent particles. The degree of orientational ordering is governed by the competition between thermal fluctuations, inter-particle interactions, and the confining electric field.

In contrast, under periodic boundary conditions, the dimers form positionally ordered hexagonal lattices while remaining globally orientationally disordered. Neither the screening parameter nor the coupling strength produces long-range orientational order, indicating a decoupling of translational and rotational ordering in bulk dimer assemblies. The system thus exhibits crystalline positional order together with an orientationally isotropic state.

Beyond its relevance to complex plasmas, the present study highlights the emergence of liquid-crystal-like behavior in a dusty plasma environment. The coexistence and competition of positional and orientational order provide a unique bridge between strongly coupled plasma physics, soft condensed matter, and liquid-crystal systems. Dusty plasmas therefore offer an attractive experimental platform for studying orientational ordering, shell transitions, defect dynamics, and collective rotational states at the individual-particle level.

The results reported here open several directions for future investigation. In particular, the possibility of orientational phase transitions, hysteresis effects, metastable states, defect-mediated ordering, and nonequilibrium rotational dynamics remains largely unexplored. Experimental studies employing shaped dust grains, external fields, or time-dependent confinement may reveal novel phases and transitions analogous to those encountered in liquid crystals and other anisotropic condensed-matter systems. Such investigations could establish complex plasmas as a versatile model system for exploring the fundamental physics of anisotropic strongly coupled matter.
Apart from this, in plasma assisted material processing system, like magnetron sputtering, typically nonspherical or shaped particles can be formed by the coagulation \cite{praburam1995cosmic}. In this work, Praburam and Goree has investigated shaped grain similar to the dimers. In this sense, the present work has the relevance in the area of material processing. 

\section*{Acknowledgments}
AD acknowledges support from the ANRF core grants
CRG/2022/002782 as well as the J C Bose Fellowship grant
ANRF/JBG/2025/000237/PS. BBS acknowledges support from the ANRF core grant CRG/2022/001470. ASK would like to acknowledge support from the Council for Scientific and Industrial Research grant 09/0086(13737)/2022-EMR-I.

\bibliography{bibliography}

@article{ichiki2004,
  title = {Melting and heating of two-dimensional Coulomb clusters in dusty plasmas},
  author = {Ichiki, R. and Ivanov, Y. and Wolter, M. and Kawai, Y. and Melzer, A.},
  journal = {Phys. Rev. E},
  volume = {70},
  issue = {6},
  pages = {066404},
  numpages = {4},
  year = {2004},
  month = {Dec},
  publisher = {American Physical Society},
  doi = {10.1103/PhysRevE.70.066404},
}

@book{nicolis1977self,
  title={Self-Organization in Nonequilibrium Systems: From Dissipative Structures to Order Through Fluctuations},
  author={Nicolis, G. and Prigogine, I.},
  isbn={9780471024019},
  lccn={76049019},
  series={A Wiley-Interscience publication},
  year={1977},
  publisher={Wiley}
}

@article { DeterministicNonperiodicFlow,
      author = "Edward N.  Lorenz",
      title = "Deterministic Nonperiodic Flow",
      journal = "Journal of Atmospheric Sciences",
      year = "1963",
      publisher = "American Meteorological Society",
      address = "Boston MA, USA",
      volume = "20",
      number = "2",
      doi = "10.1175/1520-0469(1963)020<0130:DNF>2.0.CO;2",
      pages=      "130 - 141",
}

@article{HAKEN198426,
title = {Synergetics},
journal = {Physica B+C},
volume = {127},
number = {1},
pages = {26-36},
year = {1984},
note = {Proceedings of the 4th General Conference of the Condensed Matter Division of the EPS},
issn = {0378-4363},
doi = {https://doi.org/10.1016/S0378-4363(84)80007-8},
author = {H. Haken},
}

@article{VICSEK201271,
title = {Collective motion},
journal = {Physics Reports},
volume = {517},
number = {3},
pages = {71-140},
year = {2012},
note = {Collective motion},
issn = {0370-1573},
doi = {https://doi.org/10.1016/j.physrep.2012.03.004},
author = {Tamás Vicsek and Anna Zafeiris},
}

@article{cizrok1996,
  title = {Formation of complex bacterial colonies via self-generated vortices},
  author = {Czir\'ok, Andr\'as and Ben-Jacob, Eshel and Cohen, Inon and Vicsek, Tam\'as},
  journal = {Phys. Rev. E},
  volume = {54},
  issue = {2},
  pages = {1791--1801},
  numpages = {0},
  year = {1996},
  month = {Aug},
  publisher = {American Physical Society},
  doi = {10.1103/PhysRevE.54.1791},
}

@article{sokolov2007,
  title = {Concentration Dependence of the Collective Dynamics of Swimming Bacteria},
  author = {Sokolov, Andrey and Aranson, Igor S. and Kessler, John O. and Goldstein, Raymond E.},
  journal = {Phys. Rev. Lett.},
  volume = {98},
  issue = {15},
  pages = {158102},
  numpages = {4},
  year = {2007},
  month = {Apr},
  publisher = {American Physical Society},
  doi = {10.1103/PhysRevLett.98.158102},
}

@article{cizrok2001,
  title = {Theory of periodic swarming of bacteria: Application to Proteus mirabilis},
  author = {Czir\'ok, A. and Matsushita, M. and Vicsek, T.},
  journal = {Phys. Rev. E},
  volume = {63},
  issue = {3},
  pages = {031915},
  numpages = {11},
  year = {2001},
  month = {Feb},
  publisher = {American Physical Society},
  doi = {10.1103/PhysRevE.63.031915},
}

@article{antcolonies,
author = {Detrain, Claire and C., Natan and Deneubourg, Jean-Louis},
year = {2001},
month = {05},
pages = {171-4},
title = {The influence of the physical environment on the self-organised patterns of ants},
volume = {88},
journal = {Die Naturwissenschaften},
doi = {10.1007/s001140100217}
}

@book{de1993physics,
  title={The Physics of Liquid Crystals},
  author={de Gennes, P.G. and Prost, J.},
  isbn={9780198517856},
  lccn={93009154},
  series={International Series of Monographs on Physics},
  year={1993},
  publisher={Clarendon Press}
}

@article{birdflock_reynolds,
author = {Reynolds, Craig W.},
title = {Flocks, herds and schools: A distributed behavioral model},
year = {1987},
issue_date = {July 1987},
publisher = {Association for Computing Machinery},
address = {New York, NY, USA},
volume = {21},
number = {4},
issn = {0097-8930},
doi = {10.1145/37402.37406},
journal = {SIGGRAPH Comput. Graph.},
month = aug,
pages = {25–34},
numpages = {10}
}

@article{bajec2009organizedbirds,
  title={Organized flight in birds},
  author={Bajec, Iztok Lebar and Heppner, Frank H},
  journal={Animal Behaviour},
  volume={78},
  number={4},
  pages={777--789},
  year={2009},
  publisher={Elsevier}
}

@Inbook{Beekman2008,
author="Beekman, Madeleine
and Sword, Gregory A.
and Simpson, Stephen J.",
editor="Blum, Christian
and Merkle, Daniel",
title="Biological Foundations of Swarm Intelligence",
bookTitle="Swarm Intelligence: Introduction and Applications",
year="2008",
publisher="Springer Berlin Heidelberg",
address="Berlin, Heidelberg",
pages="3--41",
isbn="978-3-540-74089-6",
doi="10.1007/978-3-540-74089-6_1",
}

@ARTICLE{szopekhoneybees,
AUTHOR={Szopek, Martina  and Stokanic, Valerin  and Radspieler, Gerald  and Schmickl, Thomas },
TITLE={Simple Physical Interactions Yield Social Self-Organization in Honeybees},
JOURNAL={Frontiers in Physics},
VOLUME={Volume 9 - 2021},
YEAR={2021},
DOI={10.3389/fphy.2021.670317},
ISSN={2296-424X},
}

@article{yadav2025bulk,
  title={Bulk and surface dominated phenomena and the formation of pentagonal structures in 2-D strongly coupled finite dust clusters},
  author={Yadav, Mamta and Katariya, Aman Singh and Sharma, Animesh and Das, Amita},
  journal={Physica D: Nonlinear Phenomena},
  pages={134821},
  year={2025},
  publisher={Elsevier}
}

@article{turing,
    author = {Turing, Alan Mathison},
    title = {The chemical basis of morphogenesis},
    journal = {Philosophical Transactions of the Royal Society of London. B, Biological Sciences},
    volume = {237},
    number = {641},
    pages = {37-72},
    year = {1952},
    month = {08},
    issn = {0080-4622},
    doi = {10.1098/rstb.1952.0012},
}

@article{prigogine&lefever,
    author = {Prigogine, I. and Lefever, R.},
    title = {Symmetry Breaking Instabilities in Dissipative Systems. II},
    journal = {The Journal of Chemical Physics},
    volume = {48},
    number = {4},
    pages = {1695-1700},
    year = {1968},
    month = {02},
}

@article{PBak1987,
  title = {Self-organized criticality: An explanation of the 1/f noise},
  author = {Bak, Per and Tang, Chao and Wiesenfeld, Kurt},
  journal = {Phys. Rev. Lett.},
  volume = {59},
  issue = {4},
  pages = {381--384},
  numpages = {0},
  year = {1987},
  month = {Jul},
  publisher = {American Physical Society},
  doi = {10.1103/PhysRevLett.59.381},
}

@article{decher1997fuzzy,
  title={Fuzzy nanoassemblies: toward layered polymeric multicomposites},
  author={Decher, Gero},
  journal={Science},
  volume={277},
  number={5330},
  pages={1232--1237},
  year={1007},
  publisher={American Association for the Advancement of Science},
  doi={10.1126/science.277.5330.1232}
}

@article{pieranski1983colloidal,
  title={Colloidal crystals},
  author={Pieranski, Pawel},
  journal={Contemporary Physics},
  volume={24},
  number={1},
  pages={25--73},
  year={1983},
  publisher={Taylor \& Francis},
  doi={10.1080/00107518308227495}
}

@article{ruben2004grid,
  title={Grid-type metal ion architectures: Functional metallosupramolecular arrays},
  author={Ruben, Mario and Rojo, Javier and Romero-Salguero, Francisco J and Uppadine, Laurence H and Lehn, Jean-Marie},
  journal={Angewandte Chemie International Edition},
  volume={43},
  number={28},
  pages={3644--3662},
  year={2004},
  publisher={Wiley Online Library},
  doi={10.1002/anie.200300636}
}

@book{Seeley1995TheWO, title={The Wisdom of the Hive: The Social Physiology of Honey Bee Colonies},
  author={Thomas D. Seeley},
  year={1995},
  publisher={Harvard University Press}
}

@book{krugman1996self,
  title={The Self Organizing Economy},
  author={Krugman, P.R.},
  isbn={9781557866998},
  lccn={95031593},
  year={1996},
  publisher={Blackwell Publishers}
}

@article{cross&hohenberg,
  title = {Pattern formation outside of equilibrium},
  author = {Cross, M. C. and Hohenberg, P. C.},
  journal = {Rev. Mod. Phys.},
  volume = {65},
  issue = {3},
  pages = {851--1112},
  numpages = {0},
  year = {1993},
  month = {Jul},
  publisher = {American Physical Society},
  doi = {10.1103/RevModPhys.65.851},
}

@article{degond2014collective,
  title={Collective dynamics and self-organization: some challenges and an example},
  author={Degond, Pierre},
  journal={ESAIM: Proceedings and Surveys},
  volume={45},
  pages={1--21},
  year={2014},
  publisher={EDP Sciences},
  doi={10.1051/proc/201445001}
}

@book{glansdorff1971thermodynamic,
  title={Thermodynamic theory of structure, stability and fluctuations},
  author={Glansdorff, Pierre and Prigogine, Ilya},
  year={1971},
  publisher={Wiley-Interscience}
}

@article{Libbrecht2005snow,
  title={The physics of snow crystals},
  author={Kenneth George Libbrecht},
  journal={Reports on Progress in Physics},
  year={2005},
  volume={68},
  pages={855 - 895},
}

@book{stanley1971introduction,
  title={Introduction to phase transitions and critical phenomena},
  author={Stanley, Harry Eugene},
  year={1971},
  publisher={Oxford University Press}
}

@book{koschmieder1993benard,
  title={B{\'e}nard cells and Taylor vortices},
  author={Koschmieder, E Lothar},
  year={1993},
  publisher={Cambridge University Press}
}

@ARTICLE{Hallet1990-qa,
  title     = "Spatial self-organization in geomorphology: from periodic
               bedforms and patterned ground to scale-invariant topography",
  author    = "Hallet, B",
  journal   = "Earth Sci. Rev.",
  publisher = "Elsevier BV",
  volume    =  29,
  number    = "1-4",
  pages     = "57--75",
  month     =  oct,
  year      =  1990,
  language  = "en"
}

@article{moefillivlev,
  title = {Complex plasmas: An interdisciplinary research field},
  author = {Morfill, Gregor E. and Ivlev, Alexei V.},
  journal = {Rev. Mod. Phys.},
  volume = {81},
  issue = {4},
  pages = {1353--1404},
  numpages = {0},
  year = {2009},
  month = {Oct},
  publisher = {American Physical Society},
  doi = {10.1103/RevModPhys.81.1353},
}

@article{Helbing2001traffic,
  title = {Traffic and related self-driven many-particle systems},
  author = {Helbing, Dirk},
  journal = {Rev. Mod. Phys.},
  volume = {73},
  issue = {4},
  pages = {1067--1141},
  numpages = {0},
  year = {2001},
  month = {Dec},
  publisher = {American Physical Society},
  doi = {10.1103/RevModPhys.73.1067},
}

@book{Epstein_Pojman_1998, place={New York}, title={An introduction to nonlinear chemical dynamics: Oscillations, waves, patterns, and chaos}, publisher={Oxford University Press}, author={Epstein, Irving R. and Pojman, John A.}, year={1998}}

@article{Whitesides2002SelfAssemblyAA,
  title={Self-Assembly at All Scales},
  author={George M. Whitesides and Bartosz A. Grzybowski},
  journal={Science},
  year={2002},
  volume={295},
  pages={2418 - 2421},
}

@article{Tompkins2014,
author = {Tompkins, Nathan and Li, Ning and Girabawe, Camille and Heymann, Michael and Ermentrout, Bard and Epstein, Irving and Fraden, Seth},
year = {2014},
month = {03},
pages = {},
title = {Testing Turing's theory of morphogenesis in chemical cells},
volume = {111},
journal = {Proceedings of the National Academy of Sciences of the United States of America},
doi = {10.1073/pnas.1322005111}
}

@article{BZRinLPD,
author = {Marchettini, Nadia and Ristori, Sandra and Rossi, Federico and Rustici, Mauro},
year = {2013},
month = {01},
pages = {55-63},
title = {An Experimental Model for Mimicking Biological Systems: The Belousov–Zhabotinsky Reaction in Lipid Membranes},
volume = {1},
journal = {International Journal of Design \& Nature and Ecodynamics},
doi = {10.2495/ECO-V1-N1-55-63}
}

@Article{BZR,
author={Field, Richard J.
and Koros, Endre
and Noyes, Richard M.},
title={Oscillations in chemical systems. II. Thorough analysis of temporal oscillation in the bromate-cerium-malonic acid system},
journal={Journal of the American Chemical Society},
year={1972},
month={Dec},
day={01},
publisher={American Chemical Society},
volume={94},
number={25},
pages={8649-8664},
issn={0002-7863},
doi={10.1021/ja00780a001},
}

@article{sato2001dynamics,
  title={Dynamics of fine particles in magnetized plasmas},
  author={Sato, Noriyoshi and Uchida, Giichiro and Kaneko, Toshiro and Shimizu, Shinya and Iizuka, Satoru},
  journal={Physics of Plasmas},
  volume={8},
  number={5},
  pages={1786--1790},
  year={2001},
  publisher={American Institute of Physics}
}

@article{KATARIYA2025134692,
title = {Diffusive transport of a 2-D magnetized dusty plasma cloud},
journal = {Physica D: Nonlinear Phenomena},
volume = {476},
pages = {134692},
year = {2025},
issn = {0167-2789},
doi = {https://doi.org/10.1016/j.physd.2025.134692},
url = {https://www.sciencedirect.com/science/article/pii/S0167278925001691},
author = {Aman Singh Katariya and Amita Das and Animesh Sharma and Bibhuti Bhusan Sahu},
}

@article{deshwal2022chaotic,
  title={Chaotic dynamics of small-sized charged Yukawa dust clusters},
  author={Deshwal, Priya and Yadav, Mamta and Prasad, Chaitanya and Sridev, Shantam and Ahuja, Yash and Maity, Srimanta and Das, Amita},
  journal={Chaos: An Interdisciplinary Journal of Nonlinear Science},
  volume={32},
  number={6},
  year={2022},
  publisher={AIP Publishing}
}

@article{plimpton1995fast,
  title={Fast parallel algorithms for short-range molecular dynamics},
  author={Plimpton, Steve},
  journal={Journal of computational physics},
  volume={117},
  number={1},
  pages={1--19},
  year={1995},
  publisher={Elsevier}
}

@article{nosenko2004shear,
  title={Shear flows and shear viscosity in a two-dimensional Yukawa system (dusty plasma)},
  author={Nosenko, V and Goree, J},
  journal={Physical review letters},
  volume={93},
  number={15},
  pages={155004},
  year={2004},
  publisher={APS}
}

@article{nose1984molecular,
  title={A molecular dynamics method for simulations in the canonical ensemble},
  author={Nos{\'e}, Sh{\=u}ichi},
  journal={Molecular physics},
  volume={52},
  number={2},
  pages={255--268},
  year={1984},
  publisher={Taylor \& Francis}
}

@article{PhysRevA.31.1695,
  title = {Canonical dynamics: Equilibrium phase-space distributions},
  author = {Hoover, William G.},
  journal = {Phys. Rev. A},
  volume = {31},
  issue = {3},
  pages = {1695--1697},
  numpages = {0},
  year = {1985},
  month = {Mar},
  publisher = {American Physical Society},
  doi = {10.1103/PhysRevA.31.1695},
}

@article{lai-lin,
  title = {Packings and defects of strongly coupled two-dimensional Coulomb clusters: Numerical simulation},
  author = {Lai, Ying-Ju and I, Lin},
  journal = {Phys. Rev. E},
  volume = {60},
  issue = {4},
  pages = {4743--4753},
  numpages = {0},
  year = {1999},
  month = {Oct},
  publisher = {American Physical Society},
  doi = {10.1103/PhysRevE.60.4743},
}

@article{astrakharchik1999properties,
  title={Properties of two-dimensional dusty plasma clusters},
  author={Astrakharchik, GE and Belousov, AI and Lozovik, Yu E},
  journal={Physics Letters A},
  volume={258},
  number={2-3},
  pages={123--130},
  year={1999},
  publisher={Elsevier}
}

@article{lisina2019dynamic,
  title = {Amplitude instability, phase transitions, and dynamic properties in two-dimensional Yukawa clusters},
  author = {Lisina, I. I. and Vaulina, \cyrchar\CYRO{}. S. and Lisin, \cyrchar\CYRE{}. \cyrchar\CYRA{}.},
  journal = {Phys. Rev. E},
  volume = {99},
  issue = {1},
  pages = {013207},
  numpages = {8},
  year = {2019},
  month = {Jan},
  publisher = {American Physical Society},
  doi = {10.1103/PhysRevE.99.013207}
}

@article{lowen1994charged,
  title={Charged rodlike colloidal suspensions: An ab initio approach},
  author={L{\"o}wen, Hartmut},
  journal={The Journal of chemical physics},
  volume={100},
  number={9},
  pages={6738--6749},
  year={1994},
  publisher={American Institute of Physics}
}

@article{molotkov2000liquid,
  title={Liquid plasma crystal: Coulomb crystallization of cylindrical macroscopic grains in a gas-discharge plasma},
  author={Molotkov, VI and Nefedov, AP and Pustyl'nik, M Yu and Torchinsky, VM and Fortov, VE and Khrapak, AG and Yoshino, K},
  journal={Journal of Experimental and Theoretical Physics Letters},
  volume={71},
  pages={102--105},
  year={2000},
  publisher={Springer}
}

@article{annaratone2001levitate,
  title = {Levitation of cylindrical particles in the sheath of an rf plasma},
  author = {Annaratone, B. M. and Khrapak, A. G. and Ivlev, A. V. and S\"ollner, G. and Bryant, P. and S\"utterlin, R. and Konopka, U. and Yoshino, K. and Zuzic, M. and Thomas, H. M. and Morfill, G. E.},
  journal = {Phys. Rev. E},
  volume = {63},
  issue = {3},
  pages = {036406},
  numpages = {6},
  year = {2001},
  month = {Feb},
  publisher = {American Physical Society},
  doi = {10.1103/PhysRevE.63.036406},
}

@article{ivlev2003rodtheory,
  title = {Rodlike particles in gas discharge plasmas: Theoretical model},
  author = {Ivlev, A. V. and Khrapak, A. G. and Khrapak, S. A. and Annaratone, B. M. and Morfill, G. and Yoshino, K.},
  journal = {Phys. Rev. E},
  volume = {68},
  issue = {2},
  pages = {026403},
  numpages = {10},
  year = {2003},
  month = {Aug},
  publisher = {American Physical Society},
  doi = {10.1103/PhysRevE.68.026403},
  }

@article{lisina2016spatial,
  title={Spatial configurations of charged rod-like particles in external electric field},
  author={Lisina, I and Lisin, E and Vaulina, O},
  journal={Physics of Plasmas},
  volume={23},
  number={3},
  year={2016},
  publisher={AIP Publishing}
}

@article{vaulina2016formation,
  title={Formation of ordered structures in systems of charged thin cylindrical grains},
  author={Vaulina, OS and Lisina, II and Lisin, EA},
  journal={Plasma Physics Reports},
  volume={42},
  pages={135--146},
  year={2016},
  publisher={Springer}
}

@article{vekselman2018quantitative,
  title={Quantitative imaging of carbon dimer precursor for nanomaterial synthesis in the carbon arc},
  author={Vekselman, V and Khrabry, A and Kaganovich, I and Stratton, B and Selinsky, RS and Raitses, Y},
  journal={Plasma Sources Science and Technology},
  volume={27},
  number={2},
  pages={025008},
  year={2018},
  publisher={IOP Publishing}
}

@article{krajnovich1995laserc2,
  title={Laser sputtering of highly oriented pyrolytic graphite at 248 nm},
  author={Krajnovich, Douglas J},
  journal={The Journal of chemical physics},
  volume={102},
  number={2},
  pages={726--743},
  year={1995},
  publisher={American Institute of Physics}
}

@article{yamagata1999optical,
  title={Optical emission study of ablation plasma plume in the preparation of diamond-like carbon films by KrF excimer laser},
  author={Yamagata, Y and Sharma, A and Narayan, J and Mayo, RM and Newman, JW and Ebihara, K},
  journal={Journal of Applied Physics},
  volume={86},
  number={8},
  pages={4154--4159},
  year={1999},
  publisher={American Institute of Physics}
}

@article{nica2020formation,
  title={Formation mechanisms of carbon dimer in excimer laser produced plasma},
  author={Nica, Petru-Edward and Ursu, Cristian},
  journal={The European Physical Journal D},
  volume={74},
  pages={1--7},
  year={2020},
  publisher={Springer}
}

@article{oohara2003pair,
  title={Pair-ion plasma generation and fullerene-dimer formation},
  author={Oohara, Wataru and Hatakeyama, Rikizo},
  journal={Thin Solid Films},
  volume={435},
  number={1-2},
  pages={280--284},
  year={2003},
  publisher={Elsevier}
}

@article{wang1997synthesis,
  title={Synthesis and X-ray structure of dumb-bell-shaped C120},
  author={Wang, Guan-Wu and Komatsu, Koichi and Murata, Yasujiro and Shiro, Motoo},
  journal={Nature},
  volume={387},
  number={6633},
  pages={583--586},
  year={1997},
  publisher={Nature Publishing Group UK London}
}

@article{shvartsburg1999ball,
  title={Ball-and-chain dimers from a hot fullerene plasma},
  author={Shvartsburg, Alexandre A and Hudgins, Robert R and Gutierrez, Rafael and Jungnickel, Gerd and Frauenheim, Thomas and Jackson, Koblar A and Jarrold, Martin F},
  journal={The Journal of Physical Chemistry A},
  volume={103},
  number={27},
  pages={5275--5284},
  year={1999},
  publisher={ACS Publications}
}

@article{hippler2017pressure,
  title={Pressure dependence of Ar $\{$$\backslash$hspace $\{$0pt$\}$$\}$ \_2\^{}+, ArTi+, and Ti $\{$$\backslash$hspace $\{$0pt$\}$$\}$ \_2\^{}+ dimer formation in a magnetron sputtering discharge},
  author={Hippler, R and Cada, M and Stranak, V and Hubicka, Z and Helm, CA},
  journal={Journal of Physics D Applied Physics},
  volume={50},
  number={44},
  pages={445205},
  year={2017}
}

@article{bogaerts1999role,
  title={Role of Ar2+ and Ar2+ ions in a direct current argon glow discharge: A numerical description},
  author={Bogaerts, Annemie and Gijbels, Renaat},
  journal={Journal of applied physics},
  volume={86},
  number={8},
  pages={4124--4133},
  year={1999},
  publisher={AIP Publishing}
}

@article{curda2023role,
  title={The role of dimers in the efficient growth of nanoparticles},
  author={Curda, Pavel and Hippler, Rainer and Cada, Martin and Kyli{\'a}n, Ond{\v{r}}ej and Stranak, Vitezslav and Hubicka, Zdenek},
  journal={Surface and Coatings Technology},
  volume={473},
  pages={130045},
  year={2023},
  publisher={Elsevier}
}

@article{hippler2018formation,
  title={Formation of $\{$Cu$\}$$\$ \_ $\{$n$\}$\^{}$\{$+$\}$(n= 1-3),$\{$Ar$\}$$\$ \_ $\{$n$\}$\^{}$\{$+$\}$(n= 1, 2), and ArCu+ ions during sputtering of a copper surface by low-energy Ar+ ion bombardment in a dilute argon atmosphere},
  author={Hippler, Rainer and Denker, Christian},
  journal={Plasma Sources Science Technology},
  volume={27},
  number={6},
  pages={065010},
  year={2018}
}

@article{praburam1995cosmic,
  title={Cosmic dust synthesis by accretion and coagulation},
  author={Praburam, G and Goree, J},
  journal={Astrophysical Journal, Part 1 (ISSN 0004-637X), vol. 441, no. 2, p. 830-838},
  volume={441},
  pages={830--838},
  year={1995}
}

@article{ivlev2000anisotropic,
  title={Anisotropic dust lattice modes},
  author={Ivlev, AV and Morfill, G},
  journal={Physical Review E},
  volume={63},
  number={1},
  pages={016409},
  year={2000},
  publisher={APS}
}

@article{thomas1996melting,
  title={Melting dynamics of a plasma crystal},
  author={Thomas, Hubertus M and Morfill, Gregor E},
  journal={Nature},
  volume={379},
  number={6568},
  pages={806--809},
  year={1996},
  publisher={Nature Publishing Group UK London}
}

@article{maity2019molecular,
  title={Molecular dynamics study of crystal formation and structural phase transition in Yukawa system for dusty plasma medium},
  author={Maity, Srimanta and Das, Amita},
  journal={Physics of Plasmas},
  volume={26},
  number={2},
  year={2019},
  publisher={AIP Publishing}
}

@article{maity2020dynamical,
  title={Dynamical states in two-dimensional charged dust particle clusters in plasma medium},
  author={Maity, Srimanta and Deshwal, Priya and Yadav, Mamta and Das, Amita},
  journal={Physical Review E},
  volume={102},
  number={2},
  pages={023213},
  year={2020},
  publisher={APS}
}

@article{hayashi1994observation,
  title={Observation of Coulomb-crystal formation from carbon particles grown},
  author={Hayashi, Yasuaki and Tachibana, Kunihide},
  journal={Jpn. J. Appl. Phys., Part},
  volume={2},
  number={33},
  pages={L804--L806},
  year={1994}
}

@article{melzer2021physics,
  title={Physics of magnetized dusty plasmas},
  author={Melzer, Andre and Kr{\"u}ger, H and Maier, D and Sch{\"u}tt, S},
  journal={Reviews of Modern Plasma Physics},
  volume={5},
  number={1},
  pages={11},
  year={2021},
  publisher={Springer}
}

@article{melzer2019cluster,
    author = {Melzer, A. and Krüger, H. and Schütt, S. and Mulsow, M.},
    title = {Finite dust clusters under strong magnetic fields},
    journal = {Physics of Plasmas},
    volume = {26},
    number = {9},
    pages = {093702},
    year = {2019},
    month = {09},
    issn = {1070-664X},
    doi = {10.1063/1.5116523},
}

@article{Feng2016,
    author = {Feng, Yan and Lin, Wei and Li, Wei and Wang, Qiaoling},
    title = {Equations of state and diagrams of two-dimensional liquid dusty plasmas},
    journal = {Physics of Plasmas},
    volume = {23},
    number = {9},
    pages = {093705},
    year = {2016},
    month = {09},
    issn = {1070-664X},
    doi = {10.1063/1.4962685},
}

@article{thomas1994plasma,
  title={Plasma crystal: Coulomb crystallization in a dusty plasma},
  author={Thomas, H and Morfill, GE and Demmel, V and Goree, J and Feuerbacher, B and M{\"o}hlmann, D},
  journal={Physical Review Letters},
  volume={73},
  number={5},
  pages={652},
  year={1994},
  publisher={APS}
}

@article{chu1994direct,
  title={Direct observation of Coulomb crystals and liquids in strongly coupled rf dusty plasmas},
  author={Chu, JH and Lin, I},
  journal={Physical review letters},
  volume={72},
  number={25},
  pages={4009},
  year={1994},
  publisher={APS}
}

@article{feng2010viscoelasticity,
  title={Viscoelasticity of 2D liquids quantified in a dusty plasma experiment},
  author={Feng, Yan and Goree, J and Liu, Bin},
  journal={Physical review letters},
  volume={105},
  number={2},
  pages={025002},
  year={2010},
  publisher={APS}
}

@article{yadav2023structure,
  title={Structure formation by electrostatic interactions in strongly coupled medium},
  author={Yadav, Mamta and Deshwal, Priya and Maity, Srimanta and Das, Amita},
  journal={Physical Review E},
  volume={107},
  number={5},
  pages={055214},
  year={2023},
  publisher={APS}
}

@article{diedrich1987observation,
  title={Observation of a phase transition of stored laser-cooled ions},
  author={Diedrich, F and Peik, E and Chen, JM and Quint, W and Walther, H},
  journal={Physical review letters},
  volume={59},
  number={26},
  pages={2931},
  year={1987},
  publisher={APS}
}

@article{wineland1987atomic,
  title={Atomic-ion Coulomb clusters in an ion trap},
  author={Wineland, David J and Bergquist, JC and Itano, Wayne M and Bollinger, JJ and Manney, CH},
  journal={Physical review letters},
  volume={59},
  number={26},
  pages={2935},
  year={1987},
  publisher={APS}
}

@article{mortensen2006observation,
  title={Observation of three-dimensional long-range order in small ion coulomb crystals in an rf trap},
  author={Mortensen, Anders and Nielsen, E and Matthey, Thierry and Drewsen, Michael},
  journal={Physical review letters},
  volume={96},
  number={10},
  pages={103001},
  year={2006},
  publisher={APS}
}

@article{melzer2003mode,
  title={Mode spectra of thermally excited two-dimensional dust Coulomb clusters},
  author={Melzer, A},
  journal={Physical Review E},
  volume={67},
  number={1},
  pages={016411},
  year={2003},
  publisher={APS}
}

@article{PhysRevB.51.7700,
  title = {Spectral properties of classical two-dimensional clusters},
  author = {Schweigert, Vitaly A. and Peeters, Fran\ifmmode \mbox{\c{c}}\else \c{c}\fi{}ois M.},
  journal = {Phys. Rev. B},
  volume = {51},
  issue = {12},
  pages = {7700--7713},
  numpages = {0},
  year = {1995},
  month = {Mar},
  publisher = {American Physical Society},
  doi = {10.1103/PhysRevB.51.7700},
}

@article{astrakharchik1999two,
  title={Two-dimensional mesoscopic dusty plasma clusters: structure and phase transitions},
  author={Astrakharchik, GE and Belousov, AI and Lozovik, Yu E},
  journal={Journal of Experimental and Theoretical Physics},
  volume={89},
  pages={696--703},
  year={1999},
  publisher={Springer}
}

@article{YADAV2024134326,
title = {Evolution of shielding cloud under oscillatory external forcing in strongly coupled ultracold neutral plasma},
journal = {Physica D: Nonlinear Phenomena},
volume = {469},
pages = {134326},
year = {2024},
issn = {0167-2789},
doi = {https://doi.org/10.1016/j.physd.2024.134326},
author = {Mamta Yadav and Aman Singh Katariya and Animesh Sharma and Amita Das},
}

@article{Konopka2000Measurement,
  author    = {Konopka, U. and Morfill, G. E. and Ratke, L.},
  title     = {Measurement of the Interaction Potential of Microspheres in a Plasma},
  journal   = {Physical Review Letters},
  year      = {2000},
  volume    = {84},
  number    = {5},
  pages     = {891--894},
  doi       = {10.1103/PhysRevLett.84.891}
}

@book{ShuklaMamun2002Intro,
  author    = {Shukla, P. K. and Mamun, A. A.},
  title     = {Introduction to Dusty Plasma Physics},
  publisher = {Institute of Physics Publishing},
  year      = {2002},
  address   = {Bristol and Philadelphia},
  doi       = {10.1201/9781420033588}
}
\end{document}